\documentclass[mnsc,nonblindrev]{informs3}
\usepackage[ruled]{algorithm2e} 

\usepackage[round]{natbib}
\usepackage{graphicx}

\usepackage{hyperref}
\usepackage{hyperref}
\definecolor{darkred}{RGB}{150,0,0}
\definecolor{darkgreen}{RGB}{0,150,0}
\definecolor{darkblue}{RGB}{0,0,200}
\hypersetup{colorlinks=true, linkcolor=darkred, citecolor=darkgreen, urlcolor=darkblue}
\usepackage{amssymb}

\usepackage{mathtools}

\usepackage[ruled]{algorithm2e}

\newtheorem{assumption}{Assumption}

\newtheorem{corollary}{Corollary}
\newtheorem{definition}{Definition}
\newtheorem{proposition}{Proposition}
\newtheorem{theorem}{Theorem}
\newtheorem{lemma}{Lemma}

\usepackage[ruled]{algorithm2e}

\usepackage{caption}
\usepackage{subcaption}

\newcommand {\Cost}{\textsc{cost}}
\newcommand {\MSE}{\textsc{MSE}}

\newcommand {\Var}{\textsc{var}}

\usepackage{bm}

\usepackage{accents}

\newcommand{\R}{{\mathbb{R}}}
\newcommand{\E}{\mathbb{E}}
\newcommand{\Pb}{\mathbb{P}}

\newcommand{\matA}{\mathcal{A}}

\newcommand{\matC}{\mathcal{C}}

\newcommand{\matF}{\mathcal{F}}

\newcommand{\matL}{\mathcal{L}}
\newcommand{\matN}{\mathcal{N}}
\newcommand{\matM}{\mathcal{M}}
\newcommand{\matP}{\mathcal{P}}
\newcommand{\matQ}{\mathcal{Q}}

\newcommand{\matS}{\mathcal{S}}

\newcommand{\matW}{\mathcal{W}}
\newcommand{\matX}{\mathcal{X}}

\newcommand{\diff}{\varepsilon}
\newcommand{\Lap}{\text{Laplace}}

\newcommand{\htheta}{\hat{\theta}}
\newcommand{\TV}{\text{TV}}
\newcommand{\KL}{\text{KL}}
\newcommand{\Central}{\mathrm{central}}
\newcommand{\Local}{\mathrm{local}}

\DeclarePairedDelimiter\floor{\lfloor}{\rfloor}

\usepackage{graphicx} %Loading the package
\graphicspath{{Common}} %Setting the graphicspat

\begin{document}

\defcitealias{AppleDP}{Apple}

\RUNAUTHOR{Fallah et al.}

\RUNTITLE{Optimal and Differentially Private Data Acquisition}

% \TITLE{Learning a common estimate via optimal differentially private data acquisition mechanisms: central vs. local architecture \thanks{We are grateful to Kunal Talwar for useful conversations and comments.}}

% \TITLE{Private data acquisition mechanisms for learning a common estimate: central vs. local architecture \thanks{We are grateful to Kunal Talwar for useful conversations and comments.}}

\TITLE{Optimal and Differentially Private Data Acquisition: Central and Local Mechanisms}

%\TITLE{Bayesian Incentive Compatible Data Acquisition Mechanism: Central and Local differential privacy\thanks{We are grateful to Kunal Talwar for useful conversations and comments.}}

% \TITLE{Optimal and differentially private data acquisition from strategic users: central and local mechanisms\thanks{We are grateful to Kunal Talwar for useful conversations and comments.}}

% \TITLE{Optimal data acquisition from strategic users: central versus local differential privacy\thanks{We are grateful to Kunal Talwar for useful conversations and comments.}}

\ARTICLEAUTHORS{%

		\AUTHOR{Alireza Fallah}
		\AFF{MIT, EECS,
			E-mail:  \URL{afallah@mit.edu}}
		\AUTHOR{Ali Makhdoumi}
		\AFF{Fuqua School of Business,
			E-mail:  \URL{ali.makhdoumi@duke.edu}}
			\AUTHOR{Azarakhsh Malekian}
		\AFF{Rotman School of Management,
			E-mail:  \URL{azarakhsh.malekian@rotman.utoronto.ca}}
			\AUTHOR{Asuman Ozdaglar}
		\AFF{MIT, EECS,
			E-mail:  \URL{asuman@mit.edu}}
} 

\ABSTRACT{
We consider a platform's problem of collecting data from privacy sensitive users to estimate an underlying parameter of interest. We formulate this question as a Bayesian-optimal mechanism design problem, in which an individual can share her (verifiable) data in exchange for a monetary reward or services, but at the same time has a (private) heterogeneous privacy cost which we quantify using differential privacy. We consider two popular differential privacy settings for providing privacy guarantees for the users: central and local. In both settings, we establish minimax lower bounds for the estimation error and derive (near) optimal estimators for given heterogeneous privacy loss levels for users. Building on this characterization, we pose the mechanism design problem as the optimal selection of an estimator and payments that will elicit truthful reporting of users' privacy sensitivities. Under a regularity condition on the distribution of privacy sensitivities we develop efficient algorithmic mechanisms to solve this problem in both privacy settings. Our mechanism in the central setting can be implemented in time $\mathcal{O}(n \log n)$ where $n$ is the number of users and our mechanism in the local setting admits a Polynomial Time Approximation Scheme (PTAS).
}

\maketitle
%%%%%%
\section{Introduction}

The data of billions of people around the world are used every day for improving search algorithms, recommendations on online platforms, personalized advertising, and the design of new drugs, services and products. With rapid advances in machine learning (ML) algorithms and further growth in data collection, these practices will become only more widespread in the years to come. A common concern with many of these data-intensive applications centers on privacy --- as a user's data is harnessed, more and more information about her behavior and preferences are uncovered and potentially utilized by platforms and advertisers.

A popular solution to the tension between privacy costs and benefits of data is to use methods such as differential privacy in order to limit the extent to which an individual's data is uncovered and exploited. The basic idea of differential privacy is to provide an upper bound on how sensitive the output of an algorithm (e.g., the vector of recommendations from an online site) is to an individual's data. Although differential privacy methods are already used by many of the tech companies, including, Apple, Google and Microsoft (see, e.g., \cite{erlingsson2014rappor} and \cite{ding2017collecting}), a key practical question remains: how do we decide how much privacy an individual will obtain? Imagine, for example, that two individuals have similar data, but one is very privacy conscious, while the other one does not think that she has any concerns of privacy. It is natural to provide different privacy levels for these two individuals when acquiring their data, but exactly how?

This paper is an attempt to answer this key question and study the impact of data market architecture on the design of mechanisms for purchasing data from privacy sensitive strategic users.
We consider a platform interested in estimating an underlying parameter using data collected from users. While users benefit from the outcome of the estimation, they are cognizant of the privacy losses they will incur and hence might be discouraged from sharing their data. User data come from some underlying population distribution where its mean is given by the parameter of interest.
We formulate this question as a mechanism design problem, in which an individual can share her data in exchange for a monetary reward or services, but at the same time has a heterogeneous privacy sensitivity that represents her cost per unit privacy loss. 
We assume a known prior on user's privacy sensitivity (which is independent of the data distribution).
While an individual's data is difficult to manipulate, her privacy preferences are easier to falsify (if monetary rewards were increasing in how privacy conscious individual is, then she might prefer to misrepresent this information). Individuals participate in the mechanism by reporting their privacy sensitivities and sharing their data. This mechanism simultaneously determines an “optimal” estimator, compensation for the users, and privacy losses an individual will incur. Thus, the mechanism endogenously determines the privacy loss levels as a function of both user sensitivities and also how their data is used in the estimation problem of the platform.

We consider two popular differential privacy settings for providing privacy guarantees for the users: central and local. In the central privacy setting, we require the output of the estimation process to be differentially private with respect to each individual's data. In the local privacy setting, we impose a differential privacy requirement with respect to the individual data of each user. Before formulating the optimal mechanism design problem, we derive optimal estimators for given heterogeneous privacy loss levels for users in the two privacy settings. We establish minimax lower bounds for the estimation error and use these bounds to characterize the form of the optimal estimator with central and local privacy guarantees. In particular,  in the central setting we show that, for a given vector of privacy losses, a linear estimator that combines a (properly designed) weighted average of the users' data points and a Laplace noise achieves the (near) optimal estimation error among all estimators that can achieve the desired privacy losses. In addition, in the local setting, we show that, for a given vector of privacy losses, first adding a Laplace noise to the data of each user and then taking a weighted average of the users' data points achieves the  optimal estimation error.

In the second part of the paper, we formulate the Bayesian-optimal mechanism design problem where the objective of the platform is to minimize the sum of the estimation error and total payment for the users. We first provide a characterization of the optimal payment as a function of the reported privacy sensitivities. This is closely related to the payment identity in Myerson's optimal auction design problem (\cite{myerson1981optimal}), but differs in that the reported privacy sensitivities of other users impacts a user's utility not only through her privacy loss level and payment but also through the overall estimation error (all users benefit from a lower estimation error). We then focus our attention to linear estimators (which were shown to be optimal for differentially private estimation given exogenous privacy loss levels). We show that under some regularity conditions on the distribution of privacy sensitivities, the problem of finding the optimal privacy levels can be cast as the solution to a non-convex optimization problem. 
In both settings, we first reformulate the platform's problem in terms of designing a pair of weight and privacy loss functions. These functions map the vector of reported privacy sensitivities to a vector of privacy losses for users and a vector of weights in the linear estimator of the platform, respectively. In the central setting, we use the structure of the problem to derive an efficient score-based algorithm for implementing our mechanism in time $\mathcal{O}(n\log n)$. In the local setting, we develop a Polynomial Time Approximation Scheme (PTAS) to solve the platform's problem.

In the last section, we compare the central and local differential privacy settings and establish that the platform achieves a (weakly) higher utility in the central privacy setting than that in the local one. 
This is because the local setting
provides a stronger privacy guarantee and hence increases the final estimator's variance, which in turn reduces the platform's utility. We also illustrate that, for a given vector of privacy sensitivities, the privacy loss level allocated to a user in the optimal local data acquisition mechanism is not necessarily higher than the central setting, and in fact, it can be strictly lower (providing better privacy guarantees).

From a technical point of view, our first technical contribution is deriving the minimax optimal private mean estimator for heterogeneous differential privacy levels. Prior to our work, the optimal estimation has been studied only for homogeneous differential privacy levels (see, e.g., \cite{duchi2013local, dwork2014algorithmic,  barber2014privacy}). Utilizing this optimal estimator, we demonstrate how existing mechanism design tools can be applied to our setting, resulting in a point-wise optimization approach using virtual values to find the optimal mechanism. Our second technical contribution involves developing efficient algorithms specifically tailored for solving non-convex point-wise optimization problems that arise in private data acquisition. This differs from the conventional mechanism design setting, where the optimal mechanism can be obtained by solving a linear program. In terms of the structural aspect, our problem deviates from the classic mechanism design, where the optimal allocation typically follows a threshold rule. Instead, in our problem, the optimal differential privacy level exhibits a continuous dependence on privacy sensitivity.

%%%%
\subsection{Related Literature}
\sloppy
Our paper builds on the growing literature on optimal data acquisition from strategic privacy conscious users. Several of these papers use differential privacy to quantify the cost users incur when sharing their data \cite{ghosh2011selling}, \cite{nissim2012privacy}, \cite{nissim2014redrawing}. A pioneering paper in this literature is \cite{ghosh2011selling}, which consider designing a mechanism for collecting data from users that explicitly  experience a cost for privacy loss. \cite{ghosh2011selling} assume that each user has a private bit and a heterogeneous privacy loss parameter and the platform's goal is to estimate the sum of user's data by using a differentially private and dominant strategy truthful mechanism. This paper considers both the case when the user data and privacy parameter are independent (as in our case) and when they are correlated. For the independent case, their mechanism results in providing a single privacy level to all users whose data are collected (because of their worst case view with a focus on dominant strategy truthful mechanisms and lack of distributional assumptions on user data or cost parameters).
For the correlated case, \cite{ghosh2011selling} provide an impossibility result for the existence of a truthful and individually rational mechanism. Several papers build on \cite{ghosh2011selling}, extending it to take it or leave it offers \cite{ligett2012take}, and strengthening the impossibility results \cite{nissim2014redrawing}.    

Another line of work tackles the open question posed by \cite{ghosh2011selling} on whether a model with distributional assumption on users' costs and Bayesian mechanism design approach could be used to develop optimal mechanism for collecting data with privacy guarantees. \cite{roth2012conducting}, \cite{chen2018optimal}, and \cite{chen2019prior} followed this approach using a randomized mechanism in which user's data is used with a probability that depends on the reported privacy costs of the users.\footnote{This is different from our mechanism in which payments and resulting privacy losses depend on the reported privacy sensitivity of all users.} These papers do not use differential privacy to model privacy costs, but rather use a menu of probability-price pairs to control the privacy loss and compensation for each user.   

Another noteworthy paper in this literature is \cite{cummings2015accuracy}, which consider data purchase from users that provide different levels of data accuracy (variance) and may strategically price access to their data. The variance in the data can represent uncertainty in data quality or intentionally added noise in order to guarantee privacy. This paper does not impose a functional form for the privacy loss in terms of a differential privacy parameter and instead allows for a flexibility in offering a menu of different variance levels (or equivalently, arbitrary costs for each level independently). 

Our paper differs from these works by assuming prior information on user privacy sensitivities, and focusing on characterizing the optimal Bayesian 
incentive compatible mechanism. We further assume that user data are drawn from the same underlying distribution. This allows the platform to put more weight on the data of a user with lower price sensitivity, leading to different privacy levels for participating users. Another important distinction of our model is our assumption that users derive utility from the accuracy of the estimation outcome which changes the privacy allocation of the optimal mechanism. Finally our paper considers different privacy architectures, central and local, and explores the different privacy guarantees provided by an optimal mechanism under these different architectures. Prior to our work \cite{Juba2021} has considered a setting in which the users benefit from a better estimation outcome. They consider a linear estimator with Laplace additive noise and show how it allows for heterogeneous privacy guarantees to different users in the central model. We depart from this paper by establishing the (near) optimal estimator, considering strategic users in reporting their privacy costs, and studying both central and local settings and their comparison
(see also \cite{pai2013privacy} for a survey).

In our paper, as well as the above papers, the platform can verify the data of users. A different stream of this literature considers a setting in which individuals have the ability to misreport their information  \cite{perote2003impossibility}, \cite{dekel2010incentive},  \cite{meir2012algorithms}, \cite{ghosh2014buying}, \cite{cai2015optimum},  \cite{liu2016learning, liu2017sequential}.

Our paper also relates to the literature that consider privacy aware mechanism design and selling strategies such as \cite{mcsherry2007mechanism}, \cite{nissim2012privacy},  \cite{abernethy2019learning},  \cite{lei2020privacy}, \cite{chen2021differential}, and \cite{chen2021privacy}. In particular, \cite{chen2021differential} consider a dynamic
personalized pricing problem with unknown nonparametric demand models under data privacy protection, while \cite{lei2020privacy}, \cite{chen2021differential}, and \cite{han2021generalized} consider parametric demand models. We note that both our research question and results are different from these papers. In particular, we study the design of optimal mechanisms for collecting data from strategic users with privacy concerns while these papers consider demand learning for personalized pricing under privacy concerns in an online learning framework and provide (tight) bounds on the regret of the optimal algorithm.

Finally, our paper relates to the literature that studies the problem of choosing the proper level of differential privacy given the goal of protecting individuals' privacy such as \cite{lee2011much}, \cite{hsu2014differential}, and \cite{mehner2021towards}. We depart from this line of work by studying the endogenous choice of differential privacy levels based on individuals' privacy sensitivity and their interactions with a platform.

More broadly, various other issues of information/data markets have been studied in the literature. In particular, \cite{horner2016selling} study the design of mechanisms for selling data and \cite{goldfarb2011online},  \cite{bergemann2015selling}, \cite{montes2019value}, and \cite{jagabathula2020inferring} study the improvements in resource allocation using personal information. The correlation among users' data and its impact on the price of data has been studied in \cite{liao2018social}, \cite{fainmesser2019digital},  \cite{acemoglu2019too}, \cite{ichihashi2021economics}, and \cite{liao2021privacy}. The impact of data tracking on the firms' competition and users has been studied in \cite{bimpikis2021data} and \cite{gur2019disclosure}. The dynamic sale of data has been studied in \cite{immorlica2021buying} and the difference between static and dynamic mechanisms has been studied in \cite{babaioff2012optimal} and \cite{drakopoulos2020providing}. \cite{bergemann2015selling} study the problem of selling cookies for targeted advertisement and study how the price of data changes with the reach of the dataset and the fragmentation
of data sales. In addition, a collection of papers, such as \cite{li2002information}, \cite{li2008confidentiality}, \cite{ha2008contracting},  \cite{shang2015information}, \cite{foster2016learning}, \cite{lobel2017optimal},  \cite{bimpikisinformation}, \cite{candogan2020optimal}, \cite{immorlica2020does},  \cite{hu2020privacy}, \cite{ashlagi2020clearing},  \cite{anunrojwong2021information}, \cite{besbes2021big}, and \cite{ashlagi2021optimal} study information-sharing in the design and analysis of markets (see \cite{bergemann2019markets} for a survey).

Finally, our paper relates to the literature on differential privacy. Initiated by the work of \cite{dwork2006our, dwork2006calibrating}, differential privacy has emerged as a prevalent framework for characterizing the privacy leakage of data oriented algorithms. Our work, in particular, is related to the private mean estimation which has been studied extensively over the past decade \cite{duchi2013local}, \cite{barber2014privacy}, \cite{karwa2017finite}, \cite{9517999}, \cite{kamath2019privately, kamath2020private}, \cite{cummings2021mean}, \cite{acharya2021differentially}.

The rest of the paper proceeds as follows. Section \ref{sec:Environment} presents the setting, describes central and local differential privacy, and provides near optimal minimax estimator with heterogeneous privacy losses. In Section \ref{sec:mechanism_formulation}, we establish how the platform's mechanism design problem turns into a point-wise optimization problem over the privacy losses. In Section \ref{sec:central}, we characterize the optimal privacy loss levels in the central privacy setting and find a polynomial time algorithm to find them. In Section \ref{sec:local}, we characterize the optimal privacy losses in the local privacy setting and establish that it admits a PTAS. Section \ref{sec:compare} compares the central and the local privacy settings. Section \ref{sec:conclusion} concludes, while the appendix includes the omitted proofs from the text.   
%%%%%%
\section{Differential privacy and platform's estimation problem} \label{sec:Environment}
%%%%%%%%%%%%%%%%%%%%%%%%%%%%%%%%%%%%%%%%%%%%%%%%
%%%%%%%%%%%%%%%%%%%%%%%%%%%%%%%%%%%%%%%%%%%%%%%%
%%%%%%%%%%%%%%%%%%%%%%%%%%%%%%%%%%%%%%%%%%%%%%%%
%%%%%%%%%%%%%%%%%%%%%%%%%%%%%%%%%%%%%%%%%%%%%%%%
%\subsection{The analyst's estimation problem and utility}
%%%%%%%%%
We consider a \emph{platform} interested in estimating an underlying parameter $\theta \in \mathbb{R}$ by collecting relevant data from a set of \emph{user}s denoted by $\mathcal{N}=\{1, \dots, n\}$. Each user $i \in \mathcal{N}$ has some personal data $X_i \in \mathcal{X}$ which is informative about $\theta$. We assume that $X_i=\theta+ Z_i$, where $(Z_1, \dots, Z_n)$ are independent and identically distributed mean zero random variables with a variance denoted by $\Var$.\footnote{The assumption that $Z_i$'s are independent and have the same variance is reasonable in the context of estimation from a population and is made to simplify the notation and analysis. Our characterization of the optimal data acquisition mechanism readily extends to a setting with correlated users' data with different variances.} Throughout the paper, for simplicity, we assume $|Z_i| \leq \frac{1}{2}$ for all $i \in \matN$.\footnote{This is without loss of generality and the analysis extends to an arbitrary bound on $|Z_i|$'s by properly adjusting the estimator used by the platform.}

Users share their data with the platform since a more accurate estimate of parameter $\theta$ is useful for their objective (e.g., identifying a treatment from collecting individual medical records). However, sharing of individual data raises privacy concerns which users are cognizant of. Failure to address these privacy concerns would discourage users from sharing their data. We model the privacy demand of users as a maximum privacy loss they can tolerate. We use the notion of differential privacy to combine optimal estimation with such privacy guarantees.

In the next section, we assume the privacy loss level each user is willing to accept is given and derive (near-)optimal estimators that achieve these levels using different privacy guarantees. In particular, the central setting provides a privacy guarantee in terms of how user data impacts the final estimate of the platform, whereas the more restrictive local setting seeks a guarantee for the individual data shared by each user.  

In section \ref{sec:central}, we endogenize the choice of the privacy loss levels by assuming a privacy sensitive user utility.

%%%%%%%%%%%%%%%%%%%%%%%%%%%%%%%%%%%%%%%%%%%%%%%%
%%%%%%%%%%%%%%%%%%%%%%%%%%%%%%%%%%%%%%%%%%%%%%%%
%%%%%%%%%%%%%%%%%%%%%%%%%%%%%%%%%%%%%%%%%%%%%%%%
%%%%%%%%%%%%%%%%%%%%%%%%%%%%%%%%%%%%%%%%%%%%%%%%
\subsection{Central and local differential privacy} \label{sec:DP}

We first formalize the differential privacy framework we use to quantify guarantees on privacy demand of users. We focus on two settings, known as {\it central} and {\it local} differential privacy.  In the central case, we assume that the users trust the platform to share their data and require a privacy guarantee for user data by limiting its impact on the output of the analyst's estimation problem. In the local case, we assume a more restrictive privacy demand on the individual data shared by each user.

We start with the definition of central differential privacy which slightly generalizes the standard definition in \cite{dwork2006our, dwork2006calibrating} by allowing different levels of privacy loss for each user.\footnote{This extension is in line with the literature that introduced \textit{personalized} or \textit{heterogeneous} differential privacy, where each user can have a different privacy loss \cite{jorgensen2015conservative, alaggan2015heterogeneous, niu2021adapdp}. In our setting users have different preferences for their privacy which motivates our definition with heterogeneous privacy loss levels.}

%%%%%%%%%%%%%%%%%%%%%%%%%%%%%%%%%%%%%%%%%%%%%%%%
\begin{definition}[Central differential privacy] \label{definition:central_DP}
Let $\bm{\diff} = (\diff_i)_{i=1}^n \in \mathbb{R}_{+}^n$. Assume $\matS, \matS' \in \matX^n$ are two datasets that differ in the $i$-th component (which represents user $i$'s data). A randomized algorithm $\matA: \matX^n \to \R$ is $\bm{\diff}$-centrally differentially private if for all measurable sets $\matW$ in $\R$,
\begin{equation*}
\Pb(\matA(\matS) \in \matW) \leq e^{\diff_i} ~ \Pb(\matA(\matS') \in \matW).
\end{equation*}
\end{definition}
%%%%%%%%%%%%%%%%%%%%%%%%%%%%%%%%%%%%%%%%%%%%%%%%

This definition implies that the algorithm's output changes with probability at most $e^{\diff_i}$ when the data of user $i$ changes. In particular, Definition \ref{definition:central_DP} is equivalent to $ e^{-\diff_i}  \leq \frac{\Pb(\matA(\matS) \in \matW)}{\Pb(\matA(\matS') \in \matW)} \leq e^{\diff_i}$ for all $i \in \mathcal{N}$, neighboring $\matS, \matS' \in \matX^n$, which only differ in user $i$'s data, and all measurable sets $\matW$ in $\R$. Therefore, $\diff_i$ can be interpreted as a variable that captures the maximum {\it privacy loss} that Algorithm $\matA$ ensures for user $i$: the smaller $\diff_i$ is, the less of an impact user $i$'s data has on the output of Algorithm $\matA$, implying a lower privacy loss (or equivalently a higher privacy guarantee) for user $i$'s data.

Local differential privacy considers the setting where the users do not trust the platform with their data. The users therefore first produce a private version of their data through a mapping before sharing it with the platform. Building on the literature on differential privacy, we refer to this mapping as a \emph{channel} (see, e.g., \cite{duchi2013local}) and define it to be locally differentially private as follows.
 
%%%%%%%%%%%%%%%%%%%%%%%%%%%%%%%%%%%%%%%%%%%%%%%%
\begin{definition}[Local differential privacy] \label{definition:local_DP}
A randomized channel $\matC: \matX \to \R$ is $\diff$-locally differentially private if for any $x, x'\in \R$ and all measurable sets $\matW$ in $\R$,
\begin{equation*}
\Pb(\matC(x) \in \matW) \leq e^{\diff} \Pb(\matC(x') \in \matW).
\end{equation*}
Let $\bm{\diff} = (\diff_i)_{i=1}^n \in \mathbb{R}_{+}^n$. An algorithm $\matA: \matX^n \to \R$ is $(\diff_i)_{i=1}^n$-locally differentially private if it takes $(\matC_i({x}_i))_{i=1}^n$ as input (as opposed to $({x}_i)_{i=1}^n$ itself), where $\matC_i$ is an $\diff_i$-locally differentially private channel.
\end{definition}
%%%%%%%%%%%%%%%%%%%%%%%%%%%%%%%%%%%%%%%%%%%%%%%%
%%%%%%%%%%%%%%%%%%%%%%%%%%%%%%%%%%%%%%%%%%%%%%%%
\begin{figure}
     \centering
     \begin{subfigure}[b]{0.42\textwidth}
         \centering
         \includegraphics[width=\textwidth]{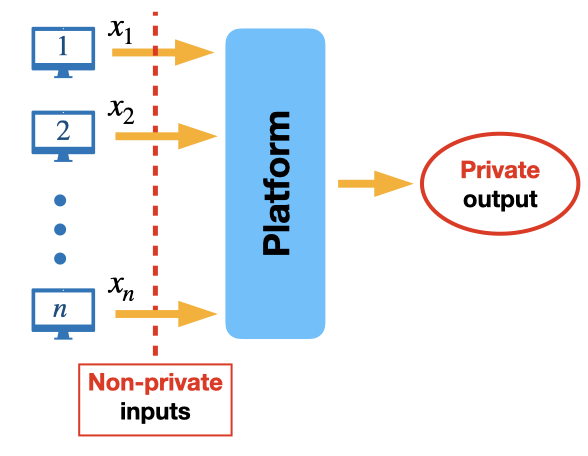}
         \caption{}
         \label{fig:central_alg}
     \end{subfigure}
     \hfill
     \begin{subfigure}[b]{0.56\textwidth}
         \centering
         \includegraphics[width=\textwidth]{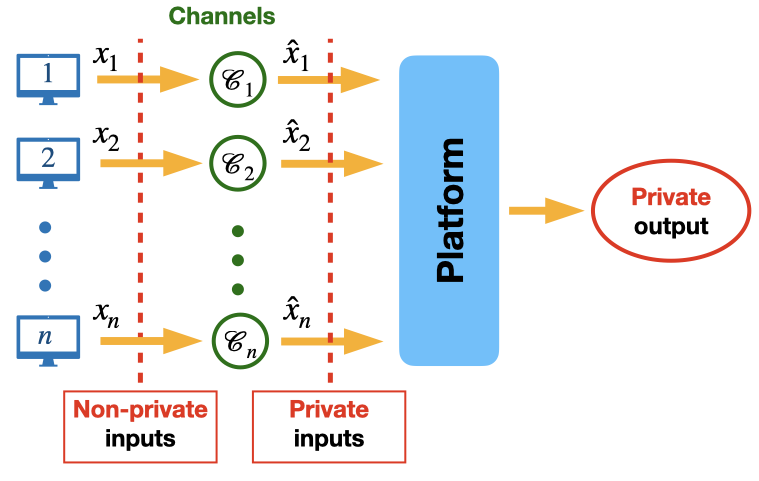}
         \caption{}
         \label{fig:local_alg}
     \end{subfigure}
     \hfill
     \label{fig:central_vs_local}
        \caption{(a) the central setting and (b) the local setting. In the local setting, in contrast to the central setting, the users privatize their data before sharing with the platform.}
\end{figure}
%%%%%%%%%%%%%%%%%%%%%%%%%%%%%%%%%%%%%%%%%%%%%%%%
%%%%%%%%%%%%%%%%%%%%%%%%%%%%%%%%%%%%%%%%%%%%%%%%

 It is worth noting that an $(\diff_i)_{i=1}^n$-locally differentially private algorithm is $(\diff_i)_{i=1}^n$-centrally differentially private according to Definition \ref{definition:central_DP} as well (see Observation 12.1 in \cite{dwork2014algorithmic}.) Figure \ref{fig:central_alg} and \ref{fig:local_alg} depict central and local differential privacy architectures, respectively.

%%%%

A point worth mentioning is that in the local privacy setting, the data is privatized directly on the user side, giving users control over its implementation. Unlike the central setting, the local setting does not rely on the platform credibly delivering the promised privacy level.

%%%%%%%%%%%%%%%%%%%%%%%%%%%%%%%%%%%%%%%%%%%%%%%%
In both central and local cases, the basic mechanism to ensure privacy is adding \textit{fine-tuned noise}. As we establish next, a Laplace mechanism which adds a zero-mean Laplace noise to the variable of interest is optimal, and therefore we adopt this throughout (we make the optimality statement precise in this section). Recall that the density of a mean-zero (one-dimensional) Laplace distribution with parameter $\eta$, denoted by $\Lap(\eta)$, is given by 
%%%%%%%
\begin{equation*}
p(z) = \frac{1}{2\eta}\exp(-|z|/\eta) \quad \text{ for all } z \in \R    
\end{equation*}
and its variance is given by $2 \eta^2$. The following lemma characterizes the differential privacy guarantees obtained by a Laplace mechanism.
%%%%%%%%%%%%%%%%%%%%%%%%%%%%%%%%%%%%%%%%%%%%%%%%
\begin{lemma}[\cite{dwork2014algorithmic}] \label{lemma:Laplace}
Consider a real-valued function $f:\matX^n \to \R$ and let $W$ be a Laplace noise with parameter $1/\diff$, i.e., $W \sim \Lap(1/\diff)$. Then, $\matA(\bm{x}) := f(\bm{x}) + W$ for any $\bm{x} \in \matX^n$, is $(\diff L_i(f))_{i=1}^n$-centrally differentially private, where $L_i(f)$ is the sensitivity of $f$ with respect to the $i$-th coordinate, and is given by
%%%%%%
\begin{equation} \label{eqn:sensitivity_def}
L_i(f) := \sup\left\{|f(\bm{x}) - f(\bm{x}')| : \text{for all } \bm{x},\bm{x}' \in \matX \text{ that only differ in the } i\text{-th coordinate} \right\}. 
\end{equation}
%%%%%%
\end{lemma}
%%%%%%%%%%%%%%%%%%%%%%%%%%%%%%%%%%%%%%%%%%%%%%%%
Next, we consider the following problem in both the central and the local differential privacy settings: Assume that the desired privacy level of users, i.e., $\diff_i$ for user $i$, is given to the platform. What is the optimal choice of the estimator in terms of expected square error? 

To answer this question, we first provide minimax lower bounds for private mean-estimation problem under both central and local definitions of differential privacy (given in Definitions \ref{definition:central_DP} and \ref{definition:local_DP}, respectively). We then prove that a linear estimator with Laplace mechanism achieves those lower bounds up to a logarithmic factor. While the private mean estimation problem has been extensively studied when privacy levels across all users are equal \cite{duchi2013local, dwork2014algorithmic,  barber2014privacy}, to the best of our knowledge, it has not been studied in our setting where the privacy levels of users are heterogeneous. 
%%%%%%%%%%%%%%%%%%%%%%%%%%%%%%%%%%%%%%%%%%%%%%%%
%%%%%%%%%%%%%%%%%%%%%%%%%%%%%%%%%%%%%%%%%%%%%%%%
%%%%%%%%%%%%%%%%%%%%%%%%%%%%%%%%%%%%%%%%%%%%%%%%
\subsection{(Near) Optimal estimation with central differential privacy} \label{sec:central_optimal}
%%%%%%
%We start by stating what we mean by minimax lower bounds. 
Let $\matP$ be a family of distributions, defined over the sample space $\matX$. Our goal is to estimate the mean $\theta: \matP \to \R$ where $\theta(P) = \E_{X \sim P}[X]$ for any $P \in \matP$. We let $X_1,\cdots , X_n$ be $n$ independent and identically distributed samples that are drawn from $P \in \matP$ and  $\bm{\diff} = (\diff_i)_{i=1}^n$ be the privacy levels. In the central setting, an estimator $\htheta(X_1,\cdots,X_n)$ is a real-valued measurable function over $\matX^n$ which estimates $\theta(P)$. We define $\matQ_{c}(\bm{\diff})$ as the class of $\bm{\diff}$-centrally differentially private estimators, according to Definition \ref{definition:central_DP}. With this notation in hand, the minimax estimation error is given by 
%%%%%%
\begin{equation} \label{eqn:minimax_central}
\matL_c(\matP, \theta, \bm{\diff}) := \inf_{\htheta \in \matQ_{c}(\bm{\diff})} \sup_{P \in \matP} 
\E_{(X_i \sim P)_{i=1}^n, \htheta} \left [ 
\left | \htheta(X_{1:n}) - \theta(P) \right |^2
\right],
\end{equation}
%%%%%%
where the expectation is taken over the randomness in both samples $X_{1:n}$ and the estimator. The supremum in the above expression is the worst-case estimation error over all distributions of the data points. Therefore, given that the platform does not know the distribution of the data points, the infimum outputs the $\bm{\diff}$-centrally differentially private estimator that minimizes this worst-case estimation error.

Our goal is to provide a lower bound on the minimax rate defined above and prove that such lower bound can be (almost) achieved by linear estimators with Laplace mechanism. To do so, let us first, formally define this class of estimators. Given the data of users $x_1,\cdots, x_n$, a linear estimator with Laplace mechanism is in the form of   
%%%%%%
\begin{align}\label{eq:estimator:central}
    \hat{\theta}= \sum_{i=1}^n w_i x_i + \Lap(1/\eta),
\end{align}
%%%%%%
where $w_i$ is the weight that the estimator allocates to the data of user $i$ with $\sum_{i=1}^n w_i = 1$. 
%%%%%%
Given this estimator, the following lemma shows that the data of each user is centrally differentially private.

\begin{lemma} \label{lemma:analyst_DP}
The estimator $\hat{\theta}$ given in \eqref{eq:estimator:central} is $(w_i \eta)_{i=1}^n$-centrally differentially private.
\end{lemma}
%%%%%%
This lemma directly follows from Lemma \ref{lemma:Laplace}. The proof of this lemma as well as other omitted proofs are presented in the appendix. 

We next establish a lower bound for the estimation error in the central setting and prove that a linear estimator with
Laplace mechanism (almost) achieves the lower bound.
%%%%%%%%%%%%%%%%%%%%%%%%%%%%%%%%
%%%%%%%%%%%%%%%%%%%%%%%%%%%%%%%%
\begin{theorem} \label{theorem:LB_Central}
Let $\bm{\diff} = (\diff_i)_{i=1}^n$ and, without loss of generality, suppose $\diff_1 \leq \cdots \leq \diff_n \leq 1$. Also, let $\matP^*$ be the family of distributions $P$ such that $|X| \leq \frac{1}{2}$ almost surely.\footnote{{The choice of upper bound $1/2$ is without loss of generality and is made to guarantee the length of the support is bounded by $1$, simplifying the equations.}} There exists a (universal) positive constant $c_l$ such that\footnote{For any $x, y \in \mathbb{R}$, we let $x \wedge y$ denote $\min\{x,y\}$.}
\begin{align}\label{eqn:central_lb}
\matL_c(\matP^*, \theta, \bm{\diff}) & \geq c_l \left ( \max_{k \in \{0,1,\cdots,n\}} \frac{1}{n-k+(\sum_{i=1}^k \diff_i)^2} \wedge 1 \right ).
\end{align}
Moreover, there exists an $\bm{\diff}$-centrally differentially private linear estimator $\htheta$ and a (universal) constant $c_u$ such that
\begin{equation} \label{eqn:central_ub}
\E_{(X_i \sim P)_{i=1}^n, \htheta} \left [ 
\left | \htheta(X_{1:n}) - \theta(P) \right |^2 \right ]
\leq c_u \max_{k \in \{0,1,\cdots,n\}} \frac{\log(n+1) }{n-k +(\sum_{i=1}^k \diff_i)^2} ,
\end{equation}
for any $P \in \matP^*$.
\end{theorem}
%%%%%%%%%%%%%%%%%%%%%%%%%%%%%%%%
We prove the lower bound by using the Le Cam's method \cite{yu1997assouad} that reduces the problem of finding lower bounds to a hypothesis testing problem between two distributions. More specifically, using this technique, we need to bound the change in the distribution of estimator's output, i.e., the distribution of $\htheta(X_{1:n})$, when the underlying data distribution changes. To bound the change in the distribution, we first notice that bounding the change in the distribution by using a single distance between the distributions does not immediately give us the desired bound. We circumvent this challenge by using a combination of two well-known distances between two distributions: Total Variation (TV) and Kullback–Leibler (KL).

We establish the upper bound by constructing a linear estimator in the form of \eqref{eq:estimator:central} that achieves the desired bound. Note that, by Lemma \ref{lemma:analyst_DP}, to have an $(\diff_i)_{i=1}^n$-centrally differentially private estimator, we should have
\begin{equation} \label{eqn:DP_constraints}
\eta w_i \leq \diff_i \text{ for all } i.    
\end{equation}

An interesting and somewhat counter-intuitive observation is that the above constraints are not necessarily all binding for the optimal estimator. In other words, the optimal estimator might end up providing higher privacy levels than reported for certain users. This means that we might achieve a lower variance for the estimator by guaranteeing better privacy levels (i.e., lower $\diff_i$'s) for certain users. The main reason for this structure is that, if we keep all the constraints active while some users ask for less privacy, this might lead to putting \textit{too much weight} on their data. In fact, the optimal estimator in the proof of Theorem \ref{theorem:LB_Central} is built by capping the weight that we assign to the data of a portion of users with the highest $\diff_i$'s, i.e., users with the lowest privacy restrictions. 
Let us elaborate this matter with an example. Suppose $(\diff_i)_{i=1}^n$ are given as
\begin{equation}\label{eqn:central_optimal_example}
\begin{aligned} 
\diff_1 = \cdots = \diff_{\floor{n-\sqrt{n}}} = \frac{1}{\sqrt{n}}, \quad  \diff_{\floor{n -\sqrt{n}} +1} = \cdots = \diff_n = 1.
\end{aligned}
\end{equation}
As shown in the proof of Theorem \ref{theorem:LB_Central}, the linear estimator 
\begin{equation}
\htheta = \sum_{i=1}^{n} \frac{1}{n} x_i + \Lap\left(\frac{1}{\sqrt{n}}\right),
\end{equation}
achieves the variance $\mathcal{O}(\frac{1}{n})$ which matches the lower bound, and hence it is optimal. Moreover, this estimator is $\frac{1}{\sqrt{n}}$-centrally differentially private with respect to every user's data, meaning it guarantees a much better level of privacy for users $\floor{n-\sqrt{n}} + 1$ to $n$.
Now let us see what happens if we consider the linear estimator that keeps all the constraints active:
\begin{equation} \label{eqn:central_suboptimal_example}
\htheta = \sum_{i=1}^n \frac{\diff_i}{\sum_{j=1}^n \diff_j} x_i + \Lap\left(\frac{1}{\sum_{j=1}^n \diff_j}\right).    
\end{equation}
The variance of this estimator is 
\begin{equation*}
\E[|\htheta - \theta|^2] = \frac{2}{(\sum_{j=1}^n \diff_j)^2}+ \sum_{i=1}^{\floor{n-\sqrt{n}}}  \frac{1/n}{(\sum_{j=1}^n \diff_j)^2} { \Var}
  +\sum_{i=\floor{n-\sqrt{n}} + 1}^n \frac{1}{(\sum_{j=1}^n \diff_j)^2} { \Var}.
\end{equation*}
Note that, $\sum_{j=1}^n \diff_j \approx 2 \sqrt{n}$, and hence, the first two terms in the right-hand side of the above expression are $\mathcal{O}(\frac{1}{n})$. However, the third term is $\Omega(\frac{1}{\sqrt{n}})$. This leads to the total variance being $\Omega(\frac{1}{\sqrt{n}})$, and thus, this estimator is suboptimal. 

%%%%%%%%%%%%%%%%%%%%%%%%%%%%%%%%
%%%%%%%%%%%%%%%%%%%%%%%%%%%%%%%%
%%%%%%%%%%%%%%%%%%%%%%%%%%%%%%%%
\subsection{Optimal estimation with local differential privacy} \label{sec:local_optimal}
%%%%%%
Here, we consider the local differential privacy setting. In this setting, and for any $i$, instead of observing $X_i$, the platform observes $\hat{X}_i := \matC_i(X_i)$, where $\matC_i : \matX \to \hat{\matX}$ is a $\diff_i$-locally differentially private channel. Hence, the estimator $\htheta$ would be defined over $\hat{\matX}^n$ and would be cast as $\htheta(\hat{X}_{1:n})$.
Also, $\matQ_{l}(\bm{\diff})$ denotes the class of mechanisms $\matM : \matX^n \to \hat{\matX}^n$ where $\matM(X_1,\cdots, X_n) = (\matC_i(X_i))_{i=1}^n$, with $\matC_i$ being an $\diff_i$-locally differentially private channel.
Under local differential privacy, the minimax rate is defined as
%%%%%%
\begin{equation} \label{eqn:minimax_local}
\matL_l(\matP, \theta, \bm{\diff}) := \inf_{\htheta, \matM \in \matQ_{l}(\bm{\diff})} \sup_{P \in \matP} 
\E_{(X_i \sim P)_{i=1}^n, \htheta} \left [ 
\left | \htheta(\hat{X}_{1:n}) - \theta(P) \right |^2
\right],
\end{equation}
%%%%%%
where the expectation is taken over the randomness in both samples $X_{1:n}$ and the estimator. Again, the supremum in the above expression is the worst-case estimation error over all distributions of the data points. Therefore, given that the platform does not know the distribution of the data points, the infimum outputs the $\bm{\diff}$-locally differentially private estimator that minimizes this worst-case estimation error.

In this case, the linear estimator with Laplace mechanism is defined as follow:
User $i$ releases an $\diff_i$-locally differentially private version of $x_i$, denoted by $\hat{x}_i$, using Laplace mechanism, i.e., $\hat{x}_i = x_i + \Lap(1/\diff_i)$. Using these private data points, we form the following estimate  
%%%%%%
\begin{align}\label{eq:estimator:local}
    \hat{\theta}= \sum_{i=1}^n w_i \hat{x}_i,
\end{align}
%%%%%%
where $w_i$ is the weight that the estimator allocates to the private data of user $i$ with $\sum_{i=1}^n w_i = 1$. 
%%%%%%%%%%%%%%%%%%%%%%%%%%%%%%%%
We next establish a lower bound for the estimation error in the local setting and prove that a
linear estimator with Laplace mechanism achieves the lower bound.
%%%%%%%%%%%%%%%%%%%%%%%%%%%%%%%%
%%%%%%%%%%%%%%%%%%%%%%%%%%%%%%%%
\begin{theorem} \label{theorem:LB_Local}
Let $\bm{\diff} = (\diff_i)_{i=1}^n$ with $\diff_i \leq 1$ for all $i$. Also, let $\matP^*$ be the family of distributions $P$ such that $|X| \leq \frac{1}{2}$ almost surely. There exists a (universal) positive constant $\ell_l$ such that
\begin{align}\label{eqn:local_lb}
\matL_l(\matP^*, \theta, \bm{\diff}) & \geq  \ell_l \left ( \frac{1}{\sum_{i=1}^n \diff_i^2} \wedge 1 \right ) . 
\end{align}
Moreover, there exists an $\bm{\diff}$-locally differentially private linear estimator $\htheta$ and a universal constant $\ell_u$ such that
\begin{equation} \label{eqn:local_ub}
\E_{(X_i \sim P)_{i=1}^n, \htheta} \left [ 
\left | \htheta(X_{1:n}) - \theta(P) \right |^2 \right ]
\leq \frac{\ell_u}{\sum_{i=1}^n \diff_i^2} ,
\end{equation}
for any $P \in \matP^*$.
\end{theorem}
%%%%%%%%%%%%%%%%%%%%%%%%%%%%%%%%%%%%%%%%%%%%
Similar to the proof of Theorem \ref{theorem:LB_Central}, we prove the lower bound by using the Le Cam's method. To establish the upper bound, similarly, we construct a linear estimator that achieves the lower bound up to a constant factor. 
%%%%%%

\section{Data acquisition mechanism with privacy guarantees} \label{sec:mechanism_formulation}
%%%%%%
In this section, we endogenize the choice of the privacy loss levels by assuming a utility function that captures different privacy sensitivities. In particular, each user $i \in \mathcal{N}$ has a type or \emph{privacy sensitivity} $c_i \in \mathbb{R}_{+}$ that represents the per unit cost of privacy loss for user $i$. We assume each $c_i$ is independently drawn from a publicly known distribution with cumulative distribution function $F_i(\cdot)$ and probability density function $f_i(\cdot)$. We also let $\mathbf{c}=(c_1, \dots, c_n)$ denote the vector of privacy sensitivities. The privacy sensitivity of each user is their private information.

We consider a mechanism whereby individuals participate by sharing their data and reporting their privacy sensitivities.\footnote{From here on, we will use the terms mechanism designer and platform interchangeably.} While users can misrepresent their privacy sensitivities, they have no capability to manipulate their data (e.g., their data is collected by the analyst when they participate or can be verified). Depending on the reported sensitivity, the analyst provides a compensation for the user in exchange for her data. This compensation may be a direct monetary payment or it may be an implicit transfer, for example, in the form of some good or service the analyst provides to the user to acquire her data. The mechanism designer simultaneously determines the privacy loss levels (which were assumed given in the previous section) and a differentially private estimator based on users' data that achieves these levels. 

Given this interaction, we next specify a data acquisition mechanism with privacy guarantees on users' data.

%%%%%%%%%%
\begin{definition}[Private data acquisition mechanism]
\textup{We call the tuple $(\hat{\theta}, \bm{\diff}, \mathbf{t})$ a \emph{private data acquisition mechanism} where
%%%%
\begin{enumerate}
\item $ \hat{\theta}:\mathcal{X}^n \times \R_{+}^n \to \R$ is a (centrally or locally) differentially private estimator that maps acquired user data $\mathbf{x} = (x_i)_{i=1}^n$ and privacy losses $\bm{\diff} = (\diff_i)_{i=1}^n$ to an estimate $\hat \theta(\mathbf{x},  \bm{\diff})$.\footnote{We assume $x_i$ is removed from $\mathbf{x}$ if user $i$ does not participate in the mechanism.}
\item For all $i \in \mathcal{N}$, $\diff_i : \R_+^n \to \R_+$ is a function that maps privacy sensitivities $\mathbf{c}$  to a privacy loss for user $i$, $\diff_i(\mathbf{c})$, with $\bm{\diff}(.)=(\diff_i(\cdot))_{i=1}^n$.
\item  For all $i \in \mathcal{N}$, $t_i : \R_+^n \to \R_+$ is a function that maps privacy sensitivities $\mathbf{c}$  to a payment for user $i$, $t_i(\mathbf{c})$, with $\bm{t}(.)=(t_i(\cdot))_{i=1}^n$.
\end{enumerate}
{ The above functions are assumed to be differentiable, with their derivatives being Riemann integrable. The minimax optimal estimators derived in Subsections \ref{sec:central_optimal} and \ref{sec:local_optimal} meet these assumptions.}
}
\end{definition}

We will study mechanisms with estimators that provide both central and local differential privacy guarantees (see Definitions \ref{definition:central_DP} and \ref{definition:local_DP}) and use the notations $\hat \theta_{\mathrm{central}}$ and $\hat \theta_{\mathrm{local}}$ to highlight the distinction.

Each user that participates in a private data acquisition mechanism $(\hat{\theta}, \bm{\diff}, \mathbf{t})$ shares her data with the platform leading to a lower estimation error. Users derive benefit from accessing this more accurate estimate (e.g., representing a new medical treatment that is of value for all users), but incur a privacy cost proportional to their privacy sensitivity $c_i$. Throughout, we find it more convenient to work with cost instead of utility. In particular, we model the user's cost from participation by the mean square error of the platform's estimate $\hat \theta$ and her privacy cost by $c_i \bm{\diff}(\mathbf{c})$. Hence, the cost function of a user $i$ with type $c_i$ who reports $c'_i$ is given by 
\begin{align}\label{eq:utility:participate}
    \Cost(c'_i, c_i; \bm{\diff}, \mathbf{t}, \hat{\theta})= \mathbb{E}_{\mathbf{c}_{-i}} \left[\MSE(c'_i, \mathbf{c}_{-i}; \bm{\diff}, \hat{\theta}) + c_i  \diff_i(\mathbf{c}_{-i}, c'_i) - t_i(\mathbf{c}_{-i}, c'_i))\right],
\end{align}
where the first term is the expected mean squared error of the estimator given by 
\begin{align*}
    \MSE(c'_i, \mathbf{c}_{-i};\bm{\diff},  \hat{\theta}) = \mathbb{E}_{\mathbf{x}}\left[ |  \hat{\theta}(\mathbf{x},  \bm{\diff})- \theta |^2 \right].
\end{align*}
Note that the privacy losses $ \bm{\diff}$ depends on reported privacy sensitivities $(c_i', \mathbf{c}_{-i})$, therefore we make the dependence of the mean square error on $(c_i', \mathbf{c}_{-i})$ explicit in our notation.
The second term of \eqref{eq:utility:participate} represents the privacy cost that the user incurs, and the third term is the payment that the user receives.

A user $i \in \mathcal{N}$ that does not participate in the mechanism does not compromise her privacy, but neither gets compensation nor enjoys the benefit of a reduced mean square error (arising from an estimate based on a collection of users' data). Therefore, the cost of a nonparticipating user becomes the mean square error of her ``best" estimate of parameter $\theta$ based on her data alone, $\hat \theta(X_i)$, given by
%%%%%%
\begin{align}\label{eq:utility:not:participate}
   \mathbb{E}_{X_i}\left[ |\hat{\theta}(X_i)- \theta |^2\right]=\mathbb{E}_{X_i}\left[ |X_i- \theta |^2\right]=\Var.
\end{align}
%%%%%%

For a given $\hat{\theta}(\cdot)$, the goal of the platform is to minimize an objective function given by 
\[\mathbb{E}_{\mathbf{c}}\left[ \MSE(\mathbf{c},\bm{\diff},  \hat{\theta})  + \sum_{i=1}^n t_i(\mathbf{c})\right],\]
{over the choices of $\diff_i(\cdot)$ and $t_i(\cdot)$ for all $i \in \mathcal{N}$. In the platform's objective, } the first term is the mean square error of estimator $\hat \theta$ given reported types $\mathbb{c}$ and resulting privacy losses $\bm{\diff}$, i.e., 
\begin{equation*}
\MSE(\mathbf{c},\bm{\diff},  \hat{\theta})  = \mathbb{E}_{\mathbf{x}} \left [ | \hat{\theta}(\mathbf{x},  \bm{\diff})- \theta |^2 \right ].
\end{equation*}
The second term is the total compensation the analyst provides to the users for truthfully reporting their privacy sensitivities and acquiring their data. In the appendix we establish that, similar to the classical mechanism design setting, \emph{revelation principle} holds and therefore the platform can focus on direct revelation mechanisms where individuals reporting their type truthfully is a (Bayesian Nash) equilibrium. \emph{Incentive compatibility}  constraints formalize this equilibrium outcome by imposing that user $i$ has no incentive to misrepresent her type when others report truthfully (i.e., reporting her type correctly is a Bayesian Nash equilibrium of the underlying incomplete information game). Similarly, {\it individual rationality} constraints ensure that the platform does not make users worse off by participating in the mechanism. 
Together with these constraints, the  mechanism designer's optimization problem can be written as
%%%%%%
\begin{align}
    \min_{\bm{\diff}(\cdot), \mathbf{t}(\cdot)}  ~~~&  \mathbb{E}_{\mathbf{c}}\left[ \MSE(\mathbf{c},\bm{\diff},  \hat{\theta})  + \sum_{i=1}^n t_i(\mathbf{c})\right]  \label{eqn:main_opt_platform} \\
    & \Cost(c_i, c_i; \bm{\diff}, \mathbf{t}, \hat{\theta}) \le  \Cost(c'_i, c_i; \bm{\diff}, \mathbf{t}, \hat{\theta}) \quad \text{ for all } i \in \mathcal{N}, c_i, c'_i \label{eq:IC} \\
    & \Cost(c_i, c_i; \bm{\diff}, \mathbf{t}, \hat{\theta}) \le  \Var \quad \text{ for all } i \in \mathcal{N}, c_i, \label{eq:IR}
\end{align}
%%%%%%
where the constraints in \eqref{eq:IC} and \eqref{eq:IR} represent the incentive compatibility and the individual rationality constraints, respectively.\footnote{We assume that the variance and the payments both appear with the same coefficient in the platform's objective. Our analysis readily extends to a setting with differing coefficients.}

%%%%%%%%%%%%%%%%%%%%%%%%%%%%
\subsection{Payment identity}
%%%%%%%%%%%%%%%%%%%%%%%%%%%%
For a given estimator $\hat{\theta}$, the platform decision comprises the privacy loss functions $\bm{\diff}(\cdot)$ and the payment functions $\mathbf{t}(\cdot)$. We next identify the payment as a function of the privacy loss functions. In this regard, we define the \emph{interim} quantities
%%%%%%
\begin{align*}
    t_i(c_i)=\mathbb{E}_{\mathbf{c}_{-i}}\left[t(c_i, \mathbf{c}_{-i}) \right] \text{ and }
    \diff_i(c_i)=\mathbb{E}_{\mathbf{c}_{-i}}\left[\diff_i(c_i, \mathbf{c}_{-i}) \right] \text{ for all } i \in \mathcal{N}, c_i.
\end{align*}
%%%%%
\begin{proposition}\label{Pro:PaymentIdentity}
For a given estimator $\hat{\theta}: \mathcal{X}^n \times \mathbb{R}_+^n  \to \mathbb{R}$, a central or local privacy data acquisition mechanism $(\hat{\theta}, \bm{\diff}, \mathbf{t})$ satisfies incentive compatibility \eqref{eq:IC} and individual rationality \eqref{eq:IR} if and only if 
\begin{align}\label{eq:Pro:PaymentIdentity}
    t_i(c_i)= \mathbb{E}_{\mathbf{c}_{-i}}\left[\MSE(\mathbf{c},\bm{\diff},  \hat{\theta}) \right] - \Var +  c_i \diff_i(c_i) +   \int_{z=c_i}^{\infty} \diff_i(z) dz + d_i,
\end{align}
for some constant $d_i \ge 0$, and $\diff_i(z)$ is {non-increasing (or equivalently, is weakly decreasing)} in $z$ for all $i \in \mathcal{N}$.
\end{proposition}
%%%%%%%%%%%%%%%%%%%%%%%%%%%%
%%%%%%%%%%%%%%%%%%%%%%%%%%%%
Proposition \ref{Pro:PaymentIdentity} determines the payment in terms of the privacy loss functions. This proposition is closely related to the payment identity in classical mechanism design (see \cite{myerson1981optimal}) and in particular single-dimensional mechanism design. In particular, by evaluating the first order condition corresponding to the incentive compatibility constraint \eqref{eq:IC}, we establish that this constraint holds if and only if 
\begin{align*}
     t_i(c_i) =   t_i(0) + \mathbb{E}_{\mathbf{c}_{-i}}\left[\MSE(c_i,\mathbf{c}_{-i},\bm{\diff},  \hat{\theta}) \right] - \mathbb{E}_{\mathbf{c}_{-i}}\left[\MSE(0,\mathbf{c}_{-i},\bm{\diff},  \hat{\theta}) \right] +  c_i \diff_i(c_i) -  \int_{z=0}^{c_i} \diff_i(z) dz
\end{align*}
and $\diff_i(z)$ is weakly decreasing in $z$. We then use the above expression in the individual rationality constraint \eqref{eq:IR} and prove
\begin{align*}
    t_i(0) =\mathbb{E}_{\mathbf{c}_{-i}}\left[\MSE(0,\mathbf{c}_{-i},\bm{\diff},  \hat{\theta}) \right] - \Var +\int_{z=0}^{\infty} \diff_i(z) dz +d_i
\end{align*}
for some $d_i \ge 0$. Equation \eqref{eq:Pro:PaymentIdentity} follows from the previous two expressions. It is worth noting  that, for a central or local privacy data acquisition mechanism $(\hat{\theta}, \bm{\diff}, \mathbf{t})$ to be optimal, we must have $d_i=0$ in \eqref{eq:Pro:PaymentIdentity}. 

{In concluding this subsection, we want to highlight that our benchmark for individual rationality (given in \eqref{eq:IR}) is that the users will not benefit from the platform's estimate if they do not participate. If we consider an alternative benchmark in which the users benefit from the platform's estimator even if they do not participate, then the payments increase, and the platform's cost decreases. However, as we show in the appendix, our characterization of the optimal privacy levels that will follow remains unchanged. 
}

%We next use this payment identity to reformulate the platform's problem.

%%%%%%%%%%%%%%%%%%%%%%%%%%%%

\subsection{Reformulating the platform's problem}

We next use Proposition \ref{Pro:PaymentIdentity} to reformulate the platform's problem in terms of only the privacy loss functions and the \emph{virtual costs}, defined as
\begin{align*}
\psi_i(c) = c + \frac{F_i(c)}{f_i(c)}, \quad \text{ for all } i \in \mathcal{N}, c \in \displaystyle \operatorname {supp}(f),
\end{align*}
{where the support of $f(\cdot)$ is defined as $\displaystyle \operatorname {supp} (f)=\{c\in \mathbb{R}_{+}\,:\,f(c)\neq 0\}$.}
%%%%%
\begin{proposition}\label{pro:refomrulation1}
For a given estimator $\hat{\theta}: \mathcal{X}^n \times \mathbb{R}_+^n \to \mathbb{R}$, the optimal privacy loss in the central or local privacy data acquisition mechanism is the solution of 
\begin{align} 
    \min_{\bm{\diff}(\cdot)}  ~~~&  \mathbb{E}_{\mathbf{c}}\left[ (n+1)\MSE(\mathbf{c},\bm{\diff}, \hat{\theta})  + \sum_{i=1}^n \diff_i(\mathbf{c}) \psi_i(c_i) \right] - n \Var \label{reformulation:central} \\
    & \diff_i(z)=\mathbb{E}_{\mathbf{c}_{-i}}\left[\diff_i(z, \mathbf{c}_{-i}) \right] \text{ is weakly decreasing in } z \text{ for all } i \in \mathcal{N}. \label{reformulation:central:1}
\end{align}
\end{proposition}
%%%%%
Proposition \ref{pro:refomrulation1}  is an analogue of Myerson's  reduction of mechanism design to virtual welfare
maximization, adapted to our data acquisition setting (\cite{myerson1981optimal}), and it follows from invoking Proposition \ref{Pro:PaymentIdentity}.

%%%%%%
\section{Privacy-concerned data acquisition in the central privacy setting} \label{sec:central}
In the rest of the paper, we will focus on linear estimators, which we showed to be near optimal for given privacy loss levels. 
Our goal in this section is to address the analyst’s mechanism design problem in the central privacy setting for the near optimal choice of estimator found in Section \ref{sec:central_optimal}:
\begin{align}\label{estimator:central:1}
\htheta_{\mathrm{central}} (x_1, \dots, x_n) := \sum_{i=1}^n w_i(\mathbf{c}) x_i + \mathrm{Laplace}\left(\frac{1}{\eta} \right)
\end{align}
such that
\begin{align*}
    \sum_{i=1}^n w_i(\mathbf{c})=1, \text{ and }
    \eta w_i(\mathbf{c}) \le \diff_i(\mathbf{c}) \text{ for all } i \in \mathcal{N}.
\end{align*}

\begin{figure}[t]
    \centering
        \includegraphics[width=.3\textwidth]{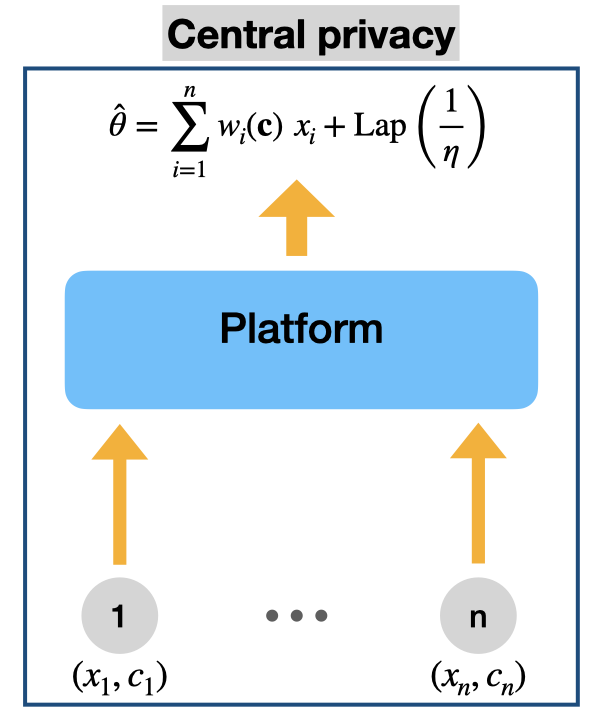}
    \caption{The interaction between the users and the platform in the central privacy setting. }\label{fig:central}
\end{figure}
%%%%%%

Figure \ref{fig:central} depicts the interaction between the platform and the users in the central privacy setting and when the platform is using the above (near) optimal choice of estimator.

%%%%%
\subsection{Characterization of the optimal central privacy data acquisition mechanism}

%%%%%
Our next theorem characterizes the optimal privacy loss function in the central privacy setting under the following assumption.
%%%
 
%%%%%% 
\begin{assumption}\label{assump:MHR}
\textup{For any $i \in \matN$, the \emph{virtual cost}  
%the support of $f_i(\cdot)$ is an interval, denoted by $[\ubar{C}_i, \bar{C}_i]$. Furthermore, 
%the function 
$\psi_i(c) = c + \frac{F_i(c)}{f_i(c)}$ is increasing in $c$.}
\end{assumption}
%%%%%%
Assumption \ref{assump:MHR} is standard in mechanism design and in particular for procurement auctions which is closer to our setting. It resembles the regularity condition adopted in mechanism design literature and holds for a variety of distributions and in particular for distributions with log-concave density functions such as uniform, exponential, and normal (see, e.g., \cite{rosling2002inventory}).
 
%%%%%%
\begin{theorem}\label{thm:central}
Suppose Assumption \ref{assump:MHR} holds. For any reported vector of privacy sensitivities $(c_1, \dots, c_n)$, the optimal privacy loss level in the central privacy data acquisition mechanism is $\diff^*_i(\mathbf{c})=y_i^*$ for $i \in \mathcal{N}$, where $(y^*_1, \dots, y^*_n)$ is the optimal solution of
\begin{align}\label{eq:thm:central}
    \min_{\mathbf{y}} &  \frac{n+1}{\left( \sum_{j=1}^n y_j \right)^2} \left( 2+ \sum_{i=1}^n   y_i ^2 ~ \Var  \right)+ \sum_{i=1}^n  \psi_i(c_i) y_i  \\
    & \text{ s.t. } y_i \ge 0, \text{ for all } i \in \mathcal{N}.\nonumber
\end{align}
Moreover, for all $i \in \mathcal{N}$ the weight of user $i$'s data in the platform's estimator is $\frac{y^*_i}{\sum_{j=1}^n y^*_j}$.
\end{theorem}
%%%%%%
Before providing the proof idea of this theorem, let us highlight the difference between our characterization and that of classic mechanism design (e.g., \cite{myerson1981optimal} or the procurement counterpart). In classic mechanism design, the designer's problem becomes a linear optimization. In our setting, however, the designer's problem is a non-linear and non-convex optimization. { An important implication of this distinction is that, contrary to classic mechanism design where the optimal mechanism typically involves a threshold rule, in this case, the optimal privacy loss level is not a threshold strategy. Instead, it is a continuous function that depends on the privacy sensitivity.}

To prove Theorem \ref{thm:central}, we first note that the mean square error of the linear estimator $\htheta_{\mathrm{central}}$ in \eqref{estimator:central:1} is given by
\begin{align*}
\MSE(\bm{c}, \bm{\diff}, \htheta_{\mathrm{central}}) =   \frac{2}{\eta^2}  + \sum_{i=1}^n w_i(\mathbf{c})^2 \Var.
\end{align*}
We next plug the above characterization into Proposition \ref{pro:refomrulation1}, and note that, if we drop the constraint \eqref{reformulation:central:1} (which is $\diff_i(c_i)=\mathbb{E}_{\mathbf{c}_{-i}}\left[\diff_i(c_i, \mathbf{c}_{-i}) \right]$ being weakly decreasing in $c_i$), it suffices to solve the following pointwise optimization problem 
\begin{align}
     \min_{\bm{\diff}(\mathbf{c}), \mathbf{w}(\mathbf{c}), \eta} &  \frac{2 (n+1)}{\eta^2}+ \sum_{i=1}^n (n+1) \Var ~ w_i(\mathbf{c})^2 +  \sum_{i=1}^n  \psi_i(c_i) \diff_i(\mathbf{c})  \nonumber \\
    & \text{ s.t. } \bm{\diff}_i(\mathbf{c}) \ge 0, \text{ for all } i \in \mathcal{N} \nonumber \\
    & \sum_{i=1}^n w_i(\mathbf{c})=1 \nonumber \\
    & \eta w_i(\mathbf{c}) \le \diff_i(\mathbf{c}) \text{ for all } i \in \mathcal{N}. \label{eq:pf:main:thm:central:1}
\end{align}
We next focus on solving the above problem. To do so, we establish that the constraints in \eqref{eq:pf:main:thm:central:1} are binding in the optimal solution and therefore this problem is equivalent to the optimization problem \eqref{eq:thm:central} given in Theorem  \ref{thm:central} statement. Finally, we conclude the proof by showing that the solution to this pointwise optimization satisfies the aforementioned constraint \eqref{reformulation:central:1} that we dropped. More specifically, we show that the $i$-th component of the optimal solution of \eqref{eq:thm:central}, under Assumption \ref{assump:MHR}, is weakly decreasing in $c_i$.

The characterization of Theorem \ref{thm:central} leads to the following observation:
\begin{corollary}\label{Cor:central:mono}
Suppose Assumption \ref{assump:MHR} holds. For any reported vector of privacy sensitivities $(c_1, \dots, c_n)$, in the optimal central data acquisition mechanism, we have  $\diff^*_i(\mathbf{c}) \ge \diff^*_j(\mathbf{c})$ for all $i,j \in \mathcal{N}$ such that $\psi_i(c_i) < \psi_j(c_j)$.
\end{corollary}
This corollary states the intuitive fact that in the optimal central data acquisition mechanism, users with higher virtual privacy sensitivities have lower (i.e., better) privacy loss levels. 

%%%%%%
\subsection{Computing the optimal privacy loss function}\label{sec:central:compute}
%%%%%%
The implementation of the optimal central privacy data acquisition mechanism involves solving problem \eqref{eq:thm:central}, which is a non-convex program. We next develop a score-based method that efficiently solves problem \eqref{eq:thm:central}.

%%%%%%
\begin{algorithm*}[t]
	%\SetAlgoNoLine
	\KwIn{The vector of privacy sensitivities $(c_1, \dots, c_n)$}
    Sort the terms $ \{\psi_i(c_i)\}_i$. Without loss of generality, let us assume 
\begin{align*}
    \psi_1(c_1) \le \dots \le  \psi_n(c_n);
\end{align*}
Let $B_0 = \tilde{B}_0 = 0$;\\
\For{$i=1$ to $n$}{
    Let
    \begin{align*}
    A_{i}= \frac{i}{2(n+1) \Var}, 
    \quad B_{i}=B_{i-1}+ \frac{\psi_i(c_i)}{2(n+1)},
    \quad \tilde{B}_{i}=\tilde{B}_{i-1}+ \frac{\psi_i(c_i)^2}{2(n+1) \Var};    
    \end{align*}
    Let 
    \begin{align*}
    OBJ_i(\lambda)= 2 (n+1) \left( \lambda A_{i} - B_{i}\right)^2 + \frac{A_{i} \lambda^2 - \tilde{B}_{i}}{2 \left(\lambda A_{i} - B_{i} \right)^2};
    \end{align*}
    Let
    \begin{align*}
    \lambda^*_i=\argmin_{\lambda}  ~ OBJ_i(\lambda) \text{ s.t. } \psi_{i}(c_{i}) \le \lambda \le \psi_{i+1}(c_{i+1}) \text{ with the convention } \psi_{n+1}(c_{n+1})=\infty;
    \end{align*}
}    
Let $i^*= \argmax_{i} \mathrm{OBJ}_i(\lambda^*_i)$; \\
\textbf{Output:} The optimal solution is given by
\begin{align*}
y^*_j =0 \text{ for } j > i^* \text{ and } y^*_j= \frac{ \lambda^*_{i^*} - \psi_j(c_j)}{2 (n+1) \Var \left(\lambda^*_{i^*} A_{i^*} - B_{i^*} \right)^2} \text{ for } j \le i^*.
\end{align*}
	\caption{Computing the optimal privacy loss in the central setting}
	\label{mech:central:compute}
\end{algorithm*}
%%%%%%
To guide the analysis, without loss of generality, we assume $ \psi_1(c_1) \le \dots \le  \psi_n(c_n)$, and define $\psi_{n+1}(c_{n+1}) = \infty$. We first rewrite problem \eqref{eq:thm:central} by introducing a variable for the summation of $y_i$'s as follows
\begin{align}\label{Eq:Pro:approximate:central:main:0}
\min_{S \ge 0 } \min_{\mathbf{y}} &  \frac{ n+1}{S^2} \left( 2+ \sum_{i=1}^n   y_i ^2 ~ \Var  \right)+ \sum_{i=1}^n  \psi_i(c_i) y_i   \\
    & \text{ s.t. } \sum_{i=1}^n y_i = S  \label{Eq:Pro:approximate:central:main:1} \\
    & y_i \ge 0, \text{ for all } i \in \mathcal{N}. \nonumber
\end{align}
For a given $S$, the optimization over $\mathbf{y}$ is a convex program. Using Karush–Kuhn–Tucker (KKT) condition (see e.g. \cite{bertsekas1997nonlinear}), the solution to this optimization problem is\footnote{For any $x \in \mathbb{R}$, we let $x^+$ denote $\max\{x,0\}$.} 
\begin{align} \label{eq:central_solutions_KKT}
 (y_1, \dots, y_n)=\left( \left(\frac{(\lambda - \psi_1(c_1) )S^2}{2 (n+1) \Var} \right)^+, \dots, \left(\frac{ ( \lambda- \psi_n(c_n)) S^2 }{2 (n+1) \Var} \right)^+ \right),
\end{align}
where $\lambda$ is such that 
\begin{align} \label{eq:S_lambda}
    \sum_{i=1}^n \left(\frac{(\lambda - \psi_i(c_i) )S^2}{2 (n+1) \Var} \right)^+ = S.
\end{align}
Using this relation, we can write $S$ as a function of $\lambda$ which allows us to rewrite the minimization problem \eqref{Eq:Pro:approximate:central:main:0} over $\lambda \in [\psi_1(c_1), \infty]$ rather than $S$. We solve this resulting minimization problem by finding the optimal $\lambda$ in the interval $ [\psi_{i}(c_{i}), \psi_{i+1}(c_i) ]$ for all $i=1, \dots, n$ and then selecting the $\lambda$ with the lowest objective function. Algorithm \ref{mech:central:compute} summarizes the above procedure and the following proposition states the formal result:
%%%%%%
\begin{proposition}\label{Pro:approximate:central}
For any vector of reported privacy sensitivities $(c_1, \dots, c_n)$, Algorithm \ref{mech:central:compute} finds the optimal privacy loss levels in the optimal central data acquisition mechanism (i.e., the solution of problem \eqref{eq:thm:central}) in time $\mathcal{O}(n \log n)$.
\end{proposition}
%%%%%%
Algorithm \ref{mech:central:compute}  needs sorting $n$ elements which requires time $\mathcal{O}(n \log n)$. We also prove that each iteration of the for loop can be done in time $\mathcal{O}(1)$, establishing that the overall running time of Algorithm \ref{mech:central:compute} is  $\mathcal{O}(n \log n)$.

As depicted in Algorithm \ref{mech:central:compute}, the virtual cost of each user determines whether the data of that user is included in the final estimator of the platform. In particular, there exists a threshold $\bar{\psi}$ such that only the data of users whose virtual cost $\psi_i(c_i)$ is below $\bar{\psi}$ are used in the estimator of the platform. This feature of the optimal data acquisition mechanism is reminiscent of the classical optimal mechanism of \cite{myerson1981optimal} with one important difference though: unlike the classical mechanism design in which the item gets allocated to a single user, here the data of multiple users are being used and that the weight of each user's data depends on her virtual cost and the entire profile of virtual costs.

%%%%%%

\section{Privacy-concerned data acquisition in the local privacy setting}\label{sec:local}
%%%%%%

In the local differential privacy setting, each user $i$ shares a differentially private version of her data with the platform who then combines them to form an estimator for the underlying parameter. In particular, first the user reports her privacy sensitivity that determines both the payment to the user and the variance of the noise to be added to the user's data. The platform then collects the ``transformed data'' of the users and combines them to form an estimation of the underlying parameter. The difference between this setting and the central privacy setting is that the data that each user shares with the platform is already differentially private. As a result, the final estimator of the platform is also differentially private (composition property of differential privacy). Therefore, the platform does not need to transform its estimator to make it differentially private and her only estimation task is finding an unbiased estimator with minimum bias. Our goal in this section is to address the analyst’s mechanism design problem in the local privacy setting for the  optimal choice of estimator found in Section \ref{sec:local_optimal}:
%%%%%%
\begin{align}\label{eq:estimator:local:badan}
    \hat{\theta}= \sum_{i=1}^n w_i \hat{x}_i, \quad \text{ where }\hat{x}_i = x_i + \Lap(1/\diff_i) \text{ for all } i \in \mathcal{N}.
\end{align}

\begin{figure}[t]
    \centering
        \includegraphics[width=.3\textwidth]{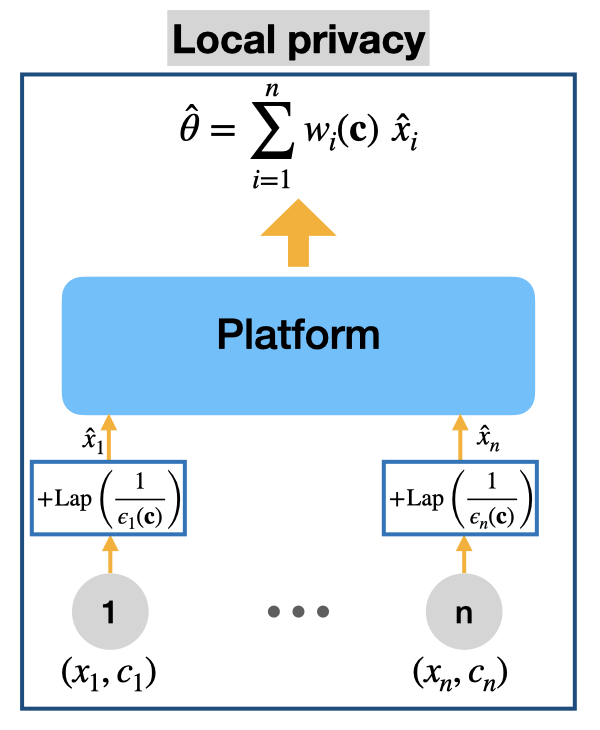}
    \caption{The interaction between the users and the platform in the local privacy setting. }\label{fig:local}
\end{figure}

Figure \ref{fig:local} depicts the interaction between the users and the platform in the local privacy setting.

%%%%%%
%%%%%%%%%%%%%%%%%%%%%%%%%%%%
\subsection{Characterization of the optimal local privacy data acquisition mechanism}
%%%%%%
Our next theorem characterizes the optimal mechanism in the local privacy setting under Assumption \ref{assump:MHR}.
%%%%%%%%%%%%%%%%%%%%%%%%%%%%%
\begin{theorem}\label{thm:local}
Suppose Assumption \ref{assump:MHR} holds. For any reported vector of privacy sensitivities $(c_1, \dots, c_n)$, the optimal privacy loss level in the local privacy data acquisition is $\diff^*_i(\mathbf{c})=y^*_i$ for $i \in \mathcal{N}$, where $(y^*_1, \dots, y^*_n)$ is the optimal solution of
\begin{align}\label{eq:thm:local}
    \min_{\mathbf{y}}  ~~~ & \frac{n+1}{ \sum_{i=1}^n \frac{1}{ \Var + \frac{2}{ y_i^2} }  } + \sum_{i=1}^n  \psi_i(c_i) y_i \\
    & \text{ s.t. } y_i \ge 0 \text{ for all } i \in \mathcal{N}.  \nonumber 
\end{align}
Moreover, for all $i \in \mathcal{N}$, the weight of user $i$'s data in the platform estimator is proportional to 
\begin{align*}
    \frac{1}{\Var + \frac{2}{ {y^*_i}^2} }.
\end{align*}
\end{theorem}
%%%%%%
To prove Theorem \ref{thm:local}, we first note that, for a given vector of privacy sensitivities $\mathbf{c}$, the mean square of the linear estimator given in \eqref{eq:estimator:local:badan} is 
\begin{align*}
  \MSE(\bm{c}, \bm{\diff}, \htheta_{\mathrm{local}}) =  \sum_{i=1}^n w_i(\mathbf{c})^2 \left( \Var + \frac{2}{\diff_i(\mathbf{c})^2}\right).
\end{align*}
Similar to the proof of Theorem \ref{thm:central}, we drop the constraint \eqref{reformulation:central:1}, and consider the following pointwise optimization:
\begin{align}
    \min_{w_i(\mathbf{c}), \diff_i(\mathbf{c})} ~~~ & \sum_{i=1}^n w_i(\mathbf{c})^2 \left( (n+1) \Var + \frac{2(n+1)}{ \diff_i(\mathbf{c})^2} \right) +  \psi_i(c_i) \diff_i(\mathbf{c}) \nonumber \\
    & \sum_{i=1}^n w_i(\mathbf{c})=1 \label{Eq:thm:local:Main:1}\\
    & w_i(\mathbf{c}) \ge 0, \diff_i(\mathbf{c}) \ge 0 \text{ for all } i \in \mathcal{N}.  \nonumber
\end{align}
We next note that the optimization over weights $(w_i(\mathbf{c}))_{i=1}^n$ subject to \eqref{Eq:thm:local:Main:1} is a quadratic optimization problem that, for a given $(\diff_i(\mathbf{c}))_{i=1}^n$, and we can solve explicitly. In particular, $w_i(\mathbf{c})$ is proportional to 
\begin{align*}
    \frac{1}{\Var + \frac{2}{ {\diff_i(\mathbf{c})}^2} } \text{ for all } i \in \mathcal{N}.
\end{align*}
Plugging in these weights, the rest of the proof follows similar to the proof of Theorem \ref{thm:central}.

The characterization of Theorem \ref{thm:local} leads to the following observation:
\begin{corollary}\label{Cor:local:mono}
Suppose Assumption \ref{assump:MHR} holds. For any reported vector of privacy sensitivities $(c_1, \dots, c_n)$, in the optimal local data acquisition mechanism, we have  $\diff^*_i(\mathbf{c}) \ge \diff^*_j(\mathbf{c})$ for all $i,j \in \mathcal{N}$ such that $\psi_i(c_i) < \psi_j(c_j)$.
\end{corollary}
This corollary, which is analogous to Corollary \ref{Cor:central:mono}, states a similar fact in the local setting: in the optimal local data acquisition mechanism, users with higher virtual privacy sensitivity have lower privacy loss levels (better privacy guarantees). 

% %%%%%%
\subsection{Computing the optimal privacy loss function}
%%%%%%
The implementation of the optimal mechanism involves solving problem \eqref{eq:thm:local}, which is a non-convex problem. Thus, using algorithms such as gradient descent might lead to finding a saddle point or a local minima rather than the global minimum. However, in what follows, we present an algorithm that takes advantage of the problem's structure and establishes that finding the global minima admits a Polynomial Time Approximation Scheme (PTAS).

To guide the analysis, without loss of generality, we assume $ \psi_1(c_1) \le \dots \le  \psi_n(c_n)$. Letting $(y^*_1, \dots, y^*_n)$ be the optimal solution of \eqref{eq:thm:local}, the first order condition implies that there exists $\lambda \in \mathbb{R}_+$ such that  
\begin{align*}
    \frac{4 y^*_i}{\left(2+ \Var {y^*_i}^2\right)^2}  =   \frac{\psi_i(c_i)}{n+1} \lambda^2, \text{ for all } y_i^* \neq 0.
\end{align*}
We first prove that if there exists $i \in \{1, \dots, n\}$ such that $y^*_{i}=0$, then  we have $y^*_j=0$ for $j > i$. We also establish that, for such $i$, we have\footnote{Equation \eqref{eq:local_solutions_KKT} holds when $\psi_i(c_i) > \psi_{i-1}(c_{i-1})$. In the proof of Proposition \ref{Pro:approximate:local}, presented in the appendix, we provide the detail for the case $\psi_i(c_i) = \psi_{i-1}(c_{i-1})$ as well.} 
\begin{align} \label{eq:local_solutions_KKT}
    y^*_j = y_{j}^{(h)}(\lambda) \text{ for } j \le i-1 \text{ and } y^*_i \in \{y_{i}^{(l)}(\lambda),y_{i}^{(h)}(\lambda)\},
\end{align}
where for any $\lambda \in \mathbb{R}_+$, $y_i^{(l)}(\lambda)$ and $y_i^{(h)}(\lambda)$ are the smallest and the largest solutions of 
\begin{align*}
  \frac{4 z}{  \left( \Var z^2 + 2 \right)^2} = \frac{\psi_i(c_i)}{n+1} \lambda^2.
\end{align*}
Therefore, the platform's problem becomes finding the optimal $i$ and the optimal $\lambda$. We search for the optimal $i$ by considering all elements of $\mathcal{N}$. We also search over the optimal $\lambda$ by considering a grid search. To form a grid for the possible optimal values of $\lambda$, we establish the following upper bound and lower bound on the optimal $\lambda$:
\begin{align*}
    \bar{y}_{i}= y^{(h)} \left( \left(\frac{(n+1) 3 \sqrt{3}}{\psi_{i-1}(c_{i-1}) 8 \sqrt{2 \Var}} \right)^{1/2} \right) \text{ and }  \underline{y}_{i}= \frac{n}{  \Var +  \left(\frac{\sqrt{2}n \left(\sum_{j=1}^n \psi_j(c_j) \right)}{(n+1)} \right)^{2/3}  }.
\end{align*}
%%%%%%
\begin{algorithm*}[t]
	\KwIn{The vector of privacy sensitivities $(c_1, \dots, c_n)$ and $\epsilon \in \mathbb{R}_{+}$}
    Sort the terms $ \psi_i(c_i)$, and without loss of generality, let us assume 
\begin{align*}
    \psi_1(c_1) \le \dots \le  \psi_n(c_n).
\end{align*}
\For{$i=1$ to $n$}{
    Let $\Delta$ be the maximum Lipschitz parameter of functions $\frac{n+1}{\lambda}$,  $y_j^{((h))}(\lambda)$, and $y^{(l)}_j(\lambda)$ over $\lambda \in [\underline{y}_{i}, \bar{y}_{i}]$;\\
    Find $$\lambda_{i} \in \mathrm{Grid}(i, \frac{\epsilon}{\Delta})=\left\{k \frac{\epsilon}{\Delta}:~ k=\lfloor \underline{y}_{i} \frac{\Delta}{\epsilon} \rfloor, \dots, \lceil \bar{y}_{i} \frac{\Delta}{\epsilon} \rceil\right\}$$ as the solution of
    \begin{align*}
    \min_{\lambda \in \mathrm{Grid}(i, \frac{\epsilon}{\Delta})} \min\left\{\frac{n+1}{\lambda} + \sum_{j=1}^{i} \psi_j(c_j) y^{(h)}_j(\lambda)  , \frac{n+1}{\lambda} + \sum_{j=1}^{i-1} \psi_j(c_j) y^{(h)}_j(\lambda) + \psi_j(c_j) y^{(l)}_j(\lambda) \right\};
\end{align*}
Let $\mathrm{OBJ}_{i}$ be the objective evaluated at $y^{(i)}_j=y_{j}^{(h)}(\lambda_i)$ for $j \le i-1$, $y^{(i)}_{j}=0$ for $j \ge i+1$, and $y^{(i)}_{i}=y_{i}^{(h)}(\lambda_i)$ if the optimal solution of the above optimization is the first term and  $y^{(i)}_{i}=y_{i}^{(l)}(\lambda_i)$, otherwise;
}
\textbf{Output:} Letting $i^*= \argmin_{i \in \mathcal{N}} \mathrm{OBJ}_{i}$, the approximate solution is $ (y_1^{(i^*)}, \dots, y_n^{(i^*)})$.
	\caption{Computing the optimal privacy loss in the local setting}
	\label{mech:local:compute}
\end{algorithm*}
%%%%%%

Algorithm \ref{mech:local:compute} summarizes the above procedure and the following proposition states the formal result:
%%%%%%
\begin{proposition}\label{Pro:approximate:local}
For any vector of reported privacy sensitivities $(c_1, \dots, c_n)$ and $\epsilon >0$, Algorithm \ref{mech:local:compute} finds privacy loss levels for the local data acquisition mechanism whose cost (i.e., the platform's objective) is at most $1+\epsilon$ of the optimal cost in time $\mathrm{poly}(n, \frac{1}{\epsilon})$.\footnote{ $\mathrm{poly}(\cdot)$ denotes a function that is polynomial in its inputs.}
\end{proposition}
Notice that the approximation factor in Proposition \ref{Pro:approximate:local} depends on the underlying parameters and therefore we have a Polynomial Time Approximation Scheme (PTAS) for finding the optimal privacy loss levels. Also, the output of Algorithm \ref{mech:local:compute} satisfies $y_i \ge y_j$ when $\psi_i(c_i) \le \psi_j(c_j)$ and therefore, as shown in Proposition \ref{Pro:PaymentIdentity}, is implementable. 

%%%%%%
{
We conclude this section by highlighting that computing the payment function \eqref{eq:Pro:PaymentIdentity} necessitates integrating over the privacy loss levels $\diff_i(\cdot)$, which does not have an explicit characterization in our setting. However, in the appendix, we demonstrate that this integral (and, therefore, the payment function) can be approximated to achieve any desired level of accuracy $\epsilon$. Consequently, this approximation yields an $\epsilon$-approximate incentive compatibility ($\epsilon$-IC) mechanism, where the incentive compatibility constraint \eqref{eq:IC} is violated by at most $\epsilon$. The concept of $\epsilon$-IC has been previously employed in the literature (see, e.g., the literature review of \cite{balseiro2022mechanism}). In the appendix, we provide the formal definition of $\epsilon$-IC and outline how the payment function \eqref{eq:Pro:PaymentIdentity} can be approximated to achieve $\epsilon$-IC.  
}

%%%%%%
\section{Data acquisition with central versus local differential privacy}\label{sec:compare}
In this section, we compare the performance of the optimal data acquisition mechanism in the central and local privacy settings. 

First, let us consider a case in which there is no restriction on the estimator, i.e., the estimator does not need to be a linear combination of users' data with a Laplace mechanism. In this case, finding the optimal value of platform's objective function in the central (local) differential privacy setting is equivalent to solving problem \eqref{eqn:main_opt_platform} for all centrally (locally) differentially private estimators. Note that, as stated in Section \ref{sec:Environment}, any $\bm{\diff}$-locally differentially private estimator is $\bm{\diff}$-centrally differentially private as well. As a result, the platform's optimal objective in the central privacy setting is always weakly smaller than her optimal objective in the local privacy setting. This is because the minimization problem in the central setting is solved over a weakly larger set of estimators.
Next, we show that the same result holds even if we restrict our focus to the class of linear estimators.
%%%%%%%%%%%%%%%%%%%%%%%%%%%%%%
%%%%%%%%%%%%%%%%%%%%%%%%%%%%%%
%%%%%%%%%%%%%%%%%%%%%%%%%%%%%%
%%%%%%%%%%%%%%%%%%%%%%%%%%%%%%
\begin{proposition}\label{Pro:compare:privacylevels}
Let $\bm{\diff} = (\diff_i)_{i=1}^n$. For any $\bm{\diff}$-locally differentially private linear estimator:
\begin{equation*}
\htheta_\Local = \sum_{i=1}^n w_i \hat{x}_i
\quad \hat{x}_i = x_i + \Lap(1/\diff_i), 
\end{equation*}
with $\sum_{i=1}^n w_i = 1$, 
there exists a $\bm{\diff}$-differentially private linear estimator $\htheta_\Central$ such that
\begin{equation} \label{eqn:central_lower}
\E[ | \htheta_\Central - \theta  |^2]    
\leq \E[ | \htheta_\Local - \theta |^2].
\end{equation}
\end{proposition}
%%%%%%%%%%%%%%%%%%%%%%%%%%%%%%
%%%%%%%%%%%%%%%%%%%%%%%%%%%%%%
%%%%%%%%%%%%%%%%%%%%%%%%%%%%%%
This result consequently implies
that, for any locally differentially private linear estimator, there exists a centrally differentially private linear estimator which delivers the same privacy loss levels with (weakly) lower estimation error. By keeping privacy loss levels unchanged, the privacy cost and the payments will also remain unchanged in the platform's objective. Hence, Proposition \ref{Pro:compare:privacylevels} implies that the platform's optimal objective function under central differential privacy setting is weakly smaller than her optimal objective function under local differential privacy. The following corollary formally states this observation. 
%%%%%%%%%%%%%%%%%%%%%%%%%%%%%%
%%%%%%%%%%%%%%%%%%%%%%%%%%%%%%
\begin{corollary}\label{cor:compare:central:local}
For any reported vector of privacy sensitivities $(c_1, \dots, c_n)$, the optimal solution of the local privacy optimization problem \eqref{eq:thm:local} is not smaller than the optimal solution of the central privacy optimization problem \eqref{eq:thm:central}.
\end{corollary}
%%%%%%%%%%%%%%%%%%%%%%%%%%%%%%
%%%%%%%%%%%%%%%%%%%%%%%%%%%%%%

\subsection{An illustrative example}
We next illustrate the difference between the performance of our proposed central privacy mechanism and our proposed local privacy mechanism in the context of a simple example. We consider two users with uniform privacy sensitivities drawn from $[1,2]$ (so that the virtual privacy sensitivity of user $i \in \{1,2\}$ becomes $2c_i-1$ for $c_i \in [1,2]$) and $\Var=1/4$.

\begin{figure}
     \centering
     \begin{subfigure}[b]{0.32 \textwidth}
         \centering
         \includegraphics[width= \textwidth]{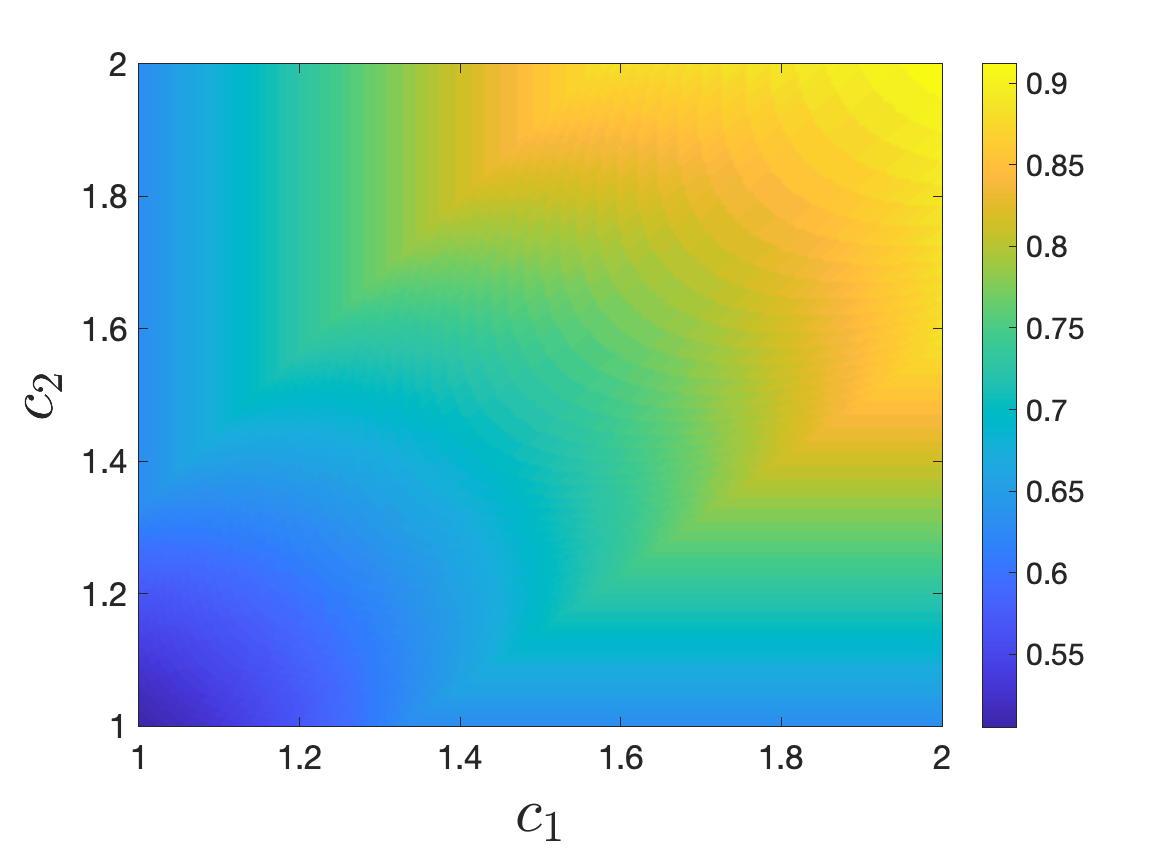}
         \caption{}
         \label{fig:central:variance}
     \end{subfigure}
     \hfill
     \begin{subfigure}[b]{0.32 \textwidth}
         \centering
         \includegraphics[width=\textwidth]{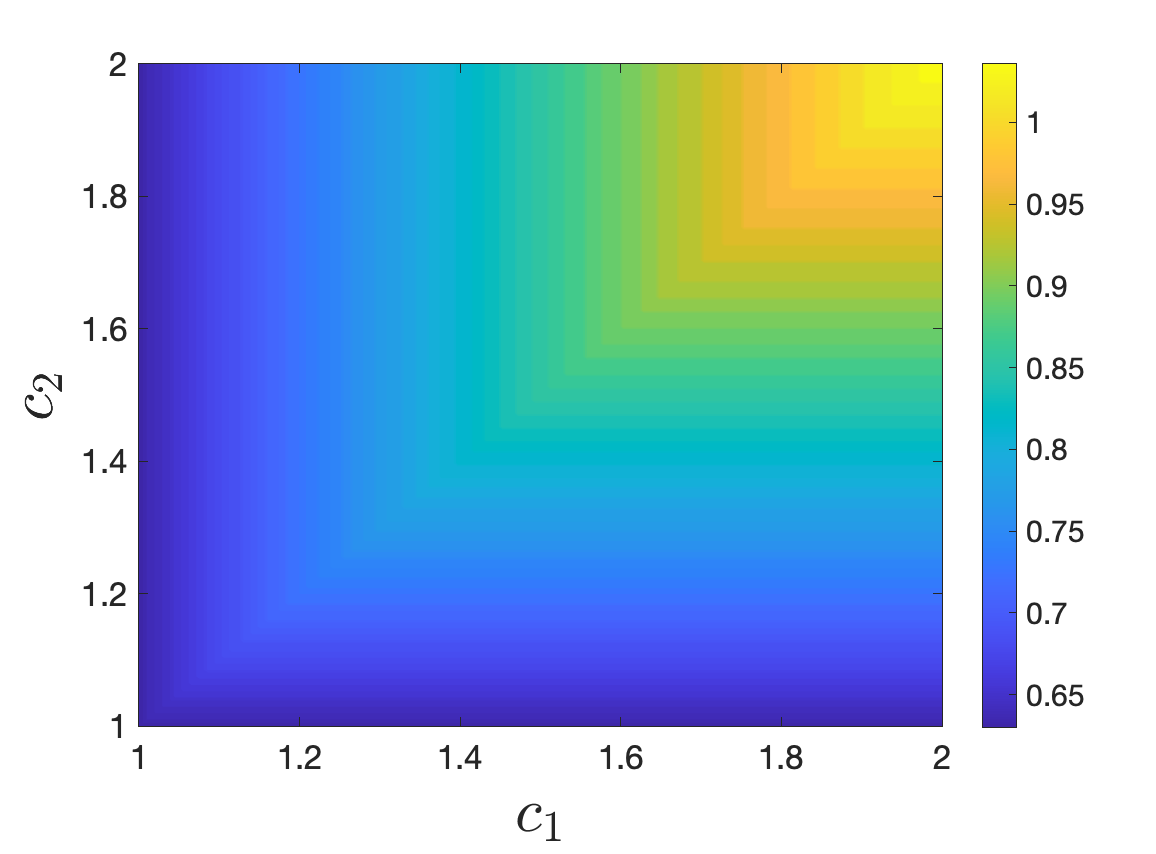}
         \caption{}
         \label{fig:local:variance}
     \end{subfigure}
     \hfill
     \begin{subfigure}[b]{0.32\textwidth}
         \centering
         \includegraphics[width=\textwidth]{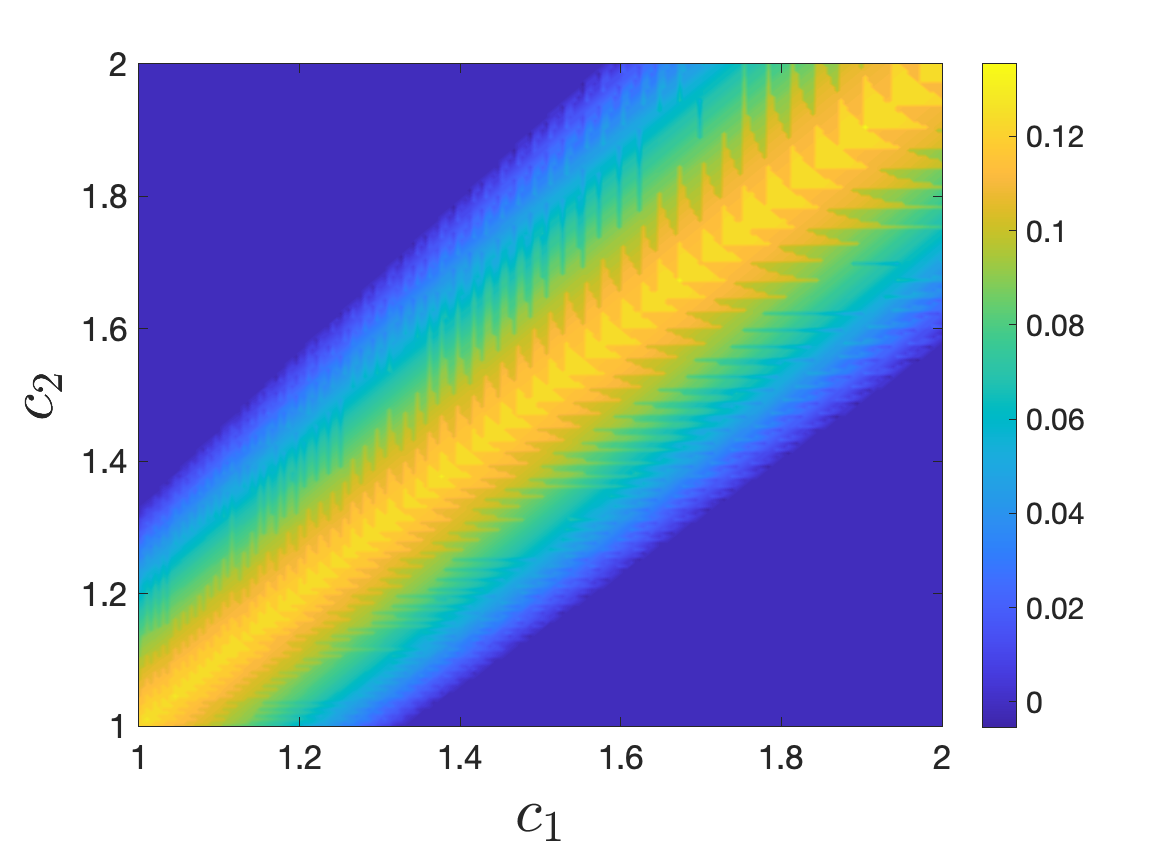}
         \caption{}
         \label{fig:diff:variance}
     \end{subfigure}
     \label{fig:variance}
        \caption{(a) the variance in the central setting, (b) the variance in the local setting, and (c) the variance in the local minus the central setting for two users with $\Var=1/4$ and uniform privacy sensitivities over $[1,2]$ as a function of the privacy sensitivities $(c_1, c_2)$.}
\end{figure}

\begin{figure}
     \centering
     \begin{subfigure}[b]{0.32\textwidth}
         \centering
         \includegraphics[width=\textwidth]{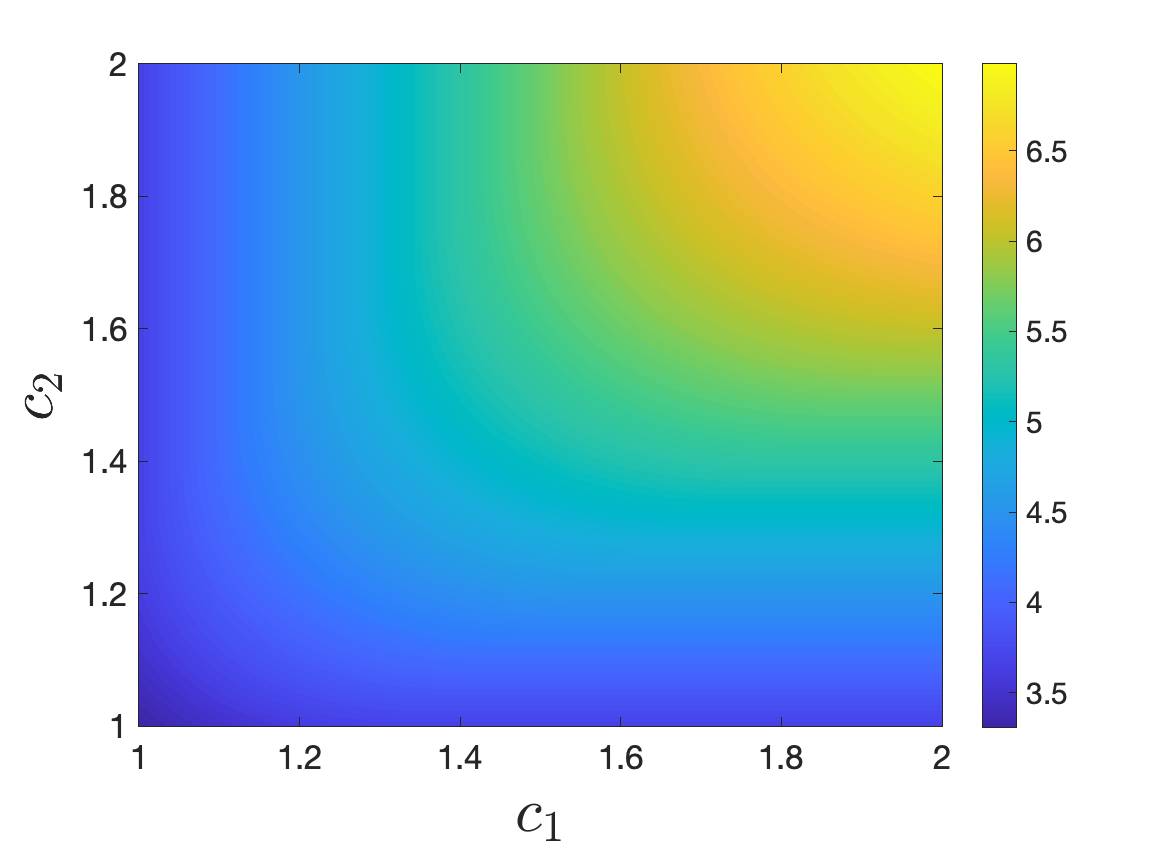}
         \caption{}
         \label{fig:central:objective}
     \end{subfigure}
     \hfill
     \begin{subfigure}[b]{0.32\textwidth}
         \centering
         \includegraphics[width=\textwidth]{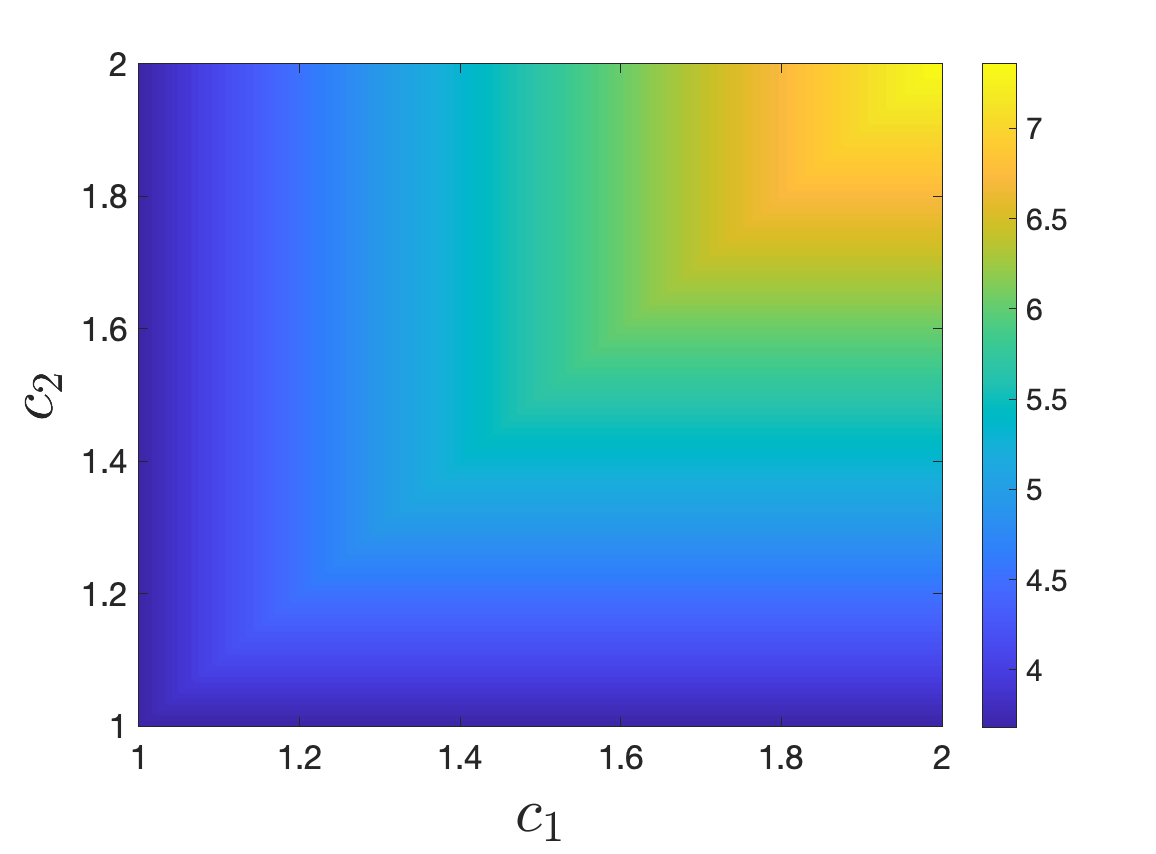}
         \caption{}
         \label{fig:local:objective}
     \end{subfigure}
     \hfill
     \begin{subfigure}[b]{0.32\textwidth}
         \centering
         \includegraphics[width=\textwidth]{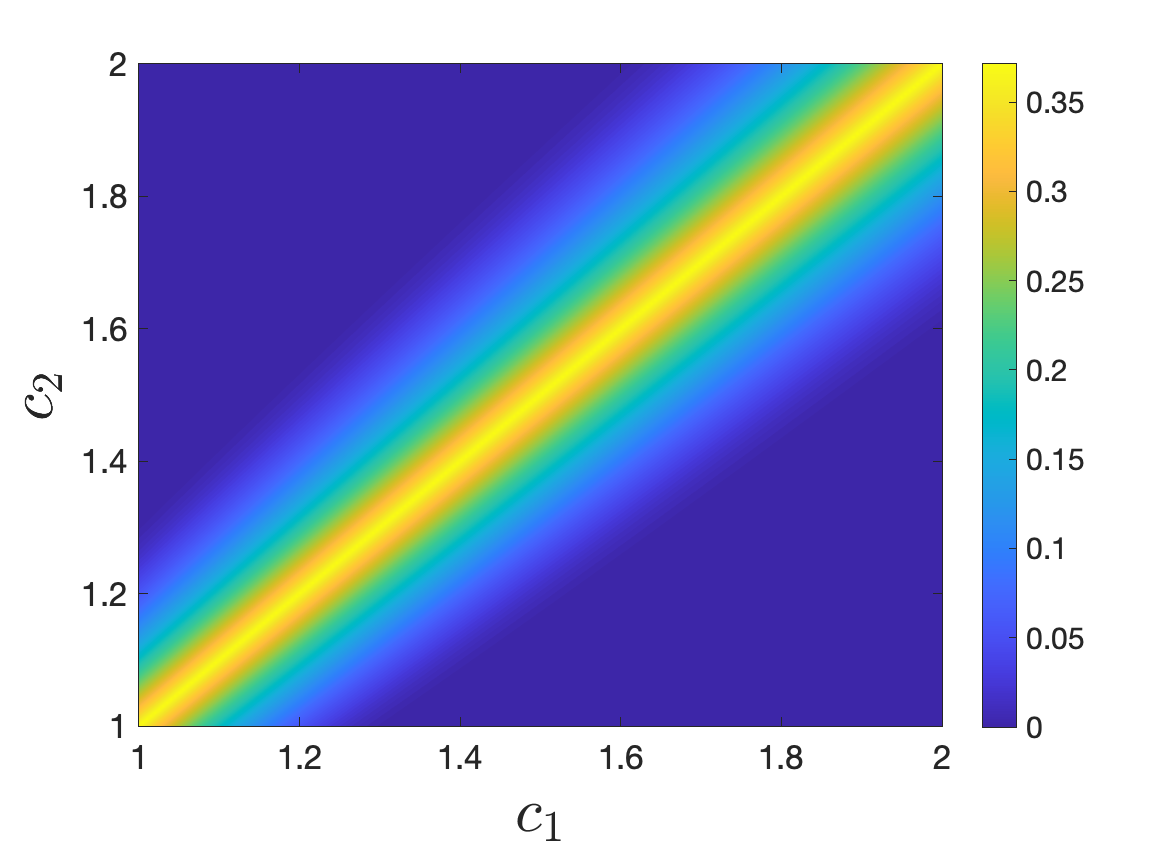}
         \caption{}
         \label{fig:diff:objective}
     \end{subfigure}
        \label{fig:objective}
        \caption{(a) the analyst's objective in the central setting, (b) the analyst's objective in the local setting, and (c) the analyst's objective in the local minus the central setting for two users with $\Var=1/4$ and uniform privacy sensitivities over $[1,2]$ as a function of the privacy sensitivities $(c_1, c_2)$.}
\end{figure}

Figures \ref{fig:central:variance}, \ref{fig:local:variance}, and  \ref{fig:diff:variance} depict the variance of the estimator for the central setting, the local setting, and their difference, respectively for all pairs of privacy sensitivities $(c_1, c_2)$. We observe that the variance in the central setting is always weakly larger than the variance in the local setting. This is because the local setting provides a stronger privacy guarantee and will hurt the variance of the final estimator. We also observe that when there is a large discrepancy between the two privacy sensitivities, the variance of the central and the local settings are equal. This is because the platform obtains all of its data from only one of the users and therefore central and local setting become identical. Further, when the two costs are very close to each (i.e., $c_1 \approx c_2$), the platform's weight for the data of each of the users in the estimator become close to each other. This implies that the variance of the local and the central setting become very close to each other.

Figures \ref{fig:central:objective}, \ref{fig:local:objective}, and \ref{fig:diff:objective} depict the platform's objective for the central setting, the local setting, and their difference, respectively for all pairs of privacy sensitivities $(c_1, c_2)$. We observe that the cost in the central setting is always weakly smaller than the cost in the local setting. This is again because the local setting provides a stronger privacy guarantee and will hurt the platform's objective. When there is a large discrepancy between the two privacy sensitivities (i.e. $|c_1-c_2| \approx 1$), the objective of the central and the local settings are equal. This is because the platform obtains all of its data from only one of the users and therefore central and local setting become identical.

\begin{figure}
     \centering
      \begin{subfigure}[b]{0.32\textwidth}
         \centering
         \includegraphics[width=\textwidth]{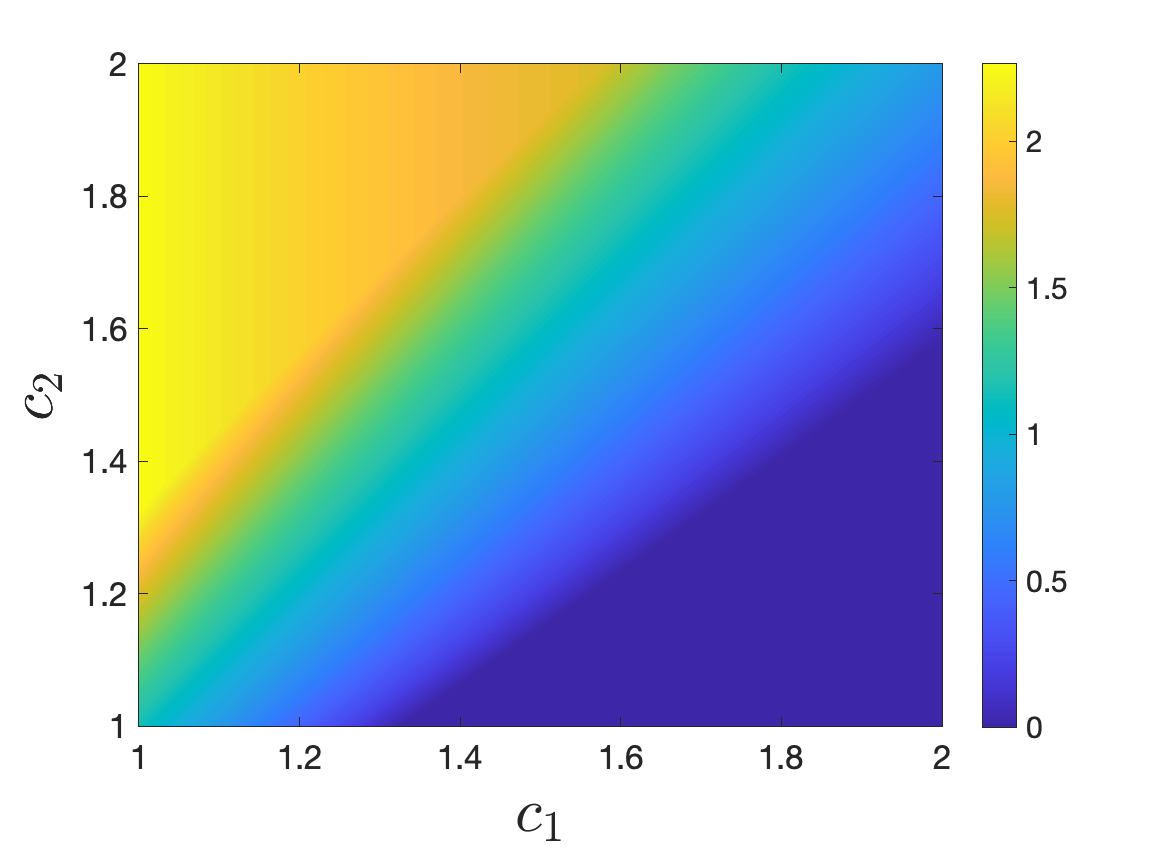}
         \caption{}
         \label{fig:allocation:central}
     \end{subfigure}
     \hfill
      \begin{subfigure}[b]{0.32\textwidth}
         \centering
         \includegraphics[width=\textwidth]{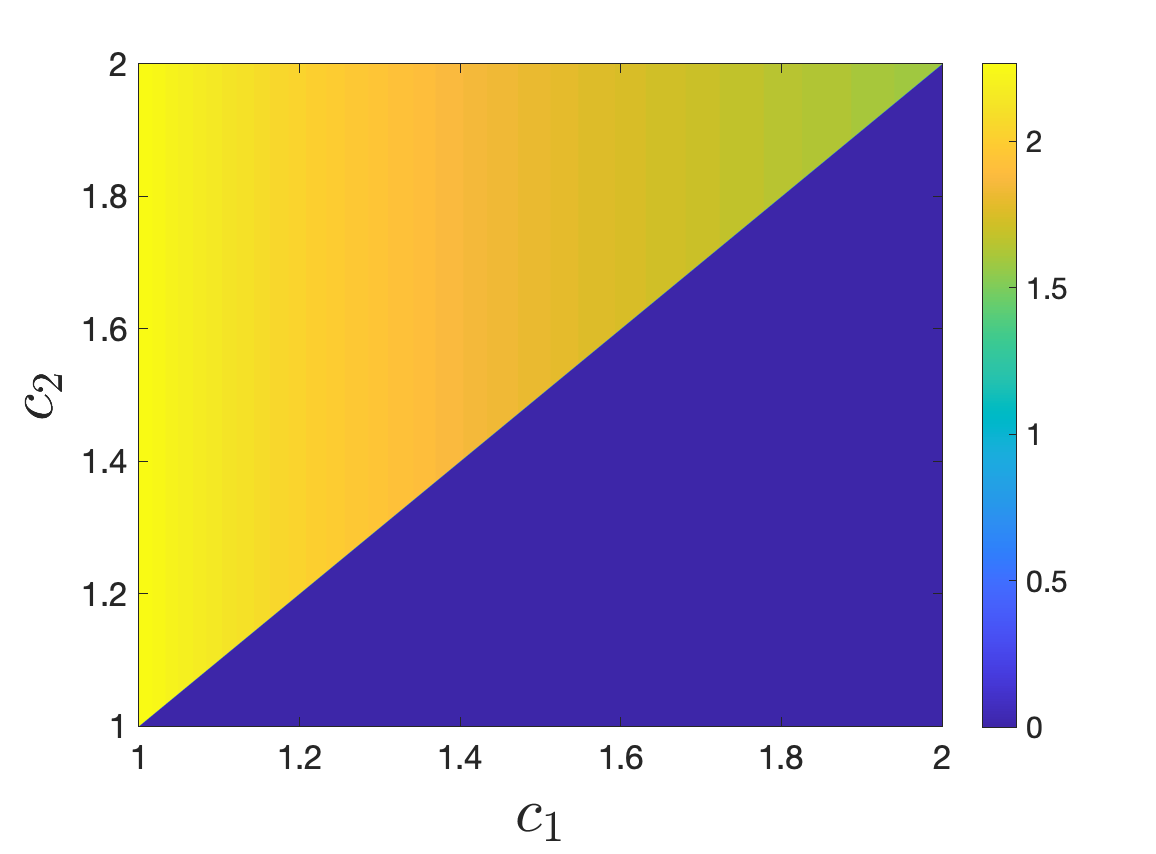}
         \caption{}
         \label{fig:allocation:local}
     \end{subfigure}
     \hfill
      \begin{subfigure}[b]{0.32\textwidth}
         \centering
         \includegraphics[width=\textwidth]{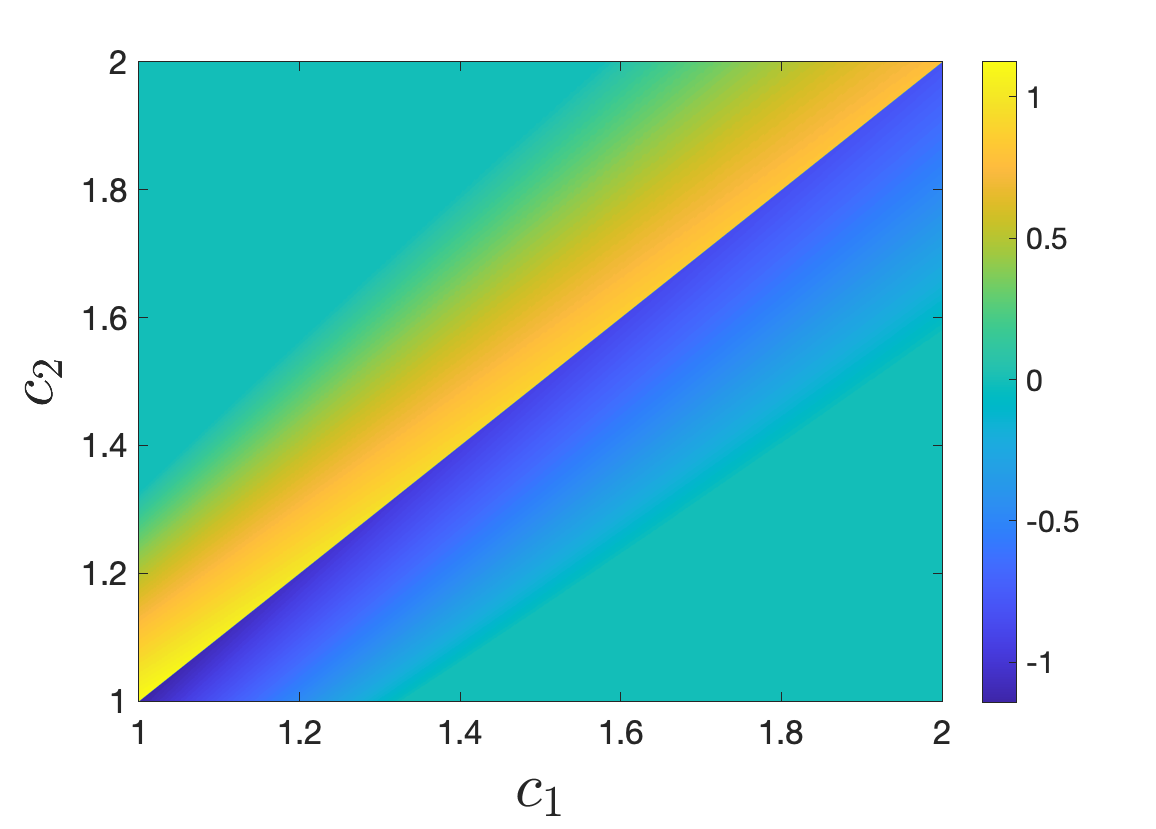}
         \caption{}
         \label{fig:allocation:diff}
     \end{subfigure}
        \label{fig:expected}
        \caption{(a) user $1$'s optimal privacy loss in the central setting, (b) user $1$'s optimal privacy loss in the local setting, and (c) the difference between user $1$'s optimal privacy loss in the local setting and the central setting as a function of $(c_1, c_2)$. Here, we have $\Var=1/4$ and the privacy sensitivities are uniform over $[1,2]$.}
\end{figure}

Figures \ref{fig:allocation:central} and  \ref{fig:allocation:local} depict the optimal allocation of user $1$ in the central and local settings, and Figure \ref{fig:allocation:diff}  depicts the optimal allocation of user $1$ in the local setting minus the central setting. We observe that, the privacy loss level of a user in the local setting can be lower (i.e., better privacy) than the central setting. We next provide a formal statement for this observation in the context of an example. 

%%%%
\subsection{Optimal privacy loss levels in the central versus local setting}
Here, we illustrate that the privacy loss level of a user in the optimal local data acquisition mechanism can be smaller than  the central setting, i.e., the optimal mechanism in the local setting may provide strictly better privacy guarantees to a user compared to the central setting. { To simplify the notation, we work with the virtual cost rather than the privacy sensitivity. Note that this is without loss of generality as we do not pose any assumption on the virtual cost (other than Assumption \ref{assump:MHR}).}
%%%%%%%%%%%%%%%%%%%%
%%%%%%%%%%%%%%%%%%%%
\begin{proposition}
\label{proposition:central_vs_local_allocation}
Assume the virtual costs of users $1, \cdots, n-1$ are all equal to $\psi_1$. We also denote the virtual cost of user $n$ by $\psi_n$. Then, for\footnote{$f(n)=\Theta(g(n))$ means that there exist $n_0$ and constants $c_1, c_2$ such that for $n \ge n_0$, we have $c_1 g(n) \le f(n) \le c_2 g(n)$.}
\begin{align*}
\psi_n \in \left[ 
\psi_1 + \Theta\left(\frac{1}{n^{2/3}}\right)
, \psi_1 + \Theta\left(\frac{1}{n^{1/3}}\right) 
\right],
\end{align*}
the optimal privacy loss level of user $n$ in the local setting is zero, while her optimal privacy level in the central setting is non-zero.
\end{proposition}
%%%
Proposition \ref{proposition:central_vs_local_allocation} follows by comparing the optimal solutions of the non-convex programs \eqref{eq:thm:central} and \eqref{eq:thm:local}. 
Note that, since the local privacy is a more stringent requirement, one may expect that, in the optimal local mechanism, the delivered privacy guarantees to users are worse (higher privacy loss levels) compared to the central setting. This proposition shows that may not be the case.
To gain an intuition, note that, as the privacy sensitivity of user $n$, i.e., $\psi_n$, increases, her privacy loss level, in both central and local settings, goes to zero. 
{ Recall that in the central estimator \eqref{sec:central_optimal}, the privacy loss level of user $n$ is denoted as $w_n \eta$. To achieve a near-zero privacy loss level, we have two possibilities: either $\eta$ approaches zero (which results in a larger estimation error due to the Laplace noise variance being $2/\eta^2$), or $w_n$ must approach zero. Similarly, in the local estimator, user $n$'s privacy loss level is represented by $\diff_n$, and if this term approaches zero, $w_n$ must also approach zero; otherwise, the estimation error will be considerably large. In summary,} to deliver such small privacy loss, in the optimal central and local mechanisms, the platform must allocate zero weight to user $n$'s data. Otherwise, the added noise in the platform's estimator makes the estimation error unbounded.

In the local setting, each user first maps her data to a private version and then shares it with the platform, meaning that, by definition, the privacy loss level of a user only depends on this mapping and not the platform's estimator. This in turn implies that the reallocation of the weights will not impact the privacy loss levels delivered to other users and hence will not change their compensations. In the central setting, however, decreasing user $n$'s weight in the optimal estimator, increases the allocated weight to other users' data (because the sum of the wights add up to one). This in turn increases their allocated privacy loss levels and hence their compensation. Therefore, in the central setting the platform is more reluctant to give up on user $n$'s data and increase other users' allocated weights, which is what we establish in Proposition \ref{proposition:central_vs_local_allocation}.

%%%%%%

%%%%%%%%%%%%%%%%%%%%%%%

%%%%%%%

\section{Conclusion and discussion}\label{sec:conclusion}
We study the design of mechanisms for acquiring data from users with privacy concerns who also benefit from a lower estimation error. We consider two architectures: (i) central privacy setting in which users share their data with the platform, incur some privacy loss and get compensated for their loss. The platform then combines the data of users and outputs an estimator that guarantees the promised heterogeneous privacy loss to  each user; and (ii) local privacy setting in which users share a differentially private version of their data with the platform, incur some privacy loss and get compensated for their loss. The platform then combines the data of users and outputs an estimator.  
 
In both cases, we first establish that a linear estimator with proper weights and added Laplace noise achieves the nearly optimal minimax bound, which is of independent interest. Building on this characterization, we then optimally solve the corresponding mechanism design problem for both settings. In the central privacy setting, we establish a polynomial time score-based algorithm that finds the optimal privacy loss levels. In the local privacy setting, however, we establish a Polynomial Time Approximation Scheme (PTAS) for finding the optimal privacy losses.
 
Finally, we compare the performance of the central and the local architectures. In particular, we show that the platform's utility in the central privacy setting is always higher than in the local privacy setting. But, there is no dominance in terms of the optimal privacy loss level, i.e., depending on her privacy sensitivity, a user may have a higher or lower privacy loss in the central setting compared to the local setting. 

{
In our analysis, we utilized a set of simplifying assumptions to aid our investigation. Here, we would like to underline these assumptions, furnish reasons for their use within the context of our application, and outline potential avenues for future exploration. As an illustrative example, we consider the purchase of medical data by emerging companies such as Hu-manity.co, where users receive compensation for sharing their medical information.

\begin{itemize}
    \item \textbf{Verifiability of data:} We made the assumption  that while users have the ability to misrepresent their privacy sensitivity, they cannot falsify the actual data itself. This assumption is applicable in scenarios where users sell the "rights" to their data, and the platform collects data, such as in the context of Hu-manity.co, where medical data from users is gathered.
    \item \textbf{Independence of data and privacy sensitivity:} We have assumed that there is no relation between users' data and their privacy sensitivity. In the context of the Hu-manity.co application, this implies that users become aware of their privacy sensitivity before their actual medical data (i.e., realized data) is revealed. This assumption arises from situations where users are unable to collect/process the data themselves. However, we recognize that in other applications, such as the sharing of financial data, this assumption may not hold. Without this assumption, there may be a sample bias in the data collected by the platform, which would require correction. We leave exploring this direction as an interesting future avenue of research.

    \item\textbf{Extensions in estimation models:} We focused on the private mean estimation task from a population, but it would be interesting to extend our results to more complex estimation models. Here are potential extensions: estimating a multi-dimensional underlying parameter denoted as $\theta$ and estimating an underlying parameter $\theta$ when customers have data $(X_i, Y_i)$ defined as $Y_i = X_i' \theta + Z_i$.

It is worth noting that extending our results to these scenarios requires establishing the equivalent counterparts of Theorems \ref{theorem:LB_Central} and \ref{theorem:LB_Local}. This involves finding the (minimax) optimal estimator while considering heterogeneous differential privacy concerns. Once such an estimator is obtained (or when the minimax optimality of the estimator is not a concern), our results on the characterization of the mechanism continue to hold. This means that similar to the derivations presented in Section \ref{sec:mechanism_formulation}, the platform's problem revolves around solving a point-wise optimization problem. However, it is important to note that, similar to our current setting, this problem can also be non-convex, necessitating the development of an efficient algorithm for its solution.
\item \textbf{Additive user utility:} In our model, each user's utility is determined as the payment received minus the mean squared error (MSE) of the estimator, minus the privacy sensitivity multiplied by their level of differential privacy. This utility form assumes two key assumptions. First, it assumes additivity and that the privacy sensitivity is directly multiplied by the privacy level. While these assumptions simplify the derivations, it is worth noting that, similar to the classic mechanism design setting, all the results extend as long as the user utility is quasi-linear. In other words, the utility is a function of the MSE, privacy sensitivity, and privacy level and is subtracted by the payment. As an example, it is possible for the MSE and payment to have user-specific known coefficients.

Second, the model assumes that there is only one privately known user parameter, which is privacy sensitivity. If the users have privately known weights for either the MSE or payment, the problem becomes a multi-dimensional mechanism design whose study is beyond the scope of this paper.

\item \textbf{Trusting the platform:} In the central setting, we assumed there is trust between users and the platform: users share their data with the platform, relying on the platform to handle the data responsibly and deliver the promised privacy level without exploiting it for other purposes (this form of credibility is present in data acquisition mechanisms and not classic auctions that are studied in   \cite{akbarpour2020credible}). This concern regarding platform credibility motivated us to also study the local privacy setting. In this setting, data is privatized directly on the user side, granting users control over the implementation of privacy measures. This local structure has been effectively implemented by various tech companies, such as Apple, which has incorporated local differential privacy techniques into their data handling processes (see, e.g., \cite{AppleDP}). 

\end{itemize}

}

\section*{Acknowledgment}
The authors thank the entire review team for the many fruitful comments and suggestions that improved both the exposition and results of the paper. We are also grateful to Kunal Talwar for useful conversations and comments. An abstract of this work appeared in the Proceedings of the 23rd ACM Conference on Economics and Computation (EC 2022), whose reviewers provided inspiring comments. Alireza Fallah acknowledges support from the Apple Scholars in AI/ML Ph.D. fellowship.
%%%%%%%%
%\newpage
\bibliography{Main_arxiv}
%%%%%%%%%%%%%%%
\newpage

\renewcommand{\theequation}{\mbox{A\arabic{equation}}}

\renewcommand{\thesection}{\mbox{A\arabic{section}}}

\setcounter{equation}{0}

\setcounter{section}{0}

\renewcommand{\thelemma}{\mbox{A\arabic{lemma}}}

\setcounter{lemma}{0}
\APPENDICES
\section{Proofs and additional details}
This appendix includes the omitted proof from the text and the additional results discussed in the text.
%%%%
\subsection{Proofs}

\subsubsection*{Proof of Lemma \ref{lemma:analyst_DP}}
Since $|Z_i| \leq \frac{1}{2}$, the difference of every two realizations of $X_i$ would be bounded by one. Therefore, the sensitivity of $\htheta$ to $x_i$, defined in \eqref{eqn:sensitivity_def}, is given by $w_i(\bm{c})$. Hence, Lemma \ref{lemma:Laplace} immediately implies the result. $\blacksquare$

%%%%%%%%%%%%%%%%%%%%%%%%%%%%%%%%%%%%%%%%%%%%%%%%%%%
%%%%%%%%%%%%%%%%%%%%%%%%%%%%%%%%%%%%%%%%%%%%%%%%%%%
%%%%%%%%%%%%%%%%%%%%%%%%%%%%%%%%%%%%%%%%%%%%%%%%%%%
%%%%%%%%%%%%%%%%%%%%%%%%%%%%%%%%%%%%%%%%%%%%%%%%%%%
\subsubsection*{Proof of Theorem \ref{theorem:LB_Central}}
For the sake of subsequent analysis, we find it helpful to recall the definition of two well-known distribution distances. Let $P$ and $Q$ be two distributions, defined over a probability space $(\Omega, \matF)$. Then,
\begin{itemize}
\item 
the total variation (TV) distance is denoted by $\|P-Q\|_{\TV}$, and is given by
\begin{equation*}
\|P-Q\|_{\TV} := \sup_{A \in \matF} |P(A) - Q(A)| = \frac{1}{2} \int_\Omega |dP - dQ|.    
\end{equation*}
%%%%%%
\item when $P$ is absolutely continuous with respect to $Q$, the Kullback–Leibler (KL) distance is denoted by $D_{\KL}(P,Q)$, and it is given by
\begin{equation*}
D_{\KL}(P,Q) := \int \log \frac{dP}{dQ} dP.
\end{equation*}
\end{itemize}
One inequality that we particularly find it helpful for analysis is the Pinsker's inequality which states that 
\begin{equation} \label{eqn:Pinsker}
\|P-Q\|_{\TV}^2 \leq \frac{1}{2} D_{\KL}(P,Q).    
\end{equation}

To prove the lower bound \eqref{eqn:central_lb}, we use Le Cam's method \cite{yu1997assouad} which is a well-known technique in deriving minimax lower bounds. The main idea of Le Cam's method is reducing the estimation problem to a testing problem. 
More formally, let $P_1$ and $P_2$ be two distributions in $\matP_k$ with
\begin{equation} \label{eqn:gamma_mean}
\gamma:= \frac{1}{2} |\theta(P_1) - \theta(P_2)|.     
\end{equation}
Furthermore, for $j \in \{1,2\}$, let $Q_j$ be the marginal distribution of $\htheta$, given that the samples $X_{1:n}$ are all drawn from $P_j$, i.e.,
\begin{equation} \label{eqn:Q_j}
Q_j(A) = \int_{\R^n} \Pb(\htheta(x_{1:n}) \in A) dP_j^n(x_{1:n}),     
\end{equation}
for any measurable set $A$. Then, by Le Cam's method, we have (see \cite{barber2014privacy} for more details)
\begin{equation} \label{eqn:LeCam_central}
 \matL_c(\matP_k, \theta, \bm{\diff}) \geq \gamma^2 \left (\frac{1}{2} - \frac{1}{2} \|Q_1 - Q_2\|_{\TV} \right ).   
\end{equation}
Next, we bound $\|Q_1 - Q_2\|_{\TV}$ in the following lemma.
%%%%%%%%%%%%%%%%%%%%%%%%%%%%%%%%%%%%%%%%%%%%%%%%%%%%
%%%%%%%%%%%%%%%%%%%%%%%%%%%%%%%%%%%%%%%%%%%%%%%%%%%%
\begin{lemma} \label{lemma:central_lb_TV}
Let $P_1$ and $P_2$ be two distributions in $\matP$ such that $P_1$ is absolutely continuous with respect to $P_2$. Consider $Q_1$ and $Q_2$ as defined in \eqref{eqn:Q_j}. Then, for any $k \in \{0,1, \cdots, n\}$, 
\begin{equation} \label{eqn:lemma_central_lb}
\|Q_1 - Q_2\|_\TV \leq 2 \|P_1 - P_2\|_\TV \sum_{i=1}^k (e^{\diff_i}-1) + \sqrt{\frac{n-k}{2} D_{\KL}(P_1,P_2)}.    
\end{equation}
\end{lemma}
%%%%%%%%%%%%%%%%%%%%%%%%%%%%%%%%%%%%%%%%%%%%%%%%%%%%
%%%%%%%%%%%%%%%%%%%%%%%%%%%%%%%%%%%%%%%%%%%%%%%%%%%%
We prove this lemma at the end of this section. Now, using this lemma, let us complete the proof of \eqref{eqn:central_lb}. Let $\delta \in [0,1/2]$, and define $P_1$ and $P_2$ as
\begin{equation} \label{eqn:P_1_2}
P_1(-1/2) = P_2(1/2) = \frac{1+\delta}{2}, \quad P_1(1/2) = P_2(-1/2) = \frac{1-\delta}{2}.    
\end{equation}
Obviously $P_1, P_2 \in \matP^*$. Also, for $i \in \{1,2\}$, $\E_{P_i}[X] = (-1)^i \delta/2$, and hence $\gamma = \delta/2$. In addition, by definition, we have
\begin{align*}
\|P_1 - P_2\|_\TV &= \frac{1}{2} \times 2 \times \left ( \frac{1+\delta}{2} -  \frac{1-\delta}{2}  \right) = \delta, \\
D_{\KL}(P_1, P_2) &= \delta \log \frac{1+\delta}{1-\delta} \leq 3 \delta^2,
\end{align*}
where the last inequality holds for $\delta \in [0,1/2]$. 
Hence, by Lemma \ref{lemma:central_lb_TV}, along with the fact that $e^{\diff_i}-1 \leq 2\diff_i$ for $\diff_i \leq 1$, we have that for 
\begin{equation*}
\|Q_1 - Q_2\|_\TV \leq 4 \delta \sum_{i=1}^k \diff_i + \delta \sqrt{\frac{3(n-k)}{2}}.    
\end{equation*}
Therefore, using \eqref{eqn:LeCam_central}, we obtain
\begin{equation}
 \matL_c(\matP_k, \theta, \bm{\diff}) \geq \frac{\delta^2}{8} \left ( 1 - \delta \left [4 \sum_{i=1}^k \diff_i + \sqrt{\frac{3(n-k)}{2}} \right ] \right ).
\end{equation}
Choosing 
\begin{equation*}
\delta = \left (8 \sum_{i=1}^k \diff_i + \sqrt{6(n-k)} \right )^{-1} \wedge \frac{1}{2},   
\end{equation*}
implies
\begin{equation}
 \matL_c(\matP_k, \theta, \bm{\diff}) \geq \frac{1}{16} \left ( \left (8 \sum_{i=1}^k \diff_i + \sqrt{6(n-k)} \right )^{-2} \wedge \frac{1}{4} \right ).
\end{equation}
Using inequality $(x+y)^2 \leq 2(x^2+y^2)$ with $x = 8 \sum_{i=1}^k \diff_i$ and $y = \sqrt{6(n-k)}$ completes the proof of \eqref{eqn:central_lb}.

Next, we show the upper bound \eqref{eqn:central_ub}. First note that, since $|X| \leq 1/2$ almost surely for any $P \in \matP$, we have $\theta(P) \leq 1/2$ for any $P \in \matP$. Hence, $\htheta = 0$, which is a linear estimator in the form of \eqref{eq:estimator:central} with $w_i = 0$ for all $i$ and $\eta = \infty$, leads to $\E[| \htheta - \theta |^2]  \leq 1/4$. Hence, it suffices to find a linear estimator $\htheta$ with Laplace mechanism such that
\begin{equation} \label{eqn:central_ub_recall}
\E[| \htheta - \theta |^2]   
\leq \mathcal{O}(1) \log(n+1) \max_{j} \frac{1}{n-j + (\sum_{i=1}^j \diff_i)^2}.
\end{equation}

To do so, let $k^*$ be the largest $k \in \{0,1,\cdots n-1\}$ such that
\begin{equation} \label{eqn:threshold_k}
\diff_{n-k} > \frac{1}{\sqrt{k+1}},    
\end{equation}
if such $k$ exists. Now, we consider two cases:

$\bullet$
First, assume such $k$ does not exists. Then, consider linear estimator
\begin{equation*}
\htheta = \sum_{i=1}^n \frac{\diff_i}{\eta} x_i + \Lap\left(\frac{1}{\eta}\right),    
\end{equation*}
with 
\begin{equation*}
\eta = \sum_{i=1}^n \diff_i.    
\end{equation*}
In this case, it is straightforward to see
\begin{equation} \label{eqn:no_k_1}
\E[ | \htheta - \theta |^2] \leq \frac{ \sum_{i=1}^n \diff_i^2 + 2}{\eta^2} 
= \frac{\sum_{i=1}^n \diff_i^2 + 2}{(\sum_{i=1}^n\diff_i)^2},
\end{equation}
where we the fact that the variance is bounded by one since the absolute value of the random variable almost surly bounded by one. Next, note that, since \eqref{eqn:threshold_k} does not hold for any $k$, we have $\diff_{n-k} \leq \frac{1}{\sqrt{k+1}}$ for any $k$, which implies that
\begin{equation*}
\sum_{i=1}^n \diff_i^2 \leq \sum_{i=1}^n \frac{1}{i} = \mathcal{O}(1) \log(n+1).    
\end{equation*}
Plugging this relation into \eqref{eqn:no_k_1} implies 
\begin{align*}
\E[| \htheta - \theta |^2] & \leq \mathcal{O}(1) \log(n+1) \frac{1}{(\sum_{i=1}^n\diff_i)^2}.
\end{align*}
This completes the proof of \eqref{eqn:central_ub_recall} since 
\begin{equation*}
\frac{1}{(\sum_{i=1}^n\diff_i)^2}    
\end{equation*}
is 
\begin{equation*}
\frac{1}{n-j + (\sum_{i=1}^j \diff_i)^2}    
\end{equation*}
with $j=n$. 

$\bullet$ Now assume there exists at least one $k$ that satisfies \eqref{eqn:threshold_k} (and hence the aforementioned $k^*$ is well-defined.) 
As a result, we have 
\begin{align}
& \diff_1 \leq \frac{1}{\sqrt{n}}, \cdots, \diff_{n-k^*-1} \leq \frac{1}{\sqrt{k^*+2}},  \label{eqn:below_k_star} \\     
& \diff_n \geq \cdots \diff_{n-k^*} > \frac{1}{\sqrt{k^*+1}}. \label{eqn:above_k_star}
\end{align}
In this case, consider the following linear estimator
\begin{equation*}
\htheta = \sum_{i=1}^{n-k^*-1} \frac{\diff_i}{\eta} x_i + \sum_{i=n-k^*}^{n} \frac{1/\sqrt{k^*+1}}{\eta} x_i 
+ \Lap\left(\frac{1}{\eta}\right),
\end{equation*}
with \begin{equation*}
\eta = \sum_{i=1}^{n-k^*-1} \diff_i + \frac{k^*+1}{\sqrt{k^*+1}} 
= \sum_{i=1}^{n-k^*-1} \diff_i + \sqrt{k^*+1}.    
\end{equation*}
First, by Lemma \ref{lemma:analyst_DP}, this estimator is 
\begin{equation*}
(\diff_1, \cdots, \diff_{n-k^*-1}, \frac{1}{\sqrt{k^*+1}}, \cdots, \frac{1}{\sqrt{k^*+1}})  
\end{equation*}
differentially private, and hence, due to \eqref{eqn:above_k_star}, it is $(\diff_i)_{i=1}^n$-differentially private as well. Second, using the fact that variance is bounded by one, we have
\begin{align}
\E[| \htheta - \theta |^2] & \leq
\frac{\sum_{i=1}^{n-k^*-1} \diff_i^2 + \frac{k^*+1}{k^*+1} + 2}{\eta^2} \nonumber \\
& \leq \frac{\sum_{i=k^*+2}^{n} \frac{1}{i} + 3}{(\sum_{i=1}^{n-k^*-1} \diff_i + \sqrt{k^*+1})^2}, \label{eqn:central_ub_k_star}
\end{align}
where the last inequality follows from \eqref{eqn:below_k_star} and definition of $\eta$. To complete the proof, note that, the numerator of \eqref{eqn:central_ub_k_star} is upper bounded by $\mathcal{O}(1) \log(n+1)$ and its denominator is lower bounded by 
\begin{equation*}
\left(\sum_{i=1}^{n-k^*-1} \diff_i\right)^2 + k^*+1 = \left(\sum_{i=1}^{j} \diff_i\right)^2 + n-j \text{ with } j=n-k^*-1.   
\end{equation*}
Hence, \eqref{eqn:central_ub_recall} holds in this case as well.
$\blacksquare$
%%%%%%%%%%%%%%%%%%%%%%%%%%%%%%%%%%%%%%%%%%%%%%%%%%%%
%%%%%%%%%%%%%%%%%%%%%%%%%%%%%%%%%%%%%%%%%%%%%%%%%%%%
\subsubsection*{Proof of Lemma \ref{lemma:central_lb_TV}}
Let $\tilde{Q}$ be the marginal distribution of $\htheta$ given that $X_1, \cdots, X_k$ are drawn from $P_1$ and $X_{k+1}, \cdots, X_n$ are drawn from $P_2$, i.e.,  
\begin{equation} \label{eqn:tilde_Q}
\tilde{Q}(A) = \int_{\R^n} \Pb(\htheta(x_{1:n}) \in A) dP_1^k(x_{1:k}) dP_2^{n-k}(x_{k+1:n}).     
\end{equation}
Note that, we have
\begin{equation} \label{eqn:triangle_TV}
\|Q_1 - Q_2\|_{\TV} \leq \|Q_1 - \tilde{Q}\|_{\TV} + \|\tilde{Q} - Q_2\|_{\TV}.    
\end{equation}
The idea is to bound the two terms on the right hand side separately. In particular, we show
\begin{align}
\|Q_1 - \tilde{Q}\|_{\TV}^2 & \leq  \frac{n-k}{2} D_{\KL}(P_1,P_2), \label{eqn:triangle_1_KL} \\
\|\tilde{Q} - Q_2\|_{\TV} & \leq 2 \|P_1 - P_2\|_\TV \sum_{i=1}^k (e^{\diff_i}-1). \label{eqn:triangle_2_TV} 
\end{align}
If we show these two bounds, then plugging them into \eqref{eqn:triangle_TV} will show Lemma \ref{lemma:central_lb_TV}. 

$\bullet$ We start by showing \eqref{eqn:triangle_1_KL}. First, note that, by data processing inequality, we have
\begin{equation*}
\|Q_1 - \tilde{Q}\|_{\TV} \leq \| P_1^n - P_1^{k} P_2^{n-k} \|_{\TV}.     
\end{equation*}
Next, using Pinsker’s inequality \eqref{eqn:Pinsker}, we obtain
\begin{align*}
{ \| P_1^n - P_1^{k} P_2^{n-k} \|_{\TV}^2} & \leq \frac{1}{2} D_{\KL}(P_1^n, P_1^{k} P_2^{n-k}) \\
&\leq \frac{n-k}{2} D_{\KL}(P_1,P_2),
\end{align*}
where the second inequality follows from the chain rule for KL-divergence. This completes the proof of \eqref{eqn:triangle_1_KL}.

$\bullet$ Next, we show \eqref{eqn:triangle_2_TV}.
By total variation distance definition, it suffices to show that, for any measurable set $A$, $|\tilde{Q}(A) - Q_2(A)|$ is upper bounded by the right hand side of \eqref{eqn:lemma_central_lb}. To see this, first, note that we have
\begin{align} \label{eqn:lb_1.5}
\tilde{Q}(A) - Q_2(A) &= \int_{\R^n} \Pb(\htheta(x_{1:n}) \in A) ( dP_1^k(x_{1:k}) - {dP_2^k}(x_{1:k})) dP_2^{n-k}(x_{k+1:n})  \nonumber \\
&= \int_{\R^{n-k}} \Delta(x_{k+1:n}) dP_2^{n-k}(x_{k+1:n}), 
\end{align}
where 
\begin{equation} \label{eqn:lb_1}
\Delta(x_{k+1:n}) := \int_{\R^k} \Pb(\htheta(x_{1:n}) \in A) ( dP_1^k(x_{1:k}) - {dP_2^k}(x_{1:k})).    
\end{equation}
To show \eqref{eqn:triangle_2_TV}, it suffices to show 
\begin{equation} \label{eqn:triangle_3_TV}
|\Delta(x_{k+1:n})| \leq 2 \|P_1 - P_2\|_\TV \sum_{i=1}^k (e^{\diff_i}-1).     
\end{equation}
To do so, first, note that $dP_1^k(x_{1:k}) - dP_2^k(x_{1:k})$ can be cast as
\begin{equation*}
dP_1^k(x_{1:k}) - dP_2^k(x_{1:k}) = 
\sum_{i=1}^k dP_2^{i-1}(x_{1:i-1}) \left (dP_1(x_i) - dP_2(x_i) \right )   dP_1^{k-i}(x_{i+1:k}).  
\end{equation*}
Plugging this into \eqref{eqn:lb_1}, we obtain
\begin{align} \label{eqn:lb_2}
\Delta(x_{k+1:n})
= \sum_{i=1}^k \int_{\R^n} \Pb(\htheta(x_{1:n}) \in A) dP_2^{i-1}(x_{1:i-1}) \left (dP_1(x_i) - dP_2(x_i) \right )  dP_1^{k-i}(x_{i+1:k}) . 
\end{align}
Let $x_{1:n}^i$ be a vector similar to $x_{1:n}$, except on $i$-th coordinate, where $x_i$ is replaced by $x'_i$. Note that, 
\begin{equation*}
\int_{\R^k} \Pb(\htheta(x_{1:n}^i) \in A) dP_2^{i-1}(x_{1:i-1}) \left (dP_1(x_i) - dP_2(x_i) \right )  dP_1^{k-i}(x_{i+1:k}) =0.
\end{equation*}
Hence, we could write \eqref{eqn:lb_2} as
\begin{align}
& \Delta(x_{k+1:n}) = \\
& \sum_{i=1}^k \int_{\R^k} \left ( \Pb(\htheta(x_{1:n}) \in A) - \Pb(\htheta(x_{1:n}^i) \in A) \right )
dP_2^{i-1}(x_{1:i-1}) \left (dP_1(x_i) - dP_2(x_i) \right )  dP_1^{k-i}(x_{i+1:k}) . \nonumber 
\end{align}
Hence, we have
\begin{align}
& |\Delta(x_{k+1:n})| \leq  \label{eqn:lb_3} \\
& \sum_{i=1}^k \int_{\R^k} \left | \Pb(\htheta(x_{1:n}) \in A) - \Pb(\htheta(x_{1:n}^i) \in A) \right |
dP_2^{i-1}(x_{1:i-1}) \left | dP_1(x_i) - dP_2(x_i) \right |  dP_1^{k-i}(x_{i+1:k}). \nonumber 
\end{align}
Note that, by differential privacy definition, we have
\begin{equation*}
(e^{-\diff_i}-1) \Pb(\htheta(x_{1:n}^i) \in A)
\leq \Pb(\htheta(x_{1:n}) \in A) - \Pb(\htheta(x_{1:n}^i) \in A) 
\leq (e^{\diff_i}-1) \Pb(\htheta(x_{1:n}^i) \in A),
\end{equation*}
which implies 
\begin{equation*}
\left | \Pb(\htheta(x_{1:n}) \in A) - \Pb(\htheta(x_{1:n}^i) \in A) \right |
\leq (e^{\diff_i}-1) \Pb(\htheta(x_{1:n}^i) \in A).
\end{equation*}
Plugging this into \eqref{eqn:lb_3}, we obtain
\begin{align}
& |\Delta(x_{k+1:n})| \leq  \label{eqn:lb_4} \\
& \sum_{i=1}^k (e^{\diff_i}-1) \int_{\R^k} \Pb(\htheta(x_{1:n}^i) \in A)
dP_2^{i-1}(x_{1:i-1}) \left | dP_1(x_i) - dP_2(x_i) \right |  dP_1^{k-i}(x_{i+1:k}). \nonumber 
\end{align}
Finally, note that 
\begin{align}
& \int_{\R^k} \Pb(\htheta(x_{1:n}^i) \in A)
dP_2^{i-1}(x_{1:i-1}) \left | dP_1(x_i) - dP_2(x_i) \right |  dP_1^{k-i}(x_{i+1:k})  \nonumber \\
& = \int_{\R^{k-1}} \Pb(\htheta(x_{1:n}^i) \in A)
dP_2^{i-1}(x_{1:i-1}) dP_1^{k-i}(x_{i+1:k}) \int_\R \left | dP_1(x_i) - dP_2(x_i) \right | \nonumber \\
& \leq 2 \|P_1 - P_2 \|_\TV,  \label{eqn:lb_5}
\end{align}
where the last inequality follows from the fact that 
\begin{equation*}
\int_{\R^{k-1}} \Pb(\htheta(x_{1:n}^i) \in A)
dP_2^{i-1}(x_{1:i-1}) dP_1^{k-i}(x_{i+1:k})    
\end{equation*}
is bounded by 1. Plugging \eqref{eqn:lb_5} into \eqref{eqn:lb_4} completes the proof of \eqref{eqn:triangle_3_TV}. 
$\blacksquare$
%%%%%%%%%%%%%%%%%%%%%%%%%%%%%%%%%%%%%%%%%%%%%%%%%%%%%%%%%%%%%%%%%
%%%%%%%%%%%%%%%%%%%%%%%%%%%%%%%%%%%%%%%%%%%%%%%%%%%%%%%%%%%%%%%%%
%%%%%%%%%%%%%%%%%%%%%%%%%%%%%%%%%%%%%%%%%%%%%%%%%%%%%%%%%%%%%%%%%
\subsubsection*{Proof of Theorem \ref{theorem:LB_Local}}
To show the lower bound \eqref{eqn:local_lb}, we again use the Le Cam's method. Here, for $j \in \{1,2\}$, we define $Q_j$ to be the marginal distribution of $\matM$, given that the samples $X_{1:n}$ are all drawn from $P_j$, i.e.,
\begin{equation} \label{eqn:Q_j_local}
Q_j(A) = \int_{\R^n} \Pb(\matM(x_{1:n}) \in A) dP_j^n(x_{1:n}),     
\end{equation}  
for any measurable set $A \subset \R^n$. Then, again, by Le Cam's method, we have (see \cite{duchi2013local} for more details)
\begin{equation} \label{eqn:LeCam_local}
 \matL_l(\matP_k, \theta, \bm{\diff}) \geq \gamma^2 \left (\frac{1}{2} - \frac{1}{2} \|Q_1 - Q_2\|_{\TV} \right ),
\end{equation}
where $\gamma$ is given by \eqref{eqn:gamma_mean}. A slight extension of Corollary 1 in \cite{duchi2013local} implies
\begin{equation*}
\|Q_1 - Q_2 \|_{\TV}^2 \leq \frac{1}{4} \left ( D_{\text{KL}}(Q_1, Q_2) + D_{\text{KL}}(Q_2, Q_1) \right) 
\leq \|P_1 - P_2\|_{\TV}^2 \sum_{i=1}^n (e^{\diff_i}-1)^2.      
\end{equation*}
Hence, for $\diff_i \leq 1$, using $e^{\diff_i}-1 \leq 2 \diff_i$, we have
\begin{equation*}
\|Q_1 - Q_2 \|_{\TV} \leq 2 \|P_1 - P_2\|_{\TV} \sqrt{\sum_{i=1}^n \diff_i^2}.    
\end{equation*}
Taking $P_1$ and $P_2$ similar to \eqref{eqn:P_1_2}, and using \eqref{eqn:LeCam_local}, we have
\begin{equation} 
 \matL_l(\matP_k, \theta, \bm{\diff}) \geq \frac{1}{8} \delta^2 \left ( 1 - 2 \delta \sqrt{\sum_{i=1}^n \diff_i^2} \right ).
\end{equation}
Choosing 
\begin{equation*}
\delta =  \frac{1}{4 \sqrt{\sum_{i=1}^n \diff_i^2}} \wedge \frac{1}{2}   
\end{equation*}
completes the proof of \eqref{eqn:local_lb}.

To show the upper bound \eqref{eqn:local_ub}, we form the following estimator
\begin{equation} \label{eqn:local_linear_estimator}
\htheta = \sum_{i=1}^n  \frac{\diff_i^2}{\sum_{j=1}^n \diff_j^2} \hat{x}_i, 
\text{ where } \hat{x}_i = x_i + \Lap\left(\frac{1}{\diff_i}\right).
\end{equation}
Clearly this estimator is $(\diff_i)_{i=1}^n$-locally differentially private. Next, note that
\begin{align}
\E \left [ | \htheta - \theta |^2 \right ] 
&= \sum_{i=1}^n \frac{\diff_i^4}{(\sum_{j=1}^n \diff_j^2)^2} \text{Var}(\hat{x}_i)  = \sum_{i=1}^n \frac{\diff_i^4}{(\sum_{j=1}^n \diff_j^2)^2} \left(\text{Var}(X) + \frac{1}{\diff_i^2}\right).
\end{align}
Using the fact that $\frac{1}{\diff_i^2} \geq 1 \geq \text{Var}(X)$, we obtain
\begin{equation}
\E \left [ | \htheta - \theta |^2 \right ] 
\leq 2 \sum_{i=1}^n \frac{\diff_i^4}{(\sum_{j=1}^n \diff_j^2)^2} \cdot \frac{1}{\diff_i^2}
= 2 \sum_{i=1}^n \frac{\diff_i^2}{(\sum_{j=1}^n \diff_j^2)^2}
=  \frac{2}{\sum_{j=1}^n \diff_j^2},
\end{equation}
which completes the proof.
 $\blacksquare$ 
%%%%%%%%%%%%%%%%%%%%%%%%%%%%%%%%%%%%%%%%%%
%%%%%%%%%%%%%%%%%%%%%%%%%%%%%%%%%%%%%%%%%%
%%%%%%%%%%%%%%%%%%%%%%%%%%%%%%%%%%%%%%%%%%
\subsubsection*{Proof of Proposition \ref{Pro:PaymentIdentity}}
Letting 
\begin{align*}
    h_i(c)=\mathbb{E}_{\mathbf{c}_{-i}}\left[\MSE(c,\mathbf{c}_{-i},\bm{\diff},  \hat{\theta}) \right],
\end{align*}
\begin{align*}
    t_i(c)=\mathbb{E}_{\mathbf{c}_{-i}}\left[t(c, \mathbf{c}_{-i}) \right], 
\end{align*}    
and
\begin{align*}
    \diff_i(c)=\mathbb{E}_{\mathbf{c}_{-i}}\left[\diff_i(c, \mathbf{c}_{-i}) \right],
\end{align*}
we can write the incentive compatibility constraint as 
\begin{align*}
    h_i(c_i) +  c_i \diff_i(c_i) - t_i (c_i) \le h_i(c'_i) + c_i \diff_i(c_i') - t_i(c'_i).
\end{align*}
Taking derivative of the right-hand side with respect to $c_i'$ and evaluating the derivative at $c'_i=c_i$ and equating it to zero leads to 
\begin{align*}
    h'_i(c_i) +  c_i \diff'_i(c_i) - t'_i(c_i)=0.
\end{align*}
By taking the integral of this expression we obtain 
\begin{align}\label{eq:Pf:Pro:PaymentIdentity:1}
    t_i(c_i) =  t_i(0) + \int_{z=0}^{c_i} \left( h'_i(z) +  z \diff'_i(z)\right) dz   = t_i(0) + h_i(c_i) - h_i(0) +  c_i \diff_i(c_i) -  \int_{z=0}^{c_i} \diff_i(z) dz.
\end{align}
We next show that the payment in \eqref{eq:Pf:Pro:PaymentIdentity:1} together with a weakly decreasing $\diff_i(z)$ guarantees that the incentive compatibility constraint. To see this, we consider two possibilities depending on whether $c'_i$ is larger or smaller than $c_i$:
\begin{itemize}
    \item For $c'_i \ge c_i$: by plugging in the payment in \eqref{eq:Pf:Pro:PaymentIdentity:1}  the incentive compatibility constraint becomes equivalent to
    \begin{align*}
        \diff_i(c'_i) (c_i - c'_i) \ge \int_{z=c'_i}^{c_i} \diff_i(z) dz,
    \end{align*}
    which holds because $\diff_i(z)$ is weakly decreasing in $z$.
    \item For $c'_i \le c_i$: again, by plugging in the payment in \eqref{eq:Pf:Pro:PaymentIdentity:1}  the incentive compatibility constraint becomes equivalent to
    \begin{align*}
        \diff_i(c'_i) (c_i - c'_i) \le \int_{z=c_i}^{c'_i} \diff_i(z) dz,
    \end{align*}
    which, again, holds because $\diff_i(z)$ is weakly decreasing in $z$. This completes one direction of the proof.
    \end{itemize}
    
To see the other direction, notice that the first order condition of the incentive compatibility implies \eqref{eq:Pf:Pro:PaymentIdentity:1}. Finally notice that the incentive compatibility implies 
    \begin{align*}
    h_i(c_i) +  c_i \diff_i(c_i) - t_i (c_i) \le h_i(c'_i) +  c_i \diff_i(c_i') - t_i(c'_i).
\end{align*}
and 
\begin{align*}
    h_i(c'_i) +  c'_i \diff_i(c'_i) - t_i (c'_i) \le h_i(c_i) +  c'_i \diff_i(c_i) - t_i(c_i).
\end{align*}
    Taking summation of these two equations results in 
    \begin{align*}
        \left( \diff_i(c_i)- \diff_i(c'_i) \right) \left( c_i - c'_i \right) \le 0,
    \end{align*}
    which shows that $\diff_i(\cdot)$ is weakly decreasing. 
    
    We next consider the individual rationality constraint. Using \eqref{eq:Pf:Pro:PaymentIdentity:1}, we can rewrite this constraint as 
    \begin{align}\label{eq:Pf:Pro:PaymentIdentity:2}
        t_i(0) \ge h_i(0) - \Var +  \int_{z=0}^{c_i} \diff_i(z) dz \quad \text{ for all } c_i
    \end{align}
    which means it only needs to hold for $c_i=\infty$. Hence, we could cast $t_i(0)$ as
    \begin{equation*}
        t_i(0) = h_i(0) - \Var +\int_{z=0}^{\infty} \diff_i(z) dz +d_i
    \end{equation*}
    for some nonnegative constant $d_i$.
    Plugging this back in \eqref{eq:Pf:Pro:PaymentIdentity:1} results in 
    \begin{align*}
        t_i(c_i)= h_i(c_i) - \Var +  c_i \diff_i(c_i) +  \int_{z=c_i}^{\infty} \diff_i(z) dz +d_i.
    \end{align*}
    This completes the proof.  $\blacksquare$
%%%%%%%%%%%%%%%%%%%%%%%%%%%%%%%%%%%%%%%%%%
%%%%%%%%%%%%%%%%%%%%%%%%%%%%%%%%%%%%%%%%%%
%%%%%%%%%%%%%%%%%%%%%%%%%%%%%%%%%%%%%%%%%%
\subsubsection*{Proof of Proposition \ref{pro:refomrulation1}}
Using the payment identity in Proposition \ref{Pro:PaymentIdentity}, we obtain 
\begin{align}\label{eq:pfthm:central:1}
    \mathbb{E}_{c_i} \left[t_i(c_i) \right] &=  \mathbb{E}[\MSE(\mathbf{c}; \bm{\diff}, \hat{\theta})] - \Var + \mathbb{E}_{c_i}[c_i \diff_i(c_i) ] +  \mathbb{E}_{c_i}\left[ \int_{z=c_i}^{\infty} \diff_i(z) dz \right] \nonumber \\
    & = \mathbb{E}[\MSE(\mathbf{c}; \bm{\diff}, \hat{\theta})] - \Var+ \int_{\mathbf{z}_{-i}}  \int_{z_i} \left(z_i \diff_i(z_i, \mathbf{z}_{-i})+ \int_{y_i=z_i}^{\infty} \diff_i(y_i, \mathbf{z}_{-i}) dy_i \right) f_i(z_i) dz_i f_{-i}(\mathbf{z}_{-i}) d\mathbf{z}_{-i} \nonumber \\
    & = \mathbb{E}[\MSE(\mathbf{c};\bm{\diff}, \hat{\theta})] - \Var +  \int_{\mathbf{z}_{-i}}  \int_{z_i} z_i \diff_i(z_i, \mathbf{z}_{-i}) f_i(z_i) dz_i f_{-i}(\mathbf{z}_{-i}) d\mathbf{z}_{-i} \nonumber \\
     & +\int_{\mathbf{z}_{-i}}  \int_{z_i=0}^{\infty} \int_{y_i=z_i}^{\infty} \diff_i(y_i, \mathbf{z}_{-i}) dy_i  f_i(z_i) dz_i f_{-i}(\mathbf{z}_{-i}) d\mathbf{z}_{-i} \nonumber \\
    & \overset{(a)}{=}
\mathbb{E}[\MSE(\mathbf{c};\bm{\diff}, \hat{\theta})] - \Var +  \int_{\mathbf{z}_{-i}}  \int_{z_i} z_i \diff_i(z_i, \mathbf{z}_{-i}) f_i(z_i) dz_i f_{-i}(\mathbf{z}_{-i}) d\mathbf{z}_{-i} \nonumber \\
     & +\int_{\mathbf{z}_{-i}}  f_{-i}(\mathbf{z}_{-i}) d\mathbf{z}_{-i} \int_{y_i=0}^{\infty} \diff_i(y_i, \mathbf{z}_{-i}) dy_i 
 \int_{z_i=0}^{y_i} f_i(z_i) dz_i  \nonumber \\
&= \mathbb{E}[\MSE(\mathbf{c};\bm{\diff}, \hat{\theta})] - \Var +  \int_{\mathbf{z}_{-i}}  \int_{z_i} z_i \diff_i(z_i, \mathbf{z}_{-i}) f_i(z_i) dz_i f_{-i}(\mathbf{z}_{-i}) d\mathbf{z}_{-i} \nonumber \\
     & +\int_{\mathbf{z}_{-i}}  f_{-i}(\mathbf{z}_{-i}) d\mathbf{z}_{-i} \int_{y_i=0}^{\infty} \diff_i(y_i, \mathbf{z}_{-i}) dy_i 
   F_i(y_i)  \nonumber \\
    & \overset{(b)}{=} \mathbb{E}[\MSE(\mathbf{c}; \bm{\diff}, \hat{\theta}) ] - \Var + \int_{\mathbf{z}}  \left(z_i+   \frac{F_{i}(z_i)}{f_{i}(z_i)} \right)  \diff_i(\mathbf{z}) f(\mathbf{z}) d\mathbf{z}, 
\end{align}
where (a) follows from changing the order of the integrals and (b) follows by a change of variable from $y_i$ to $z_i$. 
Substituting equation \eqref{eq:pfthm:central:1} in the platform's objective function, results in 
\begin{align*}
    \mathbb{E}_{\mathbf{c}}\left[  \MSE(\mathbf{c}, \bm{\diff}, \hat{\theta}) + \sum_{i=1}^n t_i(\mathbf{c})\right] =   \mathbb{E}_{\mathbf{c}}\left[ (n+1)\MSE(\mathbf{c},\bm{\diff},  \hat{\theta}) + \sum_{i=1}^n \psi(c_i) \diff_i(\mathbf{c}) \right] - n \Var.
\end{align*}
Finally, note that, by using Proposition \ref{Pro:PaymentIdentity}, the payment identity guarantees a privacy level function is decreasing if $\diff_i(\cdot)$ is decreasing. This completes the proof. $\blacksquare$
%%%%%%%%%%%%%%%%%%%%%%%%%%%%%%%%%%%%%%%%%%
%%%%%%%%%%%%%%%%%%%%%%%%%%%%%%%%%%%%%%%%%%
%%%%%%%%%%%%%%%%%%%%%%%%%%%%%%%%%%%%%%%%%%
\subsubsection*{Proof of Theorem \ref{thm:central}}
Since the variance of the data points are $\Var$, the variance of the estimator given in \eqref{estimator:central:1} (i.e., the mean square error) is 
\begin{align*}
   \frac{2}{\eta^2}  + \sum_{i=1}^n w_i(\mathbf{c})^2 \Var.
\end{align*}
By plugging this expression into the characterization of Proposition \ref{pro:refomrulation1}, we see that for any vector of reported privacy costs $(c_1, \dots, c_n)$, the point-wise optimization problem becomes
\begin{align*}
     \min_{\bm{\diff}(\mathbf{c}), \mathbf{w}(\mathbf{c}), \eta} &  \frac{2 (n+1)}{\eta^2}+ \sum_{i=1}^n (n+1) \Var ~ w_i(\mathbf{c})^2 +   \psi_i(c_i) \diff_i(\mathbf{c})  \\
    & \text{ s.t. } \bm{\diff}_i(\mathbf{c}) \ge 0, \text{ for all } i \in \mathcal{N}\\
    & \sum_{i=1}^n w_i(\mathbf{c})=1\\
    & \eta w_i(\mathbf{c}) \le \diff_i(\mathbf{c}) \text{ for all } i \in \mathcal{N}.
\end{align*}
In the optimal solution we must have $\eta w_i(\mathbf{c}) = \diff_i(\mathbf{c})$ for all $i \in \mathcal{N}$. Therefore, the above optimization is equivalent to
\begin{align*}
     \min_{\bm{\diff}(\mathbf{c}), \mathbf{w}(\mathbf{c}), \eta} &  \frac{2 (n+1)}{\eta^2}+ \sum_{i=1}^n (n+1) \Var ~ \frac{\bm{\diff}^2_i(\mathbf{c})}{\eta^2} +   \psi_i(c_i) \diff_i(\mathbf{c})  \\
    & \text{ s.t. } \bm{\diff}_i(\mathbf{c}) \ge 0, \text{ for all } i \in \mathcal{N}\\
    & \sum_{i=1}^n \frac{\bm{\diff}_i(\mathbf{c})}{\eta}=1,
\end{align*}
which in turn, by letting $\bm{\diff}_i(\mathbf{c})=y_i$ for all $i \in \mathcal{N}$, is equivalent to
\begin{align}\label{pf:eq:thm:central}
     \min_{\mathbf{y}} &  \frac{2 (n+1)}{\left( \sum_{j=1}^n y_j \right)^2}+ \sum_{i=1}^n (n+1) \Var \left(\frac{y_i}{\sum_{j=1}^n y_j} \right)^2 +   \psi_i(c_i) y_i  \\
    & \text{ s.t. } y_i \ge 0, \text{ for all } i \in \mathcal{N}. \nonumber
\end{align}
The corresponding payment to user $i$ is given by 
\begin{align*}
 & \mathbb{E}_{\mathbf{c}_{-i}}\left[\MSE(c_i, \mathbf{c}_{-i}; \bm{\diff}, \hat{\theta}_{\mathrm{central}}) \right] - \Var +  c_i \mathbb{E}_{\mathbf{c}_{-i}} \left[\diff_i(c_i, \mathbf{c}_{-i}) \right] + \  \int_{x=c_i}^{\infty} \mathbb{E}_{\mathbb{c}_{-i}} \left[\diff_i(x, \mathbf{c}_{-i}) \right] dx.
\end{align*}
By invoking Proposition \ref{Pro:PaymentIdentity}, this payment and allocation satisfy the incentive compatibility and the individual rationality constraints provided that the optimal privacy level function is weakly decreasing in the reported privacy cost which we prove next. 

Let $(y_1, \dots, y_n)$ be the solution of optimization problem \eqref{pf:eq:thm:central} for $c_1, \dots, c_n$. Now, suppose we increases one of the $c_i$'s, which, without loss of generality, we assume is the first one. Let $c'_1>c_1$ and $c'_i=c_i$ for $i=2, \dots, n$ and suppose $y'_1, \dots, y'_n$ is the corresponding optimal solution of optimization problem \eqref{pf:eq:thm:central}. The optimality condition implies that 
\begin{align*}
    & \frac{2(n+1)}{\left( \sum_{j=1}^n y_j^2 \right)^2}+\frac{(n+1)}{\left( \sum_{j=1}^n y_j^2 \right)^2} \sum_{i=1} y_i^2 \Var +  \psi_i(c_i) y_i \\
    & \le \frac{2(n+1)}{\left( \sum_{j=1}^n {y'_j}^2 \right)^2}+\frac{(n+1)}{\left( \sum_{j=1}^n {y'_j}^2 \right)^2} \sum_{i=1} {y'_i}^2 \Var +  \psi_i(c_i) y'_i 
\end{align*}
and 
\begin{align*}
    & \frac{2(n+1)}{\left( \sum_{j=1}^n {y'_j}^2 \right)^2}+\frac{(n+1)}{\left( \sum_{j=1}^n {y'_j}^2 \right)^2} \sum_{i=1} {y'_i}^2 \Var +  \psi_i(c'_i) y'_i \\
    & \le \frac{2(n+1)}{\left( \sum_{j=1}^n {y_j}^2 \right)^2}+\frac{(n+1)}{\left( \sum_{j=1}^n {y_j}^2 \right)^2} \sum_{i=1} {y_i}^2 \Var +  \psi_i(c'_i) y_i 
\end{align*}
Taking summation of both sides of these equations and using the fact that $c_i=c'_i$ for $i=2, \dots, n$, we obtain 
\begin{align*}
    (y_1- y'_1) (\psi_1(c_1)- \psi_1(c'_1)) \le  0.
\end{align*}
Assumption \ref{assump:MHR} and the above inequality establishes that the solution of problem \eqref{pf:eq:thm:central} is weakly decreasing in the privacy cost. 

Finally, notice that if the platform pays user $i$
\begin{align*}
         - \Var +  c_i \diff_i^*( \mathbf{c}) +  \int_{c_i} \diff_i^*(z, \mathbf{c}_{-i}) dz +\frac{2}{\left(\sum_{j=1}^n \diff^*_j(\mathbf{c}) \right)^2}  + \sum_{i=1}^n \left( \frac{\diff_i(\mathbf{c})}{\sum_{j=1}^n \diff_j(\mathbf{c})} \right)^2 \Var.
\end{align*}
the expected payment becomes the same as the characterization of Proposition \ref{Pro:PaymentIdentity}. This completes the proof.
 $\blacksquare$
 %%%%%%%
 \subsubsection*{Proof of Corollary \ref{Cor:central:mono}}
 If $\psi_i(c_i) = \psi_j(c_j)$, then by swapping the $i$-th and $j$-th components of the solution the objective remains the same and therefore we can always let $y^*_i \ge y^*_j$. Now, suppose $\psi_i(c_i) < \psi_j(c_j)$. We let 
%%%%%%
\begin{align*}
    \tilde{y}_{\ell}=\begin{cases}
    y^*_{\ell} \quad &\ell \neq i, j \\
    y^*_i \quad &\ell=j\\
    y^*_j \quad & \ell=i.
    \end{cases}
\end{align*}
%%%%%%
The difference of the objective function evaluated at $(\tilde{y}_1, \dots, \tilde{y}_n)$ and $(y^*_1, \dots, y^*_n)$ becomes
\begin{align*}
    \left(\psi_{i}(c_i) - \psi_i(c_j) \right) \left( y^*_{j}- y^*_i \right) \ge 0
\end{align*}
where the inequality follows from the optimality condition. Inequality $\psi_i(c_i) < \psi_j(c_j)$, implies that $y^*_i \ge y^*_j$, proving the corollary. $\blacksquare$
%%%%%%%%%%%%%%%%%%%%%%%%%%%%%%%%%%%%%%%%%%
%%%%%%%%%%%%%%%%%%%%%%%%%%%%%%%%%%%%%%%%%%
%%%%%%%%%%%%%%%%%%%%%%%%%%%%%%%%%%%%%%%%%%
\subsubsection*{Proof of Proposition \ref{Pro:approximate:central}}
Problem \eqref{eq:thm:central} is the same as 
\begin{align*}
\min_{S \ge 0 } \min_{\mathbf{y}} &  \frac{ n+1}{S^2} \left( 2+ \sum_{i=1}^n   y_i ^2 ~ \Var  \right)+ \sum_{i=1}^n  \psi_i(c_i) y_i  \\
    & \text{ s.t. } \sum_{i=1}^n y_i = S  \\
    & y_i \ge 0, \text{ for all } i \in \mathcal{N}.
\end{align*}
Let us consider the optimization over $\mathbf{y}$ for a given $S$. The Lagrangian of this problem is 
\begin{align*}
    \sum_{i=1}^n \frac{(n+1) \Var}{S^2} y_i^2 + \sum_{i=1}^n  \psi_i(c_i) y_i - \lambda \left(\sum_{i=1}^n y_i-S \right) - \sum_{i=1}^n \mu_i y_i.
\end{align*}
The KKT conditions imply that the optimal solutions $(y^*_1, \dots, y^*_n)$, $\lambda^*$ and $\mu^*_i$ satisfy
\begin{align*}
    2(n+1) \frac{\Var}{S^2} y_i^* +  \psi_i(c_i) - \lambda^* - \mu^*_i = 0 \text{ for all } i \in \mathcal{N}.
\end{align*}
If $y^*_i > 0$, we have $\mu^*_i=0$ and therefore 
\begin{align*}
    y_i^*= \frac{\left(\lambda^* - \psi_i(c_i) \right)S^2 }{2 (n+1) \Var}. 
\end{align*}
If $y_i^*=0$, we have $\mu_i^*= \psi_i(c_i) - \lambda^* \geq 0$. Hence, if, we define
\begin{align}\label{eq:pf:Pro:approximate:central:1}
 y(\lambda) = \left( \left(\frac{(\lambda - \psi_1(c_1) )S^2}{2 (n+1) \Var} \right)^+, \dots, \left(\frac{ ( \lambda- \psi_n(c_n)) S^2 }{2 (n+1) \Var} \right)^+ \right) 
 \text{ for any $\lambda$},
\end{align}
$(y_i^*)_{i=1}^n$ would be equal to $y(\lambda^*)$. Next, define
\begin{align}\label{eq:pf:Pro:approximate:central:2}
    S(\lambda) = \sum_{i=1}^n y_i(\lambda).
\end{align}
We can see that the function  $S(\lambda)$ is increasing in $\lambda$ and that $\lambda^*$ is such that $S(\lambda^*)=S$. Once we find $\lambda^*$, \eqref{eq:pf:Pro:approximate:central:1} gives the optimal solution (subject to $\sum_{i=1}^n y_i = S$). To find $\lambda^*$ we first sort the terms $\{\psi_i(c_i)\}_i$ in $\mathcal{O}(n \log n)$. Without loss of generality, let us assume 
\begin{align*}
    \psi_1(c_1) \le \dots \le  \psi_n(c_n).
\end{align*}
We let $i^* $ be the smallest element of  $\mathcal{N}$ for which 
\begin{align*}
    S( \psi_j(c_j)) \ge S.
\end{align*}
If no such element exists, then we let $i^*=n+1$. Therefore, for any $S > 0$, there exists $i^* > 1$ and $\lambda^* \in [\psi_{i^*-1}(c_{i^*-1}), \psi_{i^*}(c_{i^*})]$ such that in the optimal solution we have
\begin{align*}
    y_i=0 \text{ for } i > i^* \text{ and } y_i = \frac{\left(\lambda^* - \psi_i(c_i) \right)S^2 }{2 (n+1) \Var} \text{ for } i \leq i^*,
\end{align*}
with the convention that $\psi_{n+1}(\cdot) = \infty$. Also, using \eqref{eq:pf:Pro:approximate:central:1} and \eqref{eq:pf:Pro:approximate:central:2}, $S$ and $\lambda^*$ satisfy the following relation
\begin{align*}
    \sum_{i=1}^{i^*} \frac{ (\lambda^* - \psi_i(c_i)) S^2 }{2 (n+1) \Var} =S
\end{align*}
which results in 
\begin{align*}
    \frac{1}{S}= \lambda^* A_{i^*} - B_{i^*},
\end{align*}
with
\begin{align*}
    A_{i^*}= \frac{i^*}{2(n+1) \Var}, B_{i^*}=\sum_{i=1}^{i^*} \frac{\psi_i(c_i)}{2(n+1) \Var}.
\end{align*}
Hence, in this case, the original optimization for the given $S$ can be cast as
\begin{align}
    & \frac{2(n+1)}{S^2}+ \frac{n+1}{S^2} \sum_{i=1}^{i^*} \left(\frac{(\lambda - \psi_i(c_i))S^2}{2(n+1) \Var} \right)^2 \Var + \sum_{i=1}^{i^*} \frac{(\lambda - \psi_i(c_i))S^2}{2(n+1) \Var} \psi_i(c_i) \label{eq:pf:Pro:approximate:central:3} \\
    &=  \frac{2(n+1)}{S^2}+ \frac{S^2}{4 (n+1) \Var} \sum_{i=1}^{i^*} (\lambda^2 - \psi_i^2(c_i)) \\
    & =  2 (n+1) \left( \lambda A_{i^*} - B_{i^*}\right)^2 + \frac{1}{\left(\lambda A_{i^*} - B_{i^*} \right)^2} \left(\frac{A_{i^*}}{2} \lambda^2 - \frac{\tilde{B}_{i^*}}{2} \right), \label{eq:pf:Pro:approximate:central:4}
\end{align}
with 
\begin{equation*}
\tilde{B}_{i^*} :=  \sum_{i=1}^{i^*} \frac{\psi_i(c_i)^2}{2(n+1) \Var}.   
\end{equation*}
Note that, as $S$ moves from zero to infinity, $\lambda^*$ also moves from $\psi_1(c_1)$ to infinite. Hence, instead of minimizing \eqref{eq:pf:Pro:approximate:central:3} over $S$, we could minimize \eqref{eq:pf:Pro:approximate:central:4} over $\lambda$. To do so, it suffices to solve 
\begin{align} \label{eq:pf:Pro:approximate:central:5}
  \min_{\lambda} ~&  2 (n+1) \left( \lambda A_{i^*} - B_{i^*}\right)^2 + \frac{1}{\left(\lambda A_{i^*} - B_{i^*} \right)^2} \left(\frac{A_{i^*}}{2} \lambda^2 - \frac{\tilde{B}_{i^*}}{2} \right) \\
  & \lambda \ge \psi_{i^*}(c_{i^*}) \nonumber \\
  & \lambda \le \psi_{i^*+1}(c_{i^*+1}), \nonumber
\end{align}
for $i^* \in \{2, \cdots, n+1\}$ and pick the one with minimum value. As the last step, we establish that \eqref{eq:pf:Pro:approximate:central:5} can be solved in time $\mathcal{O}(1)$ which implies that the total optimization problem can be solved in time $\mathcal{O}(n)$. 

To do so, note that the objective function of \eqref{eq:pf:Pro:approximate:central:5} can be written as
\begin{align}
& 2 (n+1) \left( \lambda A_{i^*} - B_{i^*}\right)^2 + \frac{1}{\left(\lambda A_{i^*} - B_{i^*} \right)^2} \left(\frac{A_{i^*}}{2} \lambda^2 - \frac{\tilde{B}_{i^*}}{2} \right) \nonumber \\
& = \frac{2(n+1)\left(\lambda A_{i^*} - B_{i^*} \right)^4 + \frac{A_{i^*}}{2} \lambda^2 - \frac{\tilde{B}_{i^*}}{2}}
{\left(\lambda A_{i^*} - B_{i^*} \right)^2}. \label{eq:pf:Pro:approximate:central:6}
\end{align}
One can see that the derivative of \eqref{eq:pf:Pro:approximate:central:6} is in the form of a polynomial of degree four divided by $\left(\lambda A_{i^*} - B_{i^*} \right)^3$. Hence, the derivative of \eqref{eq:pf:Pro:approximate:central:6} has at most four roots and they all can be characterized using the formulas for roots of a degree four polynomial. Therefore, to find the solution of \eqref{eq:pf:Pro:approximate:central:5}, it suffices to compare the value of the objective function \eqref{eq:pf:Pro:approximate:central:6} at endpoints of the constraint interval $ [\psi_{i^*}(c_{i^*}), \psi_{i^*+1}(c_{i^*+1})] $  and those roots of the derivative that lie within this interval. These are at most six points and thus the optimization problem \eqref{eq:pf:Pro:approximate:central:5} can be solved in time $\mathcal{O}(1)$. $\blacksquare$
%%%%%%%%%%%%%%%%%%%%%%%%%%%%%%%%%%%%%%%%%%
%%%%%%%%%%%%%%%%%%%%%%%%%%%%%%%%%%%%%%%%%%
%%%%%%%%%%%%%%%%%%%%%%%%%%%%%%%%%%%%%%%%%%
\subsubsection*{Proof of Theorem \ref{thm:local}}
Using the payment identity in Proposition \ref{Pro:PaymentIdentity}, we obtain 
\begin{align}\label{eq:pfthm:local:1}
    \mathbb{E}_{c_i} \left[t_i(c_i) \right] = & \mathbb{E}[\MSE(\mathbf{c};  \bm{\diff}, \mathbf{w})] - \Var + \mathbb{E}_{c_i}[c_i \diff_i(c_i) ] + \ \mathbb{E}_{c_i}\left[ \int_{z=c_i} \diff_i(z) dz \right] \nonumber \\
    =& \mathbb{E}[\MSE(\mathbf{c}; \bm{\diff}, \mathbf{w})] - \Var +  \int_{\mathbf{z}_{-i}}  \int_{z_i} \left(z_i \diff_i(z_i, \mathbf{z}_{-i})+ \int_{y_i=z_i} \diff_i(y_i, \mathbf{z}_{-i}) dy_i \right) f_i(z_i) dz_i f_{-i}(\mathbf{z}_{-i}) d\mathbf{z}_{-i} \nonumber \\
    \overset{(a)}{=}& \mathbb{E}[\MSE(\mathbf{c}; \bm{\diff}, \mathbf{w})] - \Var + \ \int_{\mathbf{z}_{-i}}  \int_{z_i} \left(z_i \diff_i(z_i, \mathbf{z}_{-i})+  \diff_i(z_i, \mathbf{z}_{-i}) \frac{F_i(z_i)}{z_i} \right) f_i(z_i) dz_i f_{-i}(\mathbf{z}_{-i}) d\mathbf{z}_{-i} \nonumber \\
    =& \mathbb{E}[\MSE(\mathbf{c}; \bm{\diff}, \mathbf{w}) ] - \Var +  \int_{\mathbf{z}}  \left(z_i+   \frac{F_i(z_i)}{z_i} \right)  \diff_i(\mathbf{z}) f(\mathbf{z}) d\mathbf{z}, 
\end{align}
where (a) follows from changing the order of the integrals. Moreover, we have 
\begin{align}\label{eq:pfthm:local:2}
    \MSE(\mathbf{c};  \bm{\diff}, \mathbf{w}) =  \sum_{i=1}^n \Var w^2_i(\mathbf{c}) + \sum_{i=1}^n \frac{2 w^2_i(\mathbf{c})}{ \diff_i^2(\mathbf{c})}.
\end{align}
Substituting equations \eqref{eq:pfthm:local:1} and  \eqref{eq:pfthm:local:2} in the platform's objective function, results in
\begin{align*}
    \mathbb{E}_{\mathbf{c}}\left[  \MSE(\mathbf{c};  \bm{\diff}, \mathbf{w}) + \sum_{i=1}^n t_i(\mathbf{c})\right] = 
    \mathbb{E}_{\mathbf{c}}\left[  \sum_{i=1}^n w^2_i(\mathbf{c}) (n+1) \Var + (n+1) \frac{2 w^2_i(\mathbf{c})}{ \diff_i^2(\mathbf{c})}+  \psi(c_i) \diff_i(\mathbf{c}) \right]- n \Var.
\end{align*}
For any vector of reported privacy costs $(c_1, \dots, c_n)$, we solve the point-wise optimization problem:
\begin{align*}
    \min_{\mathbf{w}, \mathbf{y}} ~~~ & \sum_{i=1}^n w_i^2 \left( (n+1) \Var + \frac{2(n+1)}{ y_i^2} \right) +  \psi_i(c_i) y_i \\
    & \sum_{i=1}^n w_i=1 \\
    & w_i \ge 0 \text{ for all } i \in \mathcal{N}.
\end{align*}
Let us first minimize the objective over $w_i$'s. Using Cauchy-Schwarz inequality, for any given $\mathbf{y}$ we have 
\begin{align*}
   \left(  \sum_{i=1}^n w_i^2 \left( (n+1) \Var + \frac{2(n+1)}{ y_i^2} \right) \right) \left( \sum_{i=1}^n \frac{1}{ (n+1) \Var + \frac{2(n+1)}{ y_i^2} } \right) \ge \left(\sum_{i=1}^n w_i \right)^2 =1
\end{align*}
and therefore the minimum of $\sum_{i=1}^n w_i^2 \left( (n+1) \Var + \frac{2(n+1)}{ y_i^2} \right)$ becomes 
\begin{align*}
    \frac{1}{ \sum_{i=1}^n \frac{1}{ (n+1) \Var + \frac{2(n+1)}{ y_i^2} }  }.
\end{align*}
with solution 
\begin{align*}
    w_i=\frac{1}{\left( (n+1) \Var + \frac{2(n+1)}{ y_i^2} \right) \sum_{i=1}^n \frac{1}{ (n+1) \Var + \frac{2(n+1)}{ y_i^2} }} \text{ for all } i \in \mathcal{N}.
\end{align*}
Therefore, the point-wise optimization problem becomes 
\begin{align*}
    \min_{\mathbf{y}}  \frac{1}{ \sum_{i=1}^n \frac{1}{ (n+1) \Var + \frac{2(n+1)}{ y_i^2} }  } + \sum_{i=1}^n  \psi_i(c_i) y_i.
\end{align*}
A similar argument to that of Theorem \ref{thm:central} establishes that, under Assumption \ref{assump:MHR}, the optimal $x_i$ is weakly decreasing in $c_i$ and therefore the corresponding payment, noise variance, and weight function satisfy the incentive compatibility and the individual rationality. $\blacksquare$
%%%%%%%%%%%%%%%%%%%%%%%%%%%%%%%%%%

%%%%%%%%
%%%%%%%%%%%%%%%%%%%%%%%%%%%%%%%%%%%%%%%%%%
%%%%%%%%%%%%%%%%%%%%%%%%%%%%%%%%%%%%%%%%%%
\subsubsection*{Proof of Proposition \ref{Pro:approximate:local}}
Without loss of generality we assume 
\begin{align*}
 \psi_1(c_1) \le \dots \le \psi_n(c_n).   
\end{align*}
We make use of the following two lemmas in this proof.

\begin{lemma}\label{Cor:local:mono}
Suppose Assumption \ref{assump:MHR} holds. For any reported vector of privacy sensitivities $(c_1, \dots, c_n)$, in the optimal local data acquisition mechanism, we have  $\diff^*_i(\mathbf{c}) \ge \diff^*_j(\mathbf{c})$ for all $i,j \in \mathcal{N}$ such that $\psi_i(c_i) < \psi_j(c_j)$.
\end{lemma}
\emph{Proof of Lemma \ref{Cor:local:mono}:}  If $\psi_i(c_i) = \psi_j(c_j)$, then by swapping the $i$-th and $j$-th components of the solution the objective remains the same and therefore we can always let $y^*_i \ge y^*_j$. Now, suppose $\psi_i(c_i) < \psi_j(c_j)$. We let 
%%%%%%
\begin{align*}
    \tilde{y}_{\ell}=\begin{cases}
    y^*_{\ell} \quad &\ell \neq i, j \\
    y^*_i \quad &\ell=j\\
    y^*_j \quad & \ell=i.
    \end{cases}
\end{align*}
%%%%%%
The difference of the objective function evaluated at $(\tilde{y}_1, \dots, \tilde{y}_n)$ and $(y^*_1, \dots, y^*_n)$ becomes
\begin{align*}
    \left(\psi_{i}(c_i) - \psi_i(c_j) \right) \left( y^*_{j}- y^*_i \right) \ge 0
\end{align*}
where the inequality follows from the optimality condition. Inequality $\psi_i(c_i) < \psi_j(c_j)$, implies that $y^*_i \ge y^*_j$, proving the corollary. $\blacksquare$
%%%%%%

%%%%%%
\begin{lemma}\label{pf:lem:tworoots}
For any $\lambda \in \mathbb{R}_{+}$ and any $\psi_i(c_i)$ the equation
\begin{align}\label{eq:pf:lem:tworoots}
    \frac{4 z}{  \left( \Var z^2 + 2 \right)^2} = \frac{\psi_i(c_i)}{n+1} \lambda^2
\end{align}
either has no solution or at most two solutions in $\mathbb{R}_{+}$. Furthermore:
\begin{itemize}
    \item [(a)] The solutions can be found in time $\mathcal{O}(1)$. 
    \item [(b)] The smallest solution is strictly increasing in $\psi_i(c_i)$ and the largest solution is strictly decreasing in $\psi_i(c_i)$. 
\end{itemize}
\end{lemma}
%%%%%%
\emph{Proof of Lemma \ref{pf:lem:tworoots}:} The derivative of the function $\frac{4 z}{  \left( \Var z^2 + 2 \right)^2} $ with respect to $z$ is
\begin{align*}
    \frac{4}{(2+ \Var z^2)^2} \left(1- \frac{4 \Var z^2}{(2+ \Var z^2)} \right),
\end{align*}
which is positive if and only if $z \le \sqrt{\frac{2}{ 3 \Var}}$. Therefore, the function is zero at $z=0$, increases to $\frac{3 \sqrt{3}}{8 \sqrt{2 \Var}}$ at $z = \sqrt{\frac{2}{ 3 \Var}}$ and then decreases to $0$ as $z \to \infty$. Therefore, either there is no solution or there are at most two solutions. To see the proof of part (a), note that finding the solutions of \eqref{eq:pf:lem:tworoots} is equivalent to finding the roots of a degree four polynomial that can be characterized
using the formulas for roots of a degree four polynomial. Finally, to see the proof of part (b) notice that 
\begin{align*}
     \frac{4 z}{  \left( \Var z^2 + 2 \right)^2}
\end{align*}
is strictly increasing for $z \le \sqrt{\frac{2}{ 3 \Var}}$. The smallest solution of \eqref{eq:pf:lem:tworoots} is the intersection of this function over $z \le \sqrt{\frac{2}{ 3 \Var}}$ with the function level $\frac{\psi_i(c_i)}{n+1} \lambda^2$. As $\psi_i(c_i)$ increases, the intersecting $z$ strictly increases. Further, the largest solution of \eqref{eq:pf:lem:tworoots} is the intersection of this function $z \ge \sqrt{\frac{2}{ 3 \Var}}$ with the function level $\frac{\psi_i(c_i)}{n+1} \lambda^2$. As $\psi_i(c_i)$ increases, the intersecting $z$ strictly decreases. $\blacksquare$

%%%%%%
When \eqref{eq:pf:lem:tworoots} has two solutions, we let 
\begin{align*}
    y_i^{(l)}(\lambda) \text{ and } y_i^{(h)}(\lambda)
\end{align*}
be the smallest and the largest solutions, respectively. If \eqref{eq:pf:lem:tworoots} has one solution then the above two solutions coincide.

We now proceed with the proof of the proposition. The KKT condition for problem \eqref{eq:thm:local} implies that when $y_i^* \neq 0$, then  
\begin{align} \label{eqn:local_alg_KKT_recursive}
    \frac{4 y^*_i}{\psi_i(c_i)\left(2+ \Var {y^*_i}^2\right)^2}  =   \frac{1}{n+1} \left( \sum_{j=1}^n \frac{1}{\Var+ \frac{2}{{y^*_j}^2}} \right)^2. 
\end{align}
We let 
\begin{align}\label{eq:pf:fixedpoint}
    \lambda= \sum_{j=1}^n \frac{1}{\Var+ \frac{2}{{y^*_j}^2}}.
\end{align}
By using Lemma \ref{Cor:local:mono}, we know that if there exists $i^* \in \{1, \dots, n\}$ such that $y_{i^*}=0$, then  we have $y^*_i=0$ for $i > i^*$. For such $i^*$, by using Lemma \ref{pf:lem:tworoots}, we know that for $i\le i^*$, $y^*_i \in \{y_{i}^{(l)}(\lambda),y_{i}^{(h)}(\lambda)\}$.

We need to find the optimal $\lambda$ and the corresponding optimal solution. In this regard, we search over all $i^* \in \{1, \dots, n\}$ and then find the optimal $\lambda$ such that for $i > i^*$ we have $y^*_i=0$ and for $i\le i^*$, we have  $y^*_i \in \{y_{i}^{(l)}(\lambda),y_{i}^{(h)}(\lambda)\}$.

\noindent \textbf{Claim 1:} Consider $i^* \in \{1, \dots, n\}$ and an optimal solution such that for $i > i^*$ we have $y^*_i=0$ and for $i\le i^*$ we have $y^*_i \in \{y_{i}^{(l)}(\lambda),y_{i}^{(h)}(\lambda)\}$. If $\psi_{i^*}(c_{i^*}) > \psi_{i^*-1} (c_{i^*-1})$, then for all $i \le i^*-1 $, we have $y^*_i =y_{i}^{(h)}(\lambda)$.

\noindent \emph{Proof of Claim 1:} To prove this claim, we assume the contrary and reach a contradiction. In particular, suppose $y^*_i =y_{i}^{(l)}(\lambda)$ for  $i \le i^*-1 $. We can write
\begin{align*}
    y_{i}^{(l)}(\lambda)= y^*_i  \overset{(a)}{>} y^*_{i^*} \ge y_{i^*}^{(l)}(\lambda)
\end{align*}
where (a) follows from Lemma \ref{Cor:local:mono} together with $\psi_{i}(c_{i}) < \psi_{i^*}(c_{i^*}) $ (In fact, Lemma \ref{Cor:local:mono} implies $y^*_i \geq y^*_{i^*}$. However, from the proof, one could see that, since $y^*_{i^*} > 0$, the inequality would be strict.) This is a contradiction by invoking part (b) of Lemma \ref{pf:lem:tworoots}, completing the proof of Claim 1. $\blacksquare$

For the rest of the proof, we assume $ \psi_1(c_1) < \dots < \psi_n(c_n). $ We will cover the case that two or more of the $\psi_i(c_i)$'s are equal at the end. In this case,
claim 1 implies that 
\begin{align*}
    y^*_{i} = \begin{cases}
    y_{i}^{(h)}(\lambda) \quad & i < i^* \\
    y_{i}^{(l)}(\lambda) \text{ or } y_{i}^{(h)}(\lambda) \quad & i=i^*  \\
    y_{i}=0 \quad & i > i^*.
    \end{cases}
\end{align*}
This provides the solution for a given $\lambda$. We next show how we can find the (approximately) optimal $\lambda$. In this regard, we first establish a lower bound and an upper bound on $\lambda$.

\noindent \textbf{Claim 2:} Consider $i^* \in \{1, \dots, n\}$ and an optimal solution such that for $i > i^*$ we have $y^*_i=0$ and for $i\le i^*$ we have $y^*_i \in \{y_{i}^{(l)}(\lambda),y_{i}^{(h)}(\lambda)\}$. The optimal $\lambda$ satisfies 
\begin{align*}
    \lambda \in [\underline{y}_{i^*}, \bar{y}_{i^*}],
\end{align*}
where 
\begin{align*}
    \bar{y}_{i^*}= y^{(h)} \left( \left(\frac{(n+1) 3 \sqrt{3}}{\psi_{i^*-1}(c_{i^*-1}) 8 \sqrt{2 \Var}} \right)^{1/2} \right) \text{ and } \underline{y}_{i^*}= \frac{n}{  \Var +  \left(\frac{\sqrt{2}n \left(\sum_{i=1}^n \psi_i(c_i) \right)}{(n+1)} \right)^{2/3}  }.
\end{align*}
\noindent \emph{Proof of Claim 2:} As we proved in the proof of Corollary \ref{Cor:central:mono}, the maximum of $ \frac{4 z}{  \left( \Var z^2 + 2 \right)^2}$ is $\frac{3 \sqrt{3}}{8 \sqrt{2 \Var}}$. Therefore, in order to guarantee that \eqref{eq:pf:lem:tworoots} has a solution we must have 
\begin{align*}
    \lambda \le \left(\frac{(n+1) 3 \sqrt{3}}{\psi_{i^*-1}(c_{i^*-1}) 8 \sqrt{2 \Var}} \right)^{1/2}.
\end{align*}
We next derive a lower bound on $\lambda$. For $y_i=y$, the objective becomes 
\begin{align*}
    \frac{n+1}{n} \left( \Var + \frac{2}{y^2}\right) + \left(\sum_{i=1}^n \psi_i(c_i) \right) y
\end{align*}
which is a convex program whose minimum is
\begin{align*}
    \frac{n+1}{n} \left( \Var + 2 \left(\frac{2(n+1)}{n \left(\sum_{i=1}^n \psi_i(c_i) \right)} \right)^{-2/3} \right) + \left(\sum_{i=1}^n \psi_i(c_i) \right) \left(\frac{2(n+1)}{n \left(\sum_{i=1}^n \psi_i(c_i) \right)} \right)^{1/3}.
\end{align*}
Since the objective is 
\begin{align*}
    \frac{n+1}{ \lambda}  + \sum_{i=1}^{n} \psi_i(c_i) y_i
\end{align*}
and $\psi_i(c_i) \ge 0$, the optimal $\lambda$ is larger than 
\begin{align*}
    \underline{y}_{i^*}= &\frac{n+1}{\frac{n+1}{n} \left( \Var + 2 \left(\frac{2(n+1)}{n \left(\sum_{i=1}^n \psi_i(c_i) \right)} \right)^{-2/3} \right) + \left(\sum_{i=1}^n \psi_i(c_i) \right) \left(\frac{2(n+1)}{n \left(\sum_{i=1}^n \psi_i(c_i) \right)} \right)^{1/3}}\\
    % =& \frac{n}{  \Var + 2 \left(\frac{4(n+1)}{n \left(\sum_{i=1}^n \psi_i(c_i) \right)} \right)^{-2/3}  +  \left(\frac{2n}{n+1} \sum_{i=1}^n \psi_i(c_i) \right)^{2/3}}\\
    =& \frac{n}{  \Var +  \left(\frac{\sqrt{2}n \left(\sum_{i=1}^n \psi_i(c_i) \right)}{(n+1)} \right)^{2/3}  }.
\end{align*}
This completes the proof of Claim 2. $\blacksquare$

Equipped with Claims 1 and 2, we next search over the near optimal $\lambda$. Letting $\mathrm{Grid}(i^*, \frac{\epsilon}{\Delta})$ be an $\frac{\epsilon}{\Delta}$ grid of $[\underline{y}_{i^*}, \bar{y}_{i^*}]$ where $\Delta$ is the maximum Lipschitz parameter for all functions $\frac{n+1}{\lambda}$ and $y_i^{((h))}(\lambda)$ and $y^{(l)}_i(\lambda)$ over $[\underline{y}_{i^*}, \bar{y}_{i^*}]$. With this definition, the following optimization
\begin{align*}
    \min_{\lambda \in \mathrm{Grid}(i^*, \frac{\epsilon}{\Delta})} \min\left\{\frac{n+1}{\lambda} + \sum_{i=1}^{i^*} \psi_i(c_i) y^{(h)}_i(\lambda)  , \frac{n+1}{\lambda} + \sum_{i=1}^{i_1-1} \psi_i(c_i) y^{(h)}_i(\lambda) + \psi_i(c_i) y^{(l)}_i(\lambda) \right\}
\end{align*}
achieves at most $(1+ \epsilon)$ of the optimal objective. Finally, notice that $\Delta$ defined above is polynomial in $n$. Then proof completes by noting that the output of this procedure satisfies the monotonicity property in $\psi_i(c_i)$ because we do a grid search over $\lambda$ and once we find $\lambda$ the corresponding $y_i$'s are decreasing in the virtual costs.  

Finally, we conclude the proof by discussing the case that two or more of $\psi_i$'s are equal. For simplicity, we consider the case that
\begin{equation} \label{eqn:Alg_2_special}
\psi_1(c_1) < \cdots < \psi_{i}(c_i) < \psi_{i+1}(c_{i+1}) = \cdots  = \psi_{i+k}(c_{i+k})  
<  \psi_{i+k+1}(c_{i+k+ 1}) < \cdots < \psi_n(c_n).
\end{equation}
The argument here generalizes to the case when some of $\psi_i$'s are equal on two or more different values. 

For \eqref{eqn:Alg_2_special}, we need to modify the algorithm when the for loop counter reaches $i$, i.e., the case that we take $y^*_{1} = \cdots = y^*_{i} = 0$ and $y^*_{j} >0$ for $j >i$. In this case, Claim 1 would change as follow: For $i+k \leq j \leq i+1$, we have $y^*_j \in \{ y_j^{(l)}(\lambda), y_j^{(h)}(\lambda) \}$, and for $j > i+k$, we have $y^*_j = y_j^{(h)}(\lambda)$. Moreover, since $\psi_{i+1}(c_{i+1}) = \cdots  = \psi_{i+k}(c_{i+k})$, we have $ y_{i+1}^{(l)}(\lambda) = \cdots =  y_{i+k}^{(l)}(\lambda) = y_{i+1:i+k}(\lambda)$ and $ y_{i+1}^{(h)}(\lambda) = \cdots =  y_{i+k}^{(h)}(\lambda) = y_{i+1:i+k}(\lambda)$. 

Hence, we define an inner loop which considers $k+1$ cases on the number of $\{y^*_j\}_{j=i+1}^{i+k}$ that are equal to $y_{i+1:i+k}(\lambda)$. Also, when this inner for loop ends, the outer loop jumps to $i+k+1$ instead of $i+1$, and thus, the total number of iterations still remains bounded by $2n$.
$\blacksquare$
% %%%%
 \subsection*{Proof of Proposition \ref{Pro:compare:privacylevels}}
 %Without loss of generality, we could assume $w_i=0$ whenever $\diff_i=0$, as otherwise, $\E[ | \htheta_{\mathrm{local}} - \theta |^2]$ would be infinite, which immortality implies the desired result. 
Let
\begin{equation*}
\htheta_{\mathrm{central}} = \sum_{i=1}^n w_i x_i + \Lap(1/\eta),
\end{equation*}
with 
\begin{equation} \label{eqn:eta_central}
\eta = \min_{i} \frac{\diff_i}{w_i}.    
\end{equation}
First, note that, by definition, for any $i$, $\eta w_i \leq \diff_i$. Hence, by Lemma \ref{lemma:analyst_DP}, this estimator is $\bm{\diff}$-differentially private. Hence, it suffices to show \eqref{eqn:central_lower} holds. Note that
\begin{align*} 
\E[ | \htheta_{\mathrm{central}} - \theta  |^2] &= 
\Var \sum_{i=1}^n w_i^2 + \frac{2}{\eta^2}, \\
\E[ | \htheta_{\mathrm{local}} - \theta |^2] &= 
\Var \sum_{i=1}^n w_i^2 + 2 \sum_{i=1}^n \frac{w_i^2}{\diff_i^2}. 
\end{align*}
Comparing the right hand sides, to establish \eqref{eqn:central_lower}, we need to show 
\begin{equation*}
\frac{1}{\eta^2} 
\leq \sum_{i=1}^n \frac{w_i^2}{\diff_i^2}.
\end{equation*}
To do so, note that, 
\begin{equation*}
\frac{1}{\eta^2} = \left ( \frac{1}{\min_i \frac{\diff_i}{w_i}} \right )^2    
= \left ( \max_i \frac{w_i}{\diff_i} \right )^2,
\end{equation*}
which is clearly upper bounded by $\sum_{i=1}^n \frac{w_i^2}{\diff_i^2}$. Thus, the proof is complete. $\blacksquare$
%%%%%%%%%%%%%%%%%%%%%%%%%%%%%%%%%%%%%%%%%%
%%%%%%%%%%%%%%%%%%%%%%%%%%%%%%%%%%%%%%%%%%
%%%%%%%%%%%%%%%%%%%%%%%%%%%%%%%%%%%%%%%%%%

%%%%%%%%%%%%%%%%%%%%%%%%%
%%%%%%%%%%%%%%%%%%%%%%%%%%%%%%%%%%%%%%%%%%
%%%%%%%%%%%%%%%%%%%%%%%%%%%%%%%%%%%%%%%%%%
\subsection*{Proof of Proposition \ref{proposition:central_vs_local_allocation}}
We first state a more detailed version of Proposition \ref{proposition:central_vs_local_allocation}.
\addtocounter{proposition}{-1}
\begin{proposition}
Assume the virtual costs of users $1, \cdots, n-1$ are all equal to $\psi_1$. We also denote the virtual cost of user $n$ by $\psi_n$. Denote the optimal privacy loss levels in the central and local settings by $(\diff^\Central_1, \cdots, \diff^\Central_n)$ and $(\diff^\Local_1, \cdots, \diff^\Local_n)$, respectively. 
\begin{enumerate}
\item In the central setting, if $ \diff^\Central_n = 0$, then 
\begin{equation} \label{eqn:bound_Psi_central}
\psi_n \geq \psi_1 + \frac{1}{(n-1) \Var} \sqrt[3]{2 \psi_1 (n+1)^2}.    
\end{equation}
%%%%%%%%%%%%%%%%%%%%%%%
\item In the local setting, there exists a universal constant $\kappa$, independent of problem's parameters, such that, if 
\begin{equation} \label{eqn:bound_Psi_local}
\psi_n \geq \psi_1 + 
\kappa \left ( \frac{\psi^{1/3}}{n^{2/3} \Var} + \frac{\psi_1}{n} + \frac{\psi^{-1/3}}{n^{4/3} \Var^2 } \right ),    
\end{equation}
then $ \diff^\Local_n = 0$.
\end{enumerate}
Therefore, there exists $N \in \mathbb{N}$ such that for $n \ge N$ and
\begin{align}\label{eq:eqn:bound_Psi_local}
    \psi_n \in \left[ \psi_1 + 
\kappa \left ( \frac{\psi^{1/3}}{n^{2/3} \Var} + \frac{\psi_1}{n} + \frac{\psi^{-1/3}}{n^{4/3} \Var^2 } \right ), \psi_1 + \frac{1}{(n-1) \Var} \sqrt[3]{2 \psi_1 (n+1)^2} \right],
\end{align}
the optimal privacy loss level of user $n$ in the local setting is zero while her optimal privacy loss level in the central setting in non-zero.
\end{proposition}
%%%%%%%%%%
\noindent \textbf{Proof:} First, note that the optimal privacy loss levels in the central and local settings are the solutions of optimization problems \eqref{eq:thm:central} and \eqref{eq:thm:local}, respectively. We first note that in both cases, without loss of generality, we can assume $\Var = 1$ by replacing $y_i$ by $y_i \sqrt{\Var}$ and $\psi_i$ by $\psi_i \Var^{3/2}$. Therefore, without loss of generality, we could assume $\Var=1$ while studying optimization problems \eqref{eq:thm:central} and \eqref{eq:thm:local}.

We start with the central setting. The characterization of solutions \eqref{eq:central_solutions_KKT} implies that $\diff^\Central_1 = \cdots = \diff^\Central_{n-1}$.
Hence, the Lagrangian corresponding to optimization \eqref{eq:thm:central} can be rewritten as
\begin{equation}
L(y, z, \mu, \nu) := 
\frac{n+1}{\left ( (n-1) y + z \right )^2} \left ( 2 + (n-1) y^2 + z^2 \right ) 
+ (n-1) \psi_1 y + \psi_n z 
- \mu y - \nu z,  
\end{equation}
where $y$ and $z$ denote the central privacy loss level of the first $n-1$ users and the last user, respectively. 
If $ \diff^\Central_n = 0$, then there exists a tuple $(y^*, z^*, \mu^*, \nu^*)$ with $z^* = 0$ such that
\begin{align}
& \frac{\partial}{\partial y} L(y^*, z^*, \mu^*, \nu^*) = 0, \label{eqn:KKT_allocations_central_y} \\
& \frac{\partial}{\partial z} L(y^*, z^*, \mu^*, \nu^*) = 0. \label{eqn:KKT_allocations_central_z}
\end{align}
Furthermore, since $z^* = 0$ and the optimal cost is finite, we have $y^* >0$ which implies $\mu^* = 0$.
Next, note that, 
\begin{equation*}
\frac{\partial}{\partial y} L(y, z, \mu, \nu) = 
-2(n+1)(n-1)\frac{2+z^2-yz}{\left ((n-1)y+z \right)^3} + (n-1) \psi_1 - \mu.
\end{equation*}
Hence, \eqref{eqn:KKT_allocations_central_y} along with $z^* = 0$ and $\mu^* = 0$, implies
\begin{equation} \label{eqn:y_star_optimal_central}
y^* = \frac{1}{n-1} \sqrt[3]{\frac{4(n+1)}{\psi_1}}.    
\end{equation}
Also, note that
\begin{equation*}
\frac{\partial}{\partial z} L(y, z, \mu, \nu) =   
2(n+1) \frac{(n-1)yz- 2 -(n-1) y^2}{\left ((n-1)y+z \right)^3} + \psi_n - \nu.
\end{equation*}
Thus, \eqref{eqn:KKT_allocations_central_z} along with $z^* = 0$ and $\nu^* \geq 0$, implies
\begin{equation*}
\psi_n \geq 2(n+1) \frac{2 + (n-1) (y^*)^2}{(n-1)^3 (y^*)^3}    
\end{equation*}
Plugging \eqref{eqn:y_star_optimal_central} into this bound completes the proof of \eqref{eqn:bound_Psi_central}.

To show \eqref{eqn:bound_Psi_local}, it suffices to show that if $\psi_n > \psi_1$ and $\diff_n^\Local > 0$, then 
\begin{equation*}
\psi_n \leq \psi_1 + \mathcal{O}(1) 
\left ( \frac{\psi^{1/3}}{n^{2/3}} + \frac{\psi_1}{n} + \frac{1}{\psi^{1/3} n^{4/3}} \right ).    
\end{equation*}
To do so, first assume $\psi_n > \psi_1$ and $\diff_n^\Local > 0$. Then, by Lemma \ref{Cor:local:mono}, we know $\diff_n^\Local < \diff_i^\Local$ for any $i \in \{1, \cdots, n-1\}$.\footnote{It is worth mentioning that Lemma \ref{Cor:local:mono}, in fact, implies $\diff_n^\Local \leq \diff_i^\Local$. However, by reviewing its proof, one could see that the inequality should be strict, given the assumption $\diff_n^\Local > 0$.}
Next, by the characterization of solutions \eqref{eq:local_solutions_KKT}, we know $\diff^\Local_1 = \cdots = \diff^\Local_{n-1}$. With a slight abuse of notation, we denote the local privacy loss level of $n-1$ first users and the last user by $y$ and $z$, respectively, with $y^* > z^*$. 

From \eqref{eqn:local_alg_KKT_recursive} in the proof of Proposition \ref{Pro:approximate:local}, we know the following two equations hold
\begin{align}
\frac{4y^*}{\psi_1 (2+{y^*}^2)^2} &= \frac{1}{n+1} \left ( (n-1) \frac{{y^*}^2}{2+{y^*}^2} + \frac{{z^*}^2}{2+{z^*}^2} \right )^2, \label{eqn:KKT_allocations_local_y} \\
\frac{4z^*}{\psi_n (2+{z^*}^2)^2} &= \frac{1}{n+1} \left ( (n-1) \frac{{y^*}^2}{2+{y^*}^2} + \frac{{z^*}^2}{2+{z^*}^2} \right )^2. \label{eqn:KKT_allocations_local_z}  
\end{align}
We next provide upper and lower bounds on $y^*$. To do so, first, by replacing $\frac{{z^*}^2}{2+{z^*}^2}$ by $0$, we obtain
\begin{equation*}
 \frac{4y^*}{\psi_1 (2+{y^*}^2)^2}
 \geq \frac{(n-1)^2 {y^*}^4}{(n+1)(2+{y^*}^2)^2}
\end{equation*}
which implies
\begin{equation} \label{eqn:local_optimal_lower_bound}
y^* \leq \sqrt[3]{\frac{4(n+1)}{\psi_1 (n-1)^2}}.  
\end{equation}
Second, we note that $\frac{x^2}{x^2+2}$ is an increasing function of $x$ over $(0, \infty)$. Hence, given that $y^* > z^*$, replacing $\frac{{z^*}^2}{2+{z^*}^2}$ by $\frac{{y^*}^2}{2+{y^*}^2}$ leads to an upper bound for the right hand side of \eqref{eqn:KKT_allocations_local_y}. Therefore, we have
\begin{equation*}
 \frac{4y^*}{\psi_1 (2+{y^*}^2)^2}
 \leq \frac{n^2 {y^*}^4}{(n+1)(2+{y^*}^2)^2}
\end{equation*}
which implies
\begin{equation} \label{eqn:local_optimal_upper_bound}
y^* \geq \sqrt[3]{\frac{4(n+1)}{\psi_1 n^2}}.   
\end{equation}
Next, note that, we can rewrite \eqref{eqn:KKT_allocations_local_z} as
\begin{equation} \label{eqn:local_allocation_interim_1}
\frac{\sqrt{\frac{4(n+1)}{\psi_n}} \sqrt{z^*} - {z^*}^2}
{2+{z^*}^2} 
= (n-1) \frac{{y^*}^2}{{y^*}^2+2}.
\end{equation}
By replacing the left hand side by the upper bound 
\begin{equation*}
\frac{1}{2} \sqrt{\frac{4(n+1)}{\psi_n}} \sqrt{z^*},    
\end{equation*}
we obtain
\begin{equation} \label{eqn:local_allocation_interim_2}
\frac{1}{2} \sqrt{\frac{4(n+1)}{\psi_n}} \sqrt{z^*}
\geq (n-1) \frac{{y^*}^2}{{y^*}^2+2}.
\end{equation}
Next, we use the fact that $z^* < y^*$ and the upper bound on $y^*$ \eqref{eqn:local_optimal_upper_bound} to further upper bound the left hand side of \eqref{eqn:local_allocation_interim_2}. In addition, we use the lower bound on $y^*$ \ref{eqn:local_optimal_lower_bound} to lower bound the right hand side of \eqref{eqn:local_allocation_interim_2}. Taking these two steps and simplifying the equation leads to the following result:
\begin{equation} \label{eqn:local_allocation_interim_3}
\frac{(4(n+1))^{2/3}}{2 \psi_1^{1/6} (n-1)^{4/3}}
+ \sqrt{\psi_1} \frac{n^{4/3}}{(n-1)^{4/3}} \geq \sqrt{\psi_n}. 
\end{equation}
Using this inequality along with,
\begin{equation*}
n^{4/3} = (n-1)^{4/3} + \mathcal{O}(n^{1/3}),    
\end{equation*}
we obtain
\begin{equation} \label{eqn:local_allocation_interim_4}
\mathcal{O}(1) \left ( \frac{1}{\psi_1^{1/6} n^{2/3}} + \frac{\sqrt{\psi_1}}{n} \right) + \sqrt{\psi_1} 
\geq \sqrt{\psi_n}.
\end{equation}
Using the fact that $\sqrt{\psi_n} - \sqrt{\psi_1} = \frac{\psi_n - \psi_1}{\sqrt{\psi_1} + \sqrt{\psi_n}}$, we can rewrite \eqref{eqn:local_allocation_interim_4} as
\begin{equation} \label{eqn:local_allocation_interim_5}
\psi_1  + \mathcal{O}(1) (\sqrt{\psi_1} + \sqrt{\psi_n}) \left ( \frac{1}{\psi_1^{1/6} n^{2/3}} + \frac{\sqrt{\psi_1}}{n} \right)
\geq \psi_n.
\end{equation}
Upper bounding $\sqrt{\psi_n}$ on the left hand side of \eqref{eqn:local_allocation_interim_5} by using \eqref{eqn:local_allocation_interim_4} completes the proof. $\blacksquare$
\subsection{Additional results and details}\label{app:additional}
This appendix includes the detail of the discussions included in the main text.
%%%

%%%
\subsubsection*{Revelation principle for both central and local privacy settings}\label{app:Revelation:central}
 Suppose the strategy of user $i$ is a function of its privacy cost denoted by $\beta_i(c_i)$. For a given estimator $\hat{\theta}$ and mechanism $(\bm{\diff}, \mathbf{t})$, the action profile $\{\beta_i(\cdot)\}_{i=1}^n$ is an equilibrium if 
\begin{align*}
   &  \mathbb{E}_{\mathbf{c}_{-i}} \left[ \Var(\bf{\beta}_{-i}(\mathbf{c}_{-i}), \beta_i(c_i); \hat{\theta}) + c_i  \diff_i(\bf{\beta}_{-i}(\mathbf{c}_{-i}), \beta_i(c_i)) - t_i(\bf{\beta}_{-i}(\mathbf{c}_{-i}), \beta_i(c_i))\right]\\
    & \le \mathbb{E}_{\mathbf{c}_{-i}} \left[ \Var(\bf{\beta}_{-i}(\mathbf{c}_{-i}), \beta'_i(c_i);\hat{\theta}) + c_i \diff_i(\bf{\beta}_{-i}(\mathbf{c}_{-i}), \beta'_i(c_i)) - t_i(\bf{\beta}_{-i}(\mathbf{c}_{-i}), \beta'_i(c_i))\right]
\end{align*}
for all $i \in \mathcal{N}, c_i, \beta'_i(\cdot)$. 
By letting $(\tilde{\bm{\diff}}, \tilde{\mathbf{t}})$ be such that $\tilde{\diff}_i(c_1, \dots, c_n)= \diff_i(\beta_1(c_1), \dots, \beta_n(c_n))$ and $\tilde{t}_i(c_1, \dots, c_n)= t_i(\beta_1(c_1), \dots, \beta_n(c_n))$, the users will report truthfully and that the platform's objective is the same as the original mechanism. This establishes the revelation principle. $\blacksquare$
%%%%%%

{
\subsubsection*{Computing the payment function and approximate incentive compatibility}
Recall the incentive compatibility (IC) definition \eqref{eq:IC} states
\begin{equation*}
\Cost(c_i, c_i; \bm{\diff}, \mathbf{t}, \hat{\theta}) \le  \Cost(c'_i, c_i; \bm{\diff}, \mathbf{t}, \hat{\theta}) \quad \text{ for all } i \in \mathcal{N}, c_i, c'_i.    
\end{equation*}
The approximate $\epsilon$-IC definition allows for an $\epsilon$ violation of the original IC definition, i.e., 
\begin{equation} \label{eq:approx-IC}
\Cost(c_i, c_i; \bm{\diff}, \mathbf{t}, \hat{\theta}) \le  \Cost(c'_i, c_i; \bm{\diff}, \mathbf{t}, \hat{\theta}) + \epsilon \quad \text{ for all } i \in \mathcal{N}, c_i, c'_i. 
\end{equation}
The following result highlights that in both the central and local settings if we possess an algorithm that provides the estimator and privacy loss allocations for any given vector of privacy sensitivities, we can efficiently compute payment functions that satisfy $\epsilon$-IC. This means that the algorithm ensures approximate incentive compatibility with an error no greater than $\epsilon$.
\begin{lemma}\label{Lemma:Paymentcompute}
Suppose we have an algorithm that returns the estimator and the privacy loss levels for any given vector of privacy sensitivities. Then, for any $\epsilon$, we can return payment functions in polynomial time such that $\epsilon$-IC holds.
\end{lemma}
\emph{Proof of Lemma \ref{Lemma:Paymentcompute}:} Recall that the payment function is
\begin{equation*}
t_i(\mathbf{c}) = \MSE(\mathbf{c},\bm{\diff},  \hat{\theta}) - \Var + c_i \diff_i(\mathbf{c}) +  \int_{z=c_i}^{\infty} \diff_i(z,\mathbf{c}_{-i}) dz.
\end{equation*}
All the terms, except the integral, can be computed based on the algorithm's output on the estimator and privacy allocations for the vector $\mathbf{c}$. The last step is to show that we can approximate the integral efficiently. To do so, we establish that, for any $\mathbf{c}_{-i}$ and $\delta$, there exists $\bar{c}_i$ such that
\begin{equation*}
\int_{z=\bar{c}_i}^{\infty} \diff_i(z,\mathbf{c}_{-i}) dz  \leq \delta.  
\end{equation*}
To show this, first note that, even as $c_i$ increases, there is a fixed upper bound $M$ on the platform's cost in the optimization problems stated in Theorems \ref{thm:central} and \ref{thm:local}, as the platform can always ignore the data of user $i$, i.e., put $y_i = 0$ in \eqref{eq:thm:central} and \eqref{eq:thm:local}. As a result, we have
\begin{equation*}
\diff_i(z,\mathbf{c}_{-i}) \leq \frac{M}{\psi_i(z)}
 = \frac{M}{z+\frac{F_i(z)}{f_i(z)}} 
 \leq \frac{Mf_i(z)}{F_i(z)}.
\end{equation*}
Note that, for any $\delta'$, there exists $\bar{c}_i$ such that $F_i(\bar{c}_i) \geq 1-\delta'$. Therefore, we have
\begin{align*}
\int_{z=\bar{c}_i}^{\infty} \diff_i(z,\mathbf{c}_{-i}) dz
& \leq \int_{z=\bar{c}_i}^{\infty} \frac{Mf_i(z)}{F_i(z)} dz \\
& \leq \int_{z=\bar{c}_i}^{\infty} \frac{Mf_i(z)}{1-\delta'} dz \\
& \frac{M}{1-\delta'} \int_{z=\bar{c}_i}^{\infty} f_i(z) 
= \frac{M}{1-\delta'} \left (1-F_i(\bar{c}_i) \right )  
\leq \frac{M}{1-\delta'} \delta'.
\end{align*}
Setting $\delta'$ gives us the desired result.
 Now, suppose that we want an $\epsilon$-approximate of the integral. Given the above result, we can choose $\bar{c}_i$ such that 
\begin{equation*}
\int_{z=\bar{c}_i}^{\infty} \diff_i(z,\mathbf{c}_{-i}) dz  \leq \frac{\epsilon}{2}.  
\end{equation*}
Therefore, it suffices to show that we can approximate $\int_{z=c_i}^{\bar{c}_i} \diff_i(z,\mathbf{c}_{-i}) dz $ up to $\epsilon/2$ error. For any $\delta$, let $\mathcal{S}(\delta)$ be a mesh with sub-intervals of size maximum $\delta$ from $c_i$ to $\bar{c_i}$, i.e., 
\begin{equation*}
\mathcal{S}(\delta) = (I_0, I_1, \cdots, I_M),     
\end{equation*}
where 
\begin{equation*}
c_i = I_0 < I_1 < \cdots < I_M = \bar{c}_i, 
\quad \text{with } I_j - I_{j-1} \leq \delta.
\end{equation*}
Note that we have
\begin{equation*}
\sum_{j=1}^{M} (I_j - I_{j-1})\inf_{z \in {[I_{j-1}, I_{j}]}} \diff_i(z,\mathbf{c}_{-i})
\leq 
\int_{z=c_i}^{\bar{c}_i} \diff_i(z,\mathbf{c}_{-i}) dz
\leq 
\sum_{j=1}^{M} (I_j - I_{j-1}) \sup_{z \in {[I_{j-1}, I_{j}]}} \diff_i(z,\mathbf{c}_{-i}).
\end{equation*}
Notice that, as shown earlier, the optimal $\diff_i(z,\mathbf{c}_{-i})$ is decreasing in $z$ for a fixed $\mathbf{c}_{-i}$. Thus, we can rewrite the above equation as
\begin{equation*}
\sum_{j=1}^{M} (I_j - I_{j-1}) \diff_i(I_{j},\mathbf{c}_{-i})
\leq 
\int_{z=c_i}^{\bar{c}_i} \diff_i(z,\mathbf{c}_{-i}) dz
\leq 
\sum_{j=1}^{M} (I_j - I_{j-1}) \diff_i(I_{j-1},\mathbf{c}_{-i}).
\end{equation*}
Therefore, if we take $\sum_{j=0}^{M-1} (I_j - I_{j-1}) \diff_i(I_{j},\mathbf{c}_{-i})$ as an approximate of the integral, its error will be bounded by the difference of the left-hand side and the right-hand side, i.e.,
\begin{align*}
\sum_{j=1}^{M} (I_j - I_{j-1}) \diff_i(I_{j-1},\mathbf{c}_{-i}) 
& - \int_{z=c_i}^{\bar{c}_i} \diff_i(z,\mathbf{c}_{-i}) dz 
\leq 
\sum_{j=1}^{M} (I_j - I_{j-1}) \left ( \diff_i(I_{j-1},\mathbf{c}_{-i}) - \diff_i(I_{j},\mathbf{c}_{-i}) \right ) \\
& \le \delta \sum_{j=1}^{M} \left ( \diff_i(I_{j-1},\mathbf{c}_{-i}) - \diff_i(I_{j},\mathbf{c}_{-i}) \right )
= \delta \left ( \diff_i(c_i,\mathbf{c}_{-i}) - \diff_i(\bar{c}_i,\mathbf{c}_{-i}) \right ).
\end{align*}
As a result, by letting
\begin{equation*}
\delta = \frac{\epsilon}{2\left ( \diff_i(c_i,\mathbf{c}_{-i}) - \diff_i(\bar{c}_i,\mathbf{c}_{-i}) \right )},    
\end{equation*}
the exprerssion $\sum_{j=0}^{M-1} (I_j - I_{j-1}) \diff_i(I_{j},\mathbf{c}_{-i})$ becomes an $\epsilon/2$-approximate of the integral. Also, computing this sum requires solving the allocation problem $M$ times, where each time can be done in polynomial time. Finally, notice that $M$ is less than $\lceil (\bar{c}_i -c_i)/\delta \rceil$ which is order of $\Omega(\frac{1}{\epsilon})$. Hence, the above procedure establishes an FPTAS for finding the payment. $\blacksquare$
}

%%%%%
{

\subsection*{Alternative individual rationality constraint}
Here, we consider an alternative individual rationality constraint to \eqref{eq:IR} in which the users benefit from the platform's estimator even if they do not participate. In this case, constraint \eqref{eq:IR} becomes
\begin{align}\label{eq:IR:updated}
    \Cost(c_i, c_i; \bm{\diff}^{(n)}, \mathbf{t}^{(n)}, \hat{\theta}^{(n)}) \le \mathbb{E}_{\mathbf{c}_{-i}}\left[ \MSE( \mathbf{c}_{-i}, \bm{\epsilon}^{(n-1)}_{-i}, \hat{\theta}^{(n-1)}) \right],
\end{align}
where the right-hand side is the MSE of an estimator with $n-1$ data points of users in $\mathcal{N} \setminus \{i\}$, noting that without participating in the mechanism, the user does not incur any privacy cost but also does not receive any payment. Here, we use the superscript $(k)$ to show that a function has $k$ inputs. 

We next highlight how each one of our results extends to this setting. We do not repeat the proofs as they are identical to those presented earlier.

\textbf{Proposition 1':} 
For a given estimator $\hat{\theta}: \mathcal{X}^n \times \mathbb{R}_+^n  \to \mathbb{R}$ defined for all $n$, a central or local privacy data acquisition mechanism $(\hat{\theta}^{(n)}, \bm{\diff}^{(n)}, \mathbf{t}^{(n)})$ satisfies incentive compatibility \eqref{eq:IC} and individual rationality \eqref{eq:IR:updated} if and only if 
\begin{align}\label{eq:Pro:PaymentIdentity:updated}
    t_i(c_i)= \mathbb{E}_{\mathbf{c}_{-i}}\left[\MSE(\mathbf{c},\bm{\diff}^{(n)},  \hat{\theta}^{(n)}) \right] -  \mathbb{E}_{\mathbf{c}_{-i}}\left[\MSE(\mathbf{c}_{-i},\bm{\diff}^{(n-1)}_{-i},  \hat{\theta}^{(n-1)}) \right]+  c_i \diff_i(c_i) +   \int_{z=c_i}^{\infty} \diff_i^{(n)}(z) dz + d_i,
\end{align}
for some constant $d_i \ge 0$, and $\diff^{(n)}_i(z)$ is weakly decreasing) in $z$ for all $i \in \mathcal{N}$.

\textbf{Proposition 2':} For a given estimator $\hat{\theta}: \mathcal{X}^n \times \mathbb{R}_+^n \to \mathbb{R}$ defined for all $n$, the optimal privacy loss in the central or local privacy data acquisition mechanism is the solution of 
\begin{align*} 
    \min_{\{\diff_i^{(n)}(\cdot)\}_{i=1}^n}  ~~~&  \mathbb{E}_{\mathbf{c}}\left[ (n+1)\MSE(\mathbf{c},\bm{\diff}^{(n)}, \hat{\theta}^{(n)})  + \sum_{i=1}^n \diff_i(\mathbf{c}) \psi_i(c_i) \right] - \sum_{i=1}^n  \mathbb{E}_{\mathbf{c}_{-i}}  \left[ \MSE(\mathbf{c}_{-i},\bm{\diff}^{(n-1)}_{-i},  \hat{\theta}^{(n-1)}) \right] \\
    & \diff^{(n)}_i(z)=\mathbb{E}_{\mathbf{c}_{-i}}\left[\diff^{(n)}_i(z, \mathbf{c}_{-i}) \right] \text{ is weakly decreasing in } z \text{ for all } i \in \mathcal{N},
\end{align*}
where $\epsilon^{(n-1)}_{j}$ for all $i \in \mathcal{N}$ and $j \in \mathcal{N} \setminus \{i\}$ is the optimal privacy loss levels for users in $\mathcal{N} \setminus \{i\}$.

From the above proposition, it is evident that finding the optimal $\{\diff_i^{(n)}(\cdot)\}_{i=1}^n$ decouples from finding the optimal $\{\diff_i^{(k)}(\cdot)\}_{i}$ for any other $k<n$. Therefore, for both central and local settings, the characterization of the optimal privacy levels is the same as the ones given in Theorems \ref{thm:central} and \ref{thm:local}, respectively. This in turn implies that Propositions \ref{Pro:approximate:central} and \ref{Pro:approximate:local} continue to hold. The only difference between this setting and our baseline model is that here in order to compute the payments, one needs to solve for the privacy loss levels for both $n$ users and any subset of $n-1$ users. After solving these $ n+1$ optimization problems, we can then use Proposition 1' to obtain the payment for $n$ users.

}

\end{document}